\documentclass[journal]{IEEEtran}
\usepackage{graphicx,color}
\usepackage{latexsym}
\usepackage{amssymb}
\usepackage{amsmath,bm}
\usepackage{array}

\begin{document}

%%%%%%%Theorems%%%%%%%%%%%%%
%\theorembodyfont{\rmfamily}
%
%\newtheorem{theorem}{Theorem}
%\newtheorem{lemma}[theorem]{Lemma}
%\newtheorem{definition}[theorem]{Definition}
%\newtheorem{corollary}[theorem]{Corollary}
%\newtheorem{remark}[theorem]{Remark}
%\newtheorem{proposition}[theorem]{Proposition}
%
%%%%%%%%QED%%%%%%%%%%%%
\newcommand{\qed}{\hfill$\square$}
\newcommand{\suchthat}{\mbox{~s.t.~}}
\newcommand{\markov}{\leftrightarrow}
%%%%%%%%%proof environment%%%%%%%%
\newenvironment{pRoof}{%
 \noindent{\em Proof.\ }}{%
 \hspace*{\fill}\qed \\
 \vspace{2ex}}
%vectors

%\DeclareMathAlphabet{\bM}{OML}{cmm}{b}{it}

%%%%%%%%%%%%bracket%%%%%%%%%%%
\newcommand{\ket}[1]{| #1 \rangle}
\newcommand{\bra}[1]{\langle #1 |}
\newcommand{\bol}[1]{\mathbf{#1}}
\newcommand{\rom}[1]{\mathrm{#1}}
\newcommand{\san}[1]{\mathsf{#1}}
\newcommand{\mymid}{:~}
%\renewcommand{\liminf}{\underline{\lim}}
%\renewcommand{\limsup}{\overline{\lim}}
%%%%%%% argmax argmin %%%%%%%%%%%%
\newcommand{\argmax}{\mathop{\rm argmax}\limits}
\newcommand{\argmin}{\mathop{\rm argmin}\limits}

\newcommand{\Cls}{class NL}
\newcommand{\vSpa}{\vspace{0.3mm}}
\newcommand{\Prmt}{\zeta}
\newcommand{\pj}{\omega_n}

\newfont{\bg}{cmr10 scaled \magstep4}
\newcommand{\bigzerol}{\smash{\hbox{\bg 0}}}
\newcommand{\bigzerou}{\smash{\lower1.7ex\hbox{\bg 0}}}
\newcommand{\nbn}{\frac{1}{n}}
\newcommand{\ra}{\rightarrow}
\newcommand{\la}{\leftarrow}
\newcommand{\ldo}{\ldots}
\newcommand{\typi}{A_{\epsilon }^{n}}
\newcommand{\bx}{\hspace*{\fill}$\Box$}
\newcommand{\pa}{\vert}
\newcommand{\ignore}[1]{}

%
%\input{ieeecom2e.tex}
%
%---------------------------- ieeecom2e.tex--------------------------
%
%
\newcommand{\bc}{\begin{center}}  %
\newcommand{\ec}{\end{center}}
\newcommand{\befi}{\begin{figure}[h]}  %
\newcommand{\enfi}{\end{figure}}
\newcommand{\bsb}{\begin{shadebox}\begin{center}}   %
\newcommand{\esb}{\end{center}\end{shadebox}}
\newcommand{\bs}{\begin{screen}}     %
\newcommand{\es}{\end{screen}}
\newcommand{\bib}{\begin{itembox}}   %
\newcommand{\eib}{\end{itembox}}
\newcommand{\bit}{\begin{itemize}}   %
\newcommand{\eit}{\end{itemize}}
\newcommand{\defeq}{:=}

\newcommand{\Qed}{\hbox{\rule[-2pt]{3pt}{6pt}}}
\newcommand{\beq}{\begin{equation}}
\newcommand{\eeq}{\end{equation}}
\newcommand{\beqa}{\begin{eqnarray}}
\newcommand{\eeqa}{\end{eqnarray}}
\newcommand{\beqno}{\begin{eqnarray*}}
\newcommand{\eeqno}{\end{eqnarray*}}
\newcommand{\balno}{\begin{align*}}
\newcommand{\ealno}{\end{align*}}

\newcommand{\ba}{\begin{array}}
\newcommand{\ea}{\end{array}}
\newcommand{\vc}[1]{\mbox{\boldmath $#1$}}
\newcommand{\lvc}[1]{\mbox{\scriptsize \boldmath $#1$}}
\newcommand{\svc}[1]{\mbox{\scriptsize\boldmath $#1$}}

\newcommand{\wh}{\widehat}
\newcommand{\wt}{\widetilde}
\newcommand{\ts}{\textstyle}
\newcommand{\ds}{\displaystyle}
\newcommand{\scs}{\scriptstyle}
\newcommand{\vep}{\varepsilon}
\newcommand{\rhp}{\rightharpoonup}
\newcommand{\cl}{\circ\!\!\!\!\!-}
\newcommand{\bcs}{\dot{\,}.\dot{\,}}
\newcommand{\eqv}{\Leftrightarrow}
\newcommand{\leqv}{\Longleftrightarrow}
\newtheorem{co}{Corollary} 
\newtheorem{lm}{Lemma} 
\newtheorem{Ex}{Example} 
\newtheorem{Th}{Theorem}
\newtheorem{df}{Definition} 
\newtheorem{pr}{Property} 
\newtheorem{pro}{Proposition} 
\newtheorem{rem}{Remark} 

\newcommand{\lcv}{convex } 

\newcommand{\hugel}{{\arraycolsep 0mm
                    \left\{\ba{l}{\,}\\{\,}\ea\right.\!\!}}
\newcommand{\Hugel}{{\arraycolsep 0mm
                    \left\{\ba{l}{\,}\\{\,}\\{\,}\ea\right.\!\!}}
\newcommand{\HUgel}{{\arraycolsep 0mm
                    \left\{\ba{l}{\,}\\{\,}\\{\,}\vspace{-1mm}
                    \\{\,}\ea\right.\!\!}}
\newcommand{\huger}{{\arraycolsep 0mm
                    \left.\ba{l}{\,}\\{\,}\ea\!\!\right\}}}

\newcommand{\Huger}{{\arraycolsep 0mm
                    \left.\ba{l}{\,}\\{\,}\\{\,}\ea\!\!\right\}}}

\newcommand{\HUger}{{\arraycolsep 0mm
                    \left.\ba{l}{\,}\\{\,}\\{\,}\vspace{-1mm}
                    \\{\,}\ea\!\!\right\}}}

\newcommand{\hugebl}{{\arraycolsep 0mm
                    \left[\ba{l}{\,}\\{\,}\ea\right.\!\!}}
\newcommand{\Hugebl}{{\arraycolsep 0mm
                    \left[\ba{l}{\,}\\{\,}\\{\,}\ea\right.\!\!}}
\newcommand{\HUgebl}{{\arraycolsep 0mm
                    \left[\ba{l}{\,}\\{\,}\\{\,}\vspace{-1mm}
                    \\{\,}\ea\right.\!\!}}
\newcommand{\hugebr}{{\arraycolsep 0mm
                    \left.\ba{l}{\,}\\{\,}\ea\!\!\right]}}
\newcommand{\Hugebr}{{\arraycolsep 0mm
                    \left.\ba{l}{\,}\\{\,}\\{\,}\ea\!\!\right]}}

\newcommand{\HugebrB}{{\arraycolsep 0mm
                    \left.\ba{l}{\,}\\{\,}\vspace*{-1mm}\\{\,}\ea\!\!\right]}}

\newcommand{\HUgebr}{{\arraycolsep 0mm
                    \left.\ba{l}{\,}\\{\,}\\{\,}\vspace{-1mm}
                    \\{\,}\ea\!\!\right]}}

\newcommand{\hugecl}{{\arraycolsep 0mm
                    \left(\ba{l}{\,}\\{\,}\ea\right.\!\!}}
\newcommand{\Hugecl}{{\arraycolsep 0mm
                    \left(\ba{l}{\,}\\{\,}\\{\,}\ea\right.\!\!}}
\newcommand{\hugecr}{{\arraycolsep 0mm
                    \left.\ba{l}{\,}\\{\,}\ea\!\!\right)}}
\newcommand{\Hugecr}{{\arraycolsep 0mm
                    \left.\ba{l}{\,}\\{\,}\\{\,}\ea\!\!\right)}}

\newcommand{\hugepl}{{\arraycolsep 0mm
                    \left|\ba{l}{\,}\\{\,}\ea\right.\!\!}}
\newcommand{\Hugepl}{{\arraycolsep 0mm
                    \left|\ba{l}{\,}\\{\,}\\{\,}\ea\right.\!\!}}
\newcommand{\hugepr}{{\arraycolsep 0mm
                    \left.\ba{l}{\,}\\{\,}\ea\!\!\right|}}
\newcommand{\Hugepr}{{\arraycolsep 0mm
                    \left.\ba{l}{\,}\\{\,}\\{\,}\ea\!\!\right|}}

\newcommand{\MEq}[1]{\stackrel{%\mbox
{\rm (#1)}}{=}}

\newcommand{\MLeq}[1]{\stackrel{%\mbox
{\rm (#1)}}{\leq}}

\newcommand{\ML}[1]{\stackrel{%\mbox
{\rm (#1)}}{<}}

\newcommand{\MGeq}[1]{\stackrel{%\mbox
{\rm (#1)}}{\geq}}

\newcommand{\MG}[1]{\stackrel{%\mbox
{\rm (#1)}}{>}}

\newcommand{\MPreq}[1]{\stackrel{%\mbox
{\rm (#1)}}{\preceq}}

\newcommand{\MSueq}[1]{\stackrel{%\mbox
{\rm (#1)}}{\succeq}}

\newcommand{\MSubeq}[1]{\stackrel{%\mbox
{\rm (#1)}}{\subseteq}}

\newcommand{\MSupeq}[1]{\stackrel{%\mbox
{\rm (#1)}}{\supseteq}}
%                                                           %
%-----------------------------------------------------------%
%                                                           %

\newenvironment{jenumerate}
	{\begin{enumerate}\itemsep=-0.25em \parindent=1zw}{\end{enumerate}}
\newenvironment{jdescription}
	{\begin{description}\itemsep=-0.25em \parindent=1zw}{\end{description}}
\newenvironment{jitemize}
	{\begin{itemize}\itemsep=-0.25em \parindent=1zw}{\end{itemize}}
\renewcommand{\labelitemii}{$\cdot$}

\newcommand{\iro}[2]{{\color[named]{#1}#2\normalcolor}}
\newcommand{\irr}[1]{{\color[named]{Black}#1\normalcolor}}
\newcommand{\irg}[1]{{\color[named]{Green}#1\normalcolor}}
\newcommand{\irb}[1]{{\color[named]{Blue}#1\normalcolor}}
\newcommand{\irBl}[1]{{\color[named]{Black}#1\normalcolor}}
\newcommand{\irWh}[1]{{\color[named]{White}#1\normalcolor}}

\newcommand{\irY}[1]{{\color[named]{Yellow}#1\normalcolor}}
\newcommand{\irO}[1]{{\color[named]{Orange}#1\normalcolor}}
\newcommand{\irBr}[1]{{\color[named]{Purple}#1\normalcolor}}
\newcommand{\IrBr}[1]{{\color[named]{Purple}#1\normalcolor}}
\newcommand{\irBw}[1]{{\color[named]{Brown}#1\normalcolor}}
\newcommand{\irPk}[1]{{\color[named]{Magenta}#1\normalcolor}}
\newcommand{\irCb}[1]{{\color[named]{CadetBlue}#1\normalcolor}}
%\newcommand{\irDg}[1]{{\color[named]{DarkSlateGray}#1\normalcolor}}

%
%-----------------indention environment-------------------%
\newenvironment{indention}[1]{\par
\addtolength{\leftskip}{#1}\begingroup}{\endgroup\par}
%form: \begin{indention}{2.3cm}
%      \end{indention}
%---------------------------------------------------------%
%
%----------------- namelist environment-------------------%
\newcommand{\namelistlabel}[1]{\mbox{#1}\hfill} 
\newenvironment{namelist}[1]{%
\begin{list}{}
{\let\makelabel\namelistlabel
\settowidth{\labelwidth}{#1}
\setlength{\leftmargin}{1.1\labelwidth}}
}{%
\end{list}}
%form: \begin{namelist}{width}
%---------------------------------------------------------%
%
%\def\BibTeX{{\rm B\kern-.05em{\sc i\kern-.025em b}\kern-.08em
%    T\kern-.1667em\lower.7ex\hbox{E}\kern-.125emX}}
%
%\newcommand{\bfig}{\begin{figure}[f]}
%\newcommand{\efig}{\end{figure}}
%\setcounter{page}{-41}
\newcommand{\bfig}{\begin{figure}[t]}
\newcommand{\efig}{\end{figure}}
\setcounter{page}{1}

\newtheorem{theorem}{Theorem}

\newcommand{\ep}{\mbox{\rm e}}

\newcommand{\Exp}{\exp%_2
}

\newcommand{\prmtA}{\mu}
\newcommand{\prmtB}{\bar{\mu}}

\newcommand{\idenc}{{\varphi}_n}
\newcommand{\resenc}{%\tilde
{\varphi}_n}
\newcommand{\ID}{\mbox{\scriptsize ID}}
\newcommand{\TR}{\mbox{\scriptsize TR}}
\newcommand{\Av}{\mbox{\sf E}}

\newcommand{\Vl}{|}
\newcommand{\Ag}{(R,P_{X^n}|W^n)}
\newcommand{\Agv}[1]{({#1},P_{X^n}|W^n)}
\newcommand{\Avw}[1]{({#1}|W^n)}

\newcommand{\Jd}{X^nY^n}
\newcommand{\IdR}{r_n}

\newcommand{\Index}{{n,i}}

\newcommand{\cid}{C_{\mbox{\scriptsize ID}}}
\newcommand{\cida}{C_{\mbox{{\scriptsize ID,a}}}}

\newcommand{\OMega}
{\Omega^{(\mu,\alpha)}}

\newcommand{\ARgRv}{(p^{(n)},\underline{Q}^n)}

\newcommand{\loF}{\underline{F}}

\newcommand{\pmt}{\beta}

\arraycolsep 0.5mm
\date{}
%
% paper title
\title{
Exponent Function for One Helper Source 
Coding Problem at Rates outside the Rate Region
}
\author{%
%Research Memo by 
Yasutada Oohama %and Shun Watanabe
% Yasutada~Oohama,~
%\IEEEmembership{Member,~IEEE,}
\thanks{
Y. Oohama is with 
University of Electro-Communications,
1-5-1 Chofugaoka Chofu-shi, Tokyo 182-8585, Japan.
%Manuscript received xxx, 20XX; revised xxx, 20XX.
}% 
%\thanks{
%S. Watanabe is with University of Tokushima, 
%2-1 Minami Josanjima-Cho, Tokushima 770-8506, Japan.
%}
}
\markboth{
%IEEE Transactions on Information Theory,~Vol.~XX,No.~Y, 
%~Month~200X
}
{
%Oohama: 
%Converse Coding Theorems for Identification via Channels  
}
% If you want to put a publisher's ID mark on the page
% (can leave text blank if you just want to see how the
% text height on the first page will be reduced by IEEE)
%\pubid{0000--0000/00\$00.00~\copyright~2002 IEEE}
% make the title area
\maketitle

\begin{abstract}
We consider the one helper source coding problem posed and investigated 
by Ahlswede, K\"orner and Wyner. Two correlated sources are separately 
encoded and are sent to a destination where the decoder wishes to decode 
one of the two sources with an arbitrary small error probability of 
decoding.  In this system, the error probability of decoding goes to one 
as the source block length $n$ goes to infinity. This implies that we 
have a strong converse theorem for the one helper source coding problem. 
 In this paper we provide the much stronger version of this strong 
converse theorem for the one helper source coding problem. We prove that 
the error probability of decoding tends to one exponentially and derive 
an explicit lower bound of this exponent function.
\end{abstract}

\section{Introduction}

We consider the one helper source coding problem posed and investigated 
by Ahlswede, K\"orner and Wyner. Two correlated sources are separately 
encoded and are sent to a destination where the decoder wishes to decode 
one of the two sources with an arbitrary small error probability of 
decoding. In this system, the error probability of decoding goes to one 
as the source block length $n$ goes to infinity. This implies that we 
have a strong converse theorem for the one helper source coding problem. 
In this paper we provide the much stronger version of this strong 
converse theorem for the one helper source coding problem. We prove that 
the error probability of decoding tends to one exponentially and derive 
an explicit lower bound of this exponent function.

\section{Problem Formulation}

Let ${\cal X}$ and ${\cal Y}$ be finite sets and 
$\left\{(X_{t},Y_{t})\right\}_{t=1}^{\infty}$ be a stationary 
discrete memoryless source. For each $t=1,2,\cdots$, the random 
pair $(X_{t},Y_{t})$ takes values in ${\cal X}\times {\cal Y}$, 
and has a probability distribution  
$$
p_{XY}=\left\{p_{XY}(x,y)\right\}_{(x,y)\in {\cal X} \times {\cal Y}}
$$ 
We write $n$ independent copies of 
$\left\{X_{t}\right\}_{t=1}^{\infty}$ and 
$\left\{Y_{t}\right\}_{t=1}^{\infty}$, 
respectively as
$$
X^{n}=X_1,X_2,\cdots,X_{n}
\mbox{ and }Y^{n}= Y_1,Y_2,\cdots,Y_{n}.
$$
We consider a communication system depicted in Fig. 1.
Data sequences $X^{n}$ and $Y^{n}$ are separately 
encoded to 
$\varphi_1^{(n)}(X^{n})$ and $\varphi_2^{(n)}(Y^{n})$ 
and those are sent to the information processing center.
At the center the decoder function $\psi^{(n)}$ observes 
$(\varphi_1^{(n)}(X^{n}),\varphi_2^{(n)}(Y^{n}))$ to output 
the estimation $\hat{Y}^{n}$ of ${Y}^{n}$. The encoder 
functions $\varphi_1^{(n)}$ and $\varphi_2^{(n)}$ are defined by
\beq
\left.
\ba{l}
\varphi_1^{(n)}:{\cal X}^{n}\to {\cal M}_1
=\left\{\,1,2,\cdots, M_1\,\right\},
\vspace{2mm}\\
\varphi_2^{(n)}:{\cal Y}^{n}\to{\cal M}_2
=\left\{\,1,2,\cdots, M_2\,\right\},
\ea
\right\}
\label{eqn:defen1} 
\eeq

where for each $i=1,2$, $\| \varphi_i^{(n)}\|$ $(=M_i)$
stands for the range of cardinality of $\varphi_i^{(n)}$. 
The decoder function $\psi^{(n)}$ is defined by
\begin{equation}
\psi^{(n)}:{\cal M}_1 \times {\cal M}_2 
\,\to\,{\cal Y}^{n}.
\end{equation}
The error probability of decoding is 
\begin{equation}
{\rm P}_{\rm e}^{(n)}(\varphi_1^{(n)},\varphi_2^{(n)},\psi^{(n)})
=\Pr\left\{\hat{Y}^{n}\neq Y^{n}\right\}, 
\end{equation}
where $\hat{Y}^{n}
=\psi^{(n)}(
\varphi_1^{(n)}(X^{n}),
\varphi_2^{(n)}(Y^{n}))$.
%%%%%%%%%%%%%%%%%%%%%%%%%%%%%%%%%%%%%%%%%%%%%%%%%%%%%%%%
A rate pair $(R_1,R_2)$ is $\varepsilon$-{\it achievable} if 
for any $\delta>0$, there exist a positive integer 
$n_0=n_0(\varepsilon,\delta)$ and a sequence of triples 
$\{(\varphi_1^{(n)},$ $\varphi_2^{(n)},$ 
$\psi^{(n)})\}_{n\geq n_0}$ such that for $n \geq n_0$,
\begin{align*}
& \frac{1}{n}\log \| \varphi_i^{(n)}\| \leq R_i+\delta 
\mbox{ for }i=1,2, 
\\
& {\rm P}_{\rm e}^{(n)}
(\varphi_1^{(n)},\varphi_2^{(n)},\psi^{(n)})\leq \varepsilon.
\end{align*}

%\end{document}

For $\varepsilon \in (0,1)$, 
the rate region ${\cal R}_{\rm AKW }(\varepsilon| p_{XY})$ is defined by 
\begin{align*}
& {\cal R}_{\rm AKW}(\varepsilon| p_{XY})
\\
& \defeq \left\{\,(R_1,R_2):(R_1,R_2)\,
\mbox{ is $\varepsilon$-achievable for }p_{XY}\,\right\}.
\end{align*}
Furthermore, define
$$
{\cal R}_{\rm AKW}(p_{XY})
\defeq \bigcap_{\varepsilon\in (0,1)}{\cal R}_{\rm AKW}(\varepsilon| p_{XY}).
$$
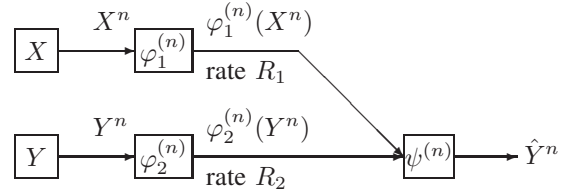
\begin{figure}[t]
\setlength{\unitlength}{0.94mm}

\begin{picture}(84,35)(4,4)
\put(10,25){\framebox(6,6){$X$}}
\put(10,10){\framebox(6,6){$Y$}}
%\put(21,46){${X^n}$}
%\put(16,43){\line(1,0){34}}
\put(21,31){${X^n}$}
\put(16,28){\vector(1,0){11}}
\put(21,16){${Y^n}$}
\put(16,13){\vector(1,0){11}}

\put(27,25){\framebox(8,6){$\varphi_1^{(n)}$}}
\put(37,31){$\varphi_1^{(n)}({X^n})$}
\put(37,24){rate $R_1$}
\put(27,10){\framebox(8,6){$\varphi_2^{(n)}$}}
\put(37,16){$\varphi_2^{(n)}({Y^n})$}
\put(37,9){rate $R_2$}

%\put(50,43){\vector(1,-1){15}}
\put(50,28){\vector(1,-1){15}}

\put(35,28){\line(1,0){15}}
\put(35,13){\vector(1,0){30}}

%\put(65,25){\framebox(6,6){$\psi_Y$}}
\put(65,10){\framebox(7,6){$\psi^{(n)}$}}

%\put(71,28){\vector(1,0){10}}
%\put(82,27){$\hat{Y}^n$}
\put(72,13){\vector(1,0){9}}
\put(82,12){$\hat{Y}^n$}

\end{picture}
%\begin{center}
%\caption
%Fig. 1.  
\caption{
One helper source coding system.
}
\label{fig:Zigzag}
%\end{center}
\noindent
\end{figure}
%
%
%\begin{theorem}
We can show that the two rate regions 
${\cal R}_{\rm AKW }( \varepsilon |$ $p_{XY})$, 
$\varepsilon \in(0,1)$ and 
${\cal R}_{\rm AKW }(p_{XY})$ 
satisfy the following property. 
%===================================================================%
%===================================================================%
\begin{pr}\label{pr:pro0aa}
$\quad$
\begin{itemize}
\item[a)] 
The regions 
${\cal R}_{\rm AKW }(\varepsilon|p_{XY})$, 
$\varepsilon \in(0,1)$, 
and ${\cal R}_{\rm AKW }($ $p_{XY})$
are closed convex sets of 
$\mathbb{R}_{+}^2$, where
\begin{align*}
\mathbb{R}_{+}^2&\defeq \{(R_1,R_2): R_1\geq 0, R_2\geq 0\}.
\end{align*}
\item[b)] 
${\cal R}_{\rm AKW}(\varepsilon|p_{XY})$ has another 
form using $(n,\vep)$-rate region 
${\cal R}_{\rm AKW}(n,\varepsilon |p_{XY})$,
the definition of which is as follows.
We set 
\begin{align*}
& {\cal R}_{\rm AKW}(n,\varepsilon |p_{XY})
=\{(R_1,R_2): %\mbox{ There exists }
\\
& \mbox{ There exists }(\varphi_1^{(n)},\varphi_2^{(n)},\psi^{(n)})
\mbox{ such that }
\\
& \frac{1}{n}\log ||\varphi_i^{(n)}||\leq R_i, i=1,2, 
\\
& {\rm P}_{\rm e}^{(n)}(\varphi_1^{(n)},\varphi_2^{(n)},\psi^{(n)})
\leq \varepsilon\}.
\end{align*} 
Using ${\cal R}_{\rm AKW}(n,$ $\varepsilon| p_{XY})$, 
${\cal R}_{\rm AKW}(\varepsilon|p_{XY})$ can be 
expressed as 
\begin{align*}
%&&
{\cal R}_{\rm AKW}(\varepsilon | p_{XY})
%\\
&={\rm cl}\left(\bigcup_{m\geq 1}
\bigcap_{n \geq m}{\cal R}_{\rm AKW}(n,\varepsilon | p_{XY})
\right).
\end{align*}

%The rate pair $(R,\Delta)$ is an $(n,\vep)$-achievable 
%rate pair if there exists a triple $(\varphi_1^{(n)},$ 
%$\varphi_2^{(n)}, \psi^{(n)})$ such that    
%The set that consists of all $\varepsilon$-achievable rate pair 
%is denoted by ${\cal R}_{\rm GMAC}(n, \varepsilon,P_1,P_2|W)$, 
%which is called 
%the $(n, \varepsilon)$-capacity region of the GMAC. 
%
\end{itemize}
\end{pr}
%===================================================================%
%===================================================================%

Proof of this property is given in Appendix \ref{sub:ApdaAaaa}.
%%%%%%%%%%%%%%%%%%%%%%%%%%%%%%%%%%%%%%%%%%%%%%%%%%%%%%%%%%%%%%%%%%%%%%%%%
%%%%%%%%%%%%%%%%%%%%%%%%%%%%%%%%%%%%%%%%%%%%%%%%%%%%%%%%%%%%%%%%%%%%%%%%%
%------------------------- Start of \ApdaAaaa --------------------------%
%%%%%%%%%%%%%%%%%%%%%%%%%%%%%%%%%%%%%%%%%%%%%%%%%%%%%%%%%%%%%%%%%%%%%%%%%
%%%%%%%%%%%%%%%%%%%%%%%%%%%%%%%%%%%%%%%%%%%%%%%%%%%%%%%%%%%%%%%%%%%%%%%%%
\newcommand{\ApdaAaaa}{
\subsection{Properties of the Rate Regions}
\label{sub:ApdaAaaa}
%}{

In this appendix we prove Property \ref{pr:pro0aa}. 
Property \ref{pr:pro0aa} part a) can easily be proved by 
the definitions of the rate distortion regions. We omit 
the proofs of this part. In the following 
argument we prove the part b). 

{\it Proof of Property \ref{pr:pro0aa} part b:} \ 
We set
\begin{align*}
\underline{\cal R}_{\rm AKW}(m, \varepsilon| p_{XY})
&=
\bigcap_{n \geq m}{\cal R}_{\rm AKW}(n,\varepsilon | p_{XY}).
\end{align*}
By the definitions of 
$\underline{\cal R}_{\rm AKW}(m,\vep|p_{XY})$ and 
${\cal R}_{\rm AKW}(\vep|p_{XY})$, 
we have that 
$ \underline{\cal R}_{\rm AKW}(m,\varepsilon|p_{XY}) 
$ $\subseteq {\cal R}_{\rm AKW}(\varepsilon|p_{XY}) 
$
for $m\geq 1$. Hence we have that 
\beq
\bigcup_{m\geq 1} \underline{\cal R}_{\rm AKW}(m,\varepsilon|p_{XY}) 
\subseteq {\cal R}_{\rm AKW}(\varepsilon|p_{XY}). 
\label{eqn:aaDs}
\eeq
We next assume that $(R_1,R_2)
\in {\cal R}_{\rm AKW}(\varepsilon|p_{XY})$.
Set
\begin{align*}
& {\cal R}_{\rm AKW}^{(\delta)}(\varepsilon|p_{XY})
\\
&\defeq \{(R_1+\delta,R_2+\delta): 
(R_1,R_2)\in {\cal R}_{\rm AKW}(\varepsilon|p_{XY})
\}
\end{align*}
Then, by the definitions of 
${\cal R}_{\rm AKW}(n,\vep$ $| p_{XY})$ 
and ${\cal R}_{\rm AKW}($ $\vep|p_{XY})$, we have that 
for any $\delta>0$, there exists $n_0(\varepsilon,\delta)$ 
such that for any $n\geq n_0(\varepsilon,\delta)$, 
$$
(R_1+\delta,R_2+\delta) \in {\cal R}_{\rm AKW}%^{(\delta)}
(n,\varepsilon|p_{XY}),
$$
which implies that 
\beqa
& &{\cal R}^{(\delta)}_{\rm AKW}(\varepsilon|p_{XY})
\subseteq  \bigcup_{n\geq n_0(\varepsilon,\delta)} 
{\cal R}_{\rm AKW}(n,\varepsilon|p_{XY})
\nonumber\\
&= & %\bigcup_{m= n_0(\delta)} 
\underline{\cal R}_{\rm AKW}(n_0(\delta),\varepsilon|p_{XY})
\nonumber\\
&\subseteq & {\rm cl}\left(\bigcup_{m \geq 1} 
\underline{\cal R}_{\rm AKW}(m,\varepsilon|p_{XY})\right).
\label{eqn:aaDsX}
\eeqa
Here we assume that there exists a pair $(R_1,R_2)$ 
belonging to ${\cal R}_{\rm AKW}(\vep | p_{XY})$ such that
\beq
(R_1,R_2)\notin 
{\rm cl}\left(
\bigcup_{m \geq 1}
\underline{\cal R}_{\rm AKW}(m,\varepsilon|p_{XY})
\right).
\label{eqn:zdff}
\eeq
Since the set in the right hand side of (\ref{eqn:zdff}) 
is a closed set, we have 
\beq
(R_1+\delta,R_2+\delta)\notin 
{\rm cl}\left(
\bigcup_{m \geq 1}
\underline{\cal R}_{\rm AKW}(m,\varepsilon|p_{XY})
\right)
\label{eqn:aas}
\eeq
for some small $\delta>0$. On the other hand we have 
$(R_1+\delta,R_2+\delta)$ 
$\in {\cal R}^{(\delta)}_{\rm AKW}(\varepsilon|p_{XY})$, 
which contradicts (\ref{eqn:aaDsX}). Thus we 
have 
\begin{align}
& \bigcup_{m\geq 1}
\underline{\cal R}_{\rm AKW}(m,\varepsilon|p_{XY})
\nonumber\\
&\subseteq{\cal R}_{\rm AKW}(\varepsilon|p_{XY})
%\nonumber\\
%&
\subseteq 
%& 
{\rm cl}\left(
\bigcup_{m \geq 1}
\underline{{\cal R}}_{\rm AKW}(m,\varepsilon|p_{XY})
\right).
\quad \label{eqn:aaaQw}
\end{align}
Note here that ${\cal R}_{\rm AKW}$$(\varepsilon$$|p_{XY})$ 
is a closed set. Then from (\ref{eqn:aaaQw}), we conclude that
\begin{align*}
{\cal R}_{\rm AKW}(\varepsilon|W)
&= 
{\rm cl}\left(
\bigcup_{m \geq 1}
\underline{{\cal R}}_{\rm AKW}(m,\varepsilon|p_{XY})
\right)
\\
&=
{\rm cl}\left(
\bigcup_{m \geq 1}
\bigcap_{n \geq m}
{\cal R}_{\rm AKW}(n,\varepsilon|p_{XY})
\right),
\end{align*}
completing the proof.
\hfill\IEEEQED
%%%%%%%%%%%%%%%%%%%%%%%%%%%%%%%%%%%%%%%%%%%%%%%%%%%%%%%%%%%%%%%%%%%%%%%%%
%%%%%%%%%%%%%%%%%%%%%%%%%%%%%%%%%%%%%%%%%%%%%%%%%%%%%%%%%%%%%%%%%%%%%%%%%
%%%%%%%%%%%%%%%%%%%%%%%%%%%%%%%%%%%%%%%%%%%%%%%%%%%%%%%%%%%%%%%%%%%%%%%%%
%--------------------------- End of \ApdaAaaa --------------------------%
%%%%%%%%%%%%%%%%%%%%%%%%%%%%%%%%%%%%%%%%%%%%%%%%%%%%%%%%%%%%%%%%%%%%%%%%%
%%%%%%%%%%%%%%%%%%%%%%%%%%%%%%%%%%%%%%%%%%%%%%%%%%%%%%%%%%%%%%%%%%%%%%%%%
}%%%%%%%%%%%%%%%%%%%%%%%%%%%%%%%%%%%%%%%%%%%%%%%%%%%%%%%%%%%%%%%%%%%%%%%%

It is well known that ${\cal R}_{\rm AKW}(p_{XY})$ 
was determined by Ahlswede, K\"orner and Wyner.
To describe their result we introduce an auxiliary random variable 
$U$ taking values in a finite set ${\cal U}$.  
We assume that the joint distribution 
of $(U,X,Y)$ is 
$$ 
p_{U{X}{Y}}(u,x,y)=p_{U}(u)
p_{{X}|U}(x|u)p_{Y|X}(y|x).
$$
The above condition is equivalent to $U \markov X \leftrightarrow Y$. 
Define the set of probability distribution $p=p_{UXY}$
by
\begin{align*}
&{\cal P}(p_{XY})
\defeq 
\{p_{UXY}: \pa {\cal U} \pa \leq \pa {\cal X} \pa+1,
U \markov  X\markov Y \}.
\end{align*}
Set
\begin{align*}
{\cal R}(p)
&\defeq 
\ba[t]{l}
\{(R_1,R_2): R_1,R_2 \geq 0\,,
\vSpa\\
\ba{rcl}
R_1 & \geq & I_p({X};{U}), R_2  \geq H_p({Y}|{U})\},
\ea
\ea
\\
{\cal R}(p_{XY})&\defeq \bigcup_{p \in {\cal P}(p_{XY})}
{\cal R}(p).
\end{align*}
We can show that the region ${\cal R}(p_{XY})$ satisfies 
the following property.
\begin{pr}\label{pr:pro0}  
$\quad$
\begin{itemize}
\item[a)] 
The region ${\cal R}(p_{XY})$ is a closed convex 
subset of $\mathbb{R}_{+}^2$.
%%%%%%%%%%%%%%%%%%%%%%%%%%%%%%%%%%%%%%%%%%%%%%%%%%%%%%%%%%%%%%%%%%
%, where%%%%%%%%%%%%%%%%%%%%%%%%%%%%%%%%%%%%%%%%%%%%%%%%%%%%%%%%%%
%\begin{align*}%%%%%%%%%%%%%%%%%%%%%%%%%%%%%%%%%%%%%%%%%%%%%%%%%%%%%%%%%%%
%\mathbb{R}_{+}^2&\defeq &\{(R_1,R_2): R_1 \geq 0,R_2 \geq 0\}.%%%
%\end{align*}%%%%%%%%%%%%%%%%%%%%%%%%%%%%%%%%%%%%%%%%%%%%%%%%%%%%%%%%%%%
%%%%%%%%%%%%%%%%%%%%%%%%%%%%%%%%%%%%%%%%%%%%%%%%%%%%%%%%%%%%%%%%%%
\item[b)] For any $p_{XY}$, we have 
\beq
\min_{(R_1,R_2)\in {\cal R}(p_{XY})}(R_1+R_2)=H_p(Y).
\label{eqn:SdEEE}
\eeq
The minimun 
is attained by $(R_1,R_2)=(0,H_p(Y))$. 
This result implies that 
\begin{align*}
& {\cal R}(p_{XY}) \subseteq 
%\\
%& 
\{(R_1,R_2): R_1+R_2 \geq H_p(Y)\} \cap \mathbb{R}_{+}^2.
\end{align*} 
Furthermore, the point $(0,H_p(Y))$ always belongs to 
${\cal R}(p_{XY})$. 
%In general, the lower boundary of 
%${\cal R}(p_{XY})$ contains a line segment of slope $-1$ going 
%through the point $(0,H_p(Y))$ but this line segment may reduce 
%to the point $(0, H_p(Y))$.
\end{itemize}
\end{pr}

\begin{figure}[t]
\bc
\includegraphics[width=5.6cm]{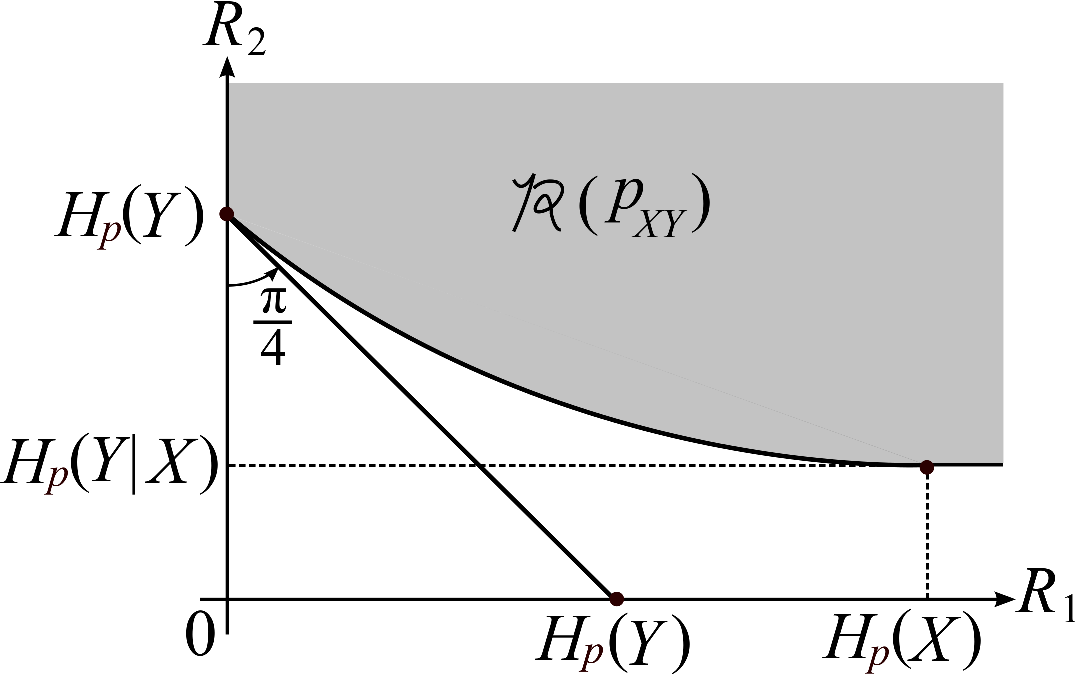}
\caption{
A typical shape of ${\cal R}(p_{XY})$.
}
\label{fig:AKWRegDMS} 
\ec
\end{figure}

Property \ref{pr:pro0} part a) is a well known property.
Proof of Property \ref{pr:pro0} part b) is easy. 
Proofs of Property \ref{pr:pro0} parts a) and b) are omitted. 
A typical shape 
of the rate region ${\cal R}(p_{XY})$ is shown in Fig. \ref{fig:AKWRegDMS}.

%%%%%%%%%%%%%%%%%%%%%%%%%%%%%%%%%%%%%%%%%%%%%%%%%%%%%%%%%%%%%%%%%
%%%%%%%%%%%%%%%%%%%%%%%%%%%%%%%%%%%%%%%%%%%%%%%%%%%%%%%%%%%%%%%%%

The rate region ${\cal R}_{\rm AKW}(p_{XY})$ was determined by 
Ahlswede and K\"orner \cite{ak75} and Wyner \cite{w0}. 
Their result is the following.
\begin{Th}[Ahlswede, K\"orner \cite{ak75} and Wyner \cite{w0}]
\label{th:akw0}
\begin{align*}
& {\cal R}_{\rm AKW}(p_{XY})={\cal R}(p_{XY}).
\end{align*}
\end{Th}

On the converse coding theorem Ahlswede {\it et al.} \cite{agk76}
obtained the following. 
\begin{Th}[Ahlswede et al. \cite{agk76}]
\label{th:agk}
For each fixed $\varepsilon $ $\in (0,1)$, 
we have 
\begin{align*}
& {\cal R}_{\rm AKW}(\varepsilon | p_{XY})={\cal R}(p_{XY}).
\end{align*}
\end{Th}

Gu and Effors \cite{GuEf09} examined a speed 
of convergence for ${\rm P}_{\rm e}^{(n)}$ 
to tend to 1 as $n\to \infty$ by carefully checking 
the proof of Ahlswede {\it et al.}
\cite{agk76}. However they could not obtain 
a result on an explicit form of the exponent 
function with respect to the code length $n$. 

Our aim is to find an explicit form of the exponent function 
for the error probability of decoding to tend to one as 
$n \to \infty$ when $(R_1,R_2) \notin {\cal R}_{\rm AKW}(p_{XY})$. 
To examine this quantity, we define the following quantity. Set
\begin{align*}
&  {\rm P}_{\rm c}^{(n)}(\varphi_1^{(n)},\varphi_2^{(n)},\psi^{(n)})
\defeq 
1-{\rm P}_{\rm e}^{(n)}(\varphi_1^{(n)},\varphi_2^{(n)},\psi^{(n)}),
\\
&  
G^{(n)}(R_1,R_2|p_{XY})
\\
&\defeq
\min_{\scs 
(\varphi_1^{(n)},\varphi_2^{(n)},\psi^{(n)}):
    \atop{\scs 
    (1/n)\log \|\varphi_i^{(n)}\|\leq R_i,i=1,2 
    }
}
\hspace*{-4mm}
\left(-\frac{1}{n}\right)
\log {\rm P}_{\rm c}^{(n)}(\varphi_1^{(n)},\varphi_2^{(n)},\psi^{(n)}).
\\
& 
G(R_1,R_2|p_{XY}) \defeq 
\lim_{n \to \infty}G^{(n)}(R_1,R_2|p_{XY}),
\\
& {\cal G}(p_{XY}) \defeq \{(R_1,R_2,G): 
G\geq G(R_1,R_2|p_{XY})\}. 
\end{align*}
By time sharing we have that 
\begin{align}
& G^{(n+m)}\left(\left.
\frac{n R_1+m R_1^{\prime}}{n+m},
\frac{n R_2+m R_2^{\prime}}{n+m}\right|p_{XY}\right) 
\nonumber
\\
&\leq \frac{nG^{(n)}(R_1,R_2|p_{XY}) 
+mG^{(m)}(R_1^{\prime},R_2^{\prime}|p_{XY})}{n+m}. 
\label{eqn:aaZ} 
\end{align}
Choosing $R=R^\prime$ in (\ref{eqn:aaZ}), we 
obtain the following subadditivity property
on $\{G^{(n)}(R_1,R_2|p_{XY})$ $\}_{n\geq 1}$: 
\begin{align*}
& G^{(n+m)}(R_1,R_2|p_{XY}) 
\\
&\leq \frac{nG^{(n)}(R_1,R_2|p_{XY}) 
+mG^{(m)}(R_1,R_2|p_{XY})}{n+m},
\end{align*}
from which we have that $G^{(n)}(R_{1},R_2|p_{XY})$ exists 
and satisfies the following: 
\begin{align*}
&\lim_{n\to\infty}G^{(n)}(R_{1},R_2|p_{XY}) 
=\inf_{n\geq 1}G^{(n)}(R_1,R_2|p_{XY}).
\end{align*}
The exponent function $G(R_1,R_2|p_{XY})$ is a convex function 
of $(R_1,R_2)$. In fact, from (\ref{eqn:aaZ}), we have that  
for any $\alpha \in [0,1]$
\begin{align*}
& G(\alpha R_1+\bar{\alpha}R_1^{\prime},
     \alpha R_2+\bar{\alpha}R_2^{\prime}|p_{XY})
\\
&\leq 
\alpha G(R_1,R_2|p_{XY})
+\bar{\alpha} G( R_1^{\prime},R_2^{\prime}|p_{XY}).
\end{align*}
The region ${\cal G}(p_{XY})$ is also a closed convex set. 
Our main aim is to find an explicit characterization of 
${\cal G}(p_{XY})$. In this paper we derive an explicit 
outer bound of ${\cal G}$ $(p_{XY})$ whose section by 
the plane $G=0$ coincides with ${\cal R}_{\rm AKW}(p_{XY})$.

\section{Main Result}

In this section we state our main result. 
We first  explain that the region ${\cal R}(p_{XY})$ can 
be expressed with a family of supporting hyperplanes. 
To describe this result we define 
a set of probability distributions on 
${\cal U}$ $\times{\cal X}$ $\times{\cal Y}$ by
\begin{align*}
{\cal P}_{\rm sh}(p_{XY})
&\defeq 
\{p=p_{UXY}: \pa {\cal U} \pa \leq \pa {\cal X} \pa,
  U \markov  X\markov Y \}.
\end{align*}
For $\mu\geq 0$, define
\begin{align*}
& R^{(\mu)}(p_{XY})
\defeq 
\min_{p \in {\cal P}_{\rm sh}(p_{XY})}
\left\{{\prmtA} I_p(X;U)+ \bar{\mu}H_p(Y|U)\right\}.
\end{align*}
Furthermore, define
\begin{align*}
& {\cal R}_{\rm sh}(p_{XY})
  \defeq \bigcap_{\mu \in [0,1]}\{(R_1,R_2):
\ba[t]{l}
{\prmtA} R_1+ \bar{\mu}R_2 \\ 
\geq R^{(\mu)}(p_{XY})\}.
\ea 
\end{align*}

Then we have the following property.
\begin{pr}\label{pr:pro0z} $\quad$
\begin{itemize}
\item[a)] The bound $|{\cal U}|\leq |{\cal X}|$
is sufficient to describe 
$R^{(\mu)}($ $p_{XY})$. 

\item[b)] For every $\mu\in [0,1]$, we have 
\begin{align}
&\min_{(R_1,R_2)\in {\cal R}(p_{XY})}
\{\mu R_1+\bar{\mu}R_2\}=R^{(\mu)}(p_{XY}).
\label{eqn:SDAzz}
\end{align}

%\item[c)] For every $\mu\geq 1$, we have 
%$R^{(\mu)}(p_{XY})=H_p(Y).$

\item[c)] For any $p_{XY}$ we have
\beq
 {\cal R}_{\rm sh}(p_{XY})%=\underline{\cal R}_{\rm sh}(p_{XY})
={\cal R}(p_{XY}).
\label{eqn:PropEqB}
\eeq
\end{itemize}
\end{pr}

%Property \ref{pr:pro1} part a) follows from 
%Lemma \ref{lm:CardLmd} in Appendix \ref{sub:ApdaAAA}. 
%Proof of this lemma is given in this appendix.
%

Property \ref{pr:pro0z} part a) is stated as Lemma \ref{lm:CardLm} 
in Appendix \ref{sub:ApdaAAA}. Proof of this lemma is given in this appendix.
%%%%%%%%%%%%%%%%%%%%%%%%%%%%%%%%%%%%%%%%%%%%%%%%%%%%%%%%%%%%%%%%%%%%%%%%%%%%%%%
%------------------------------- Start of \ApdaAAA----------------------------%
%%%%%%%%%%%%%%%%%%%%%%%%%%%%%%%%%%%%%%%%%%%%%%%%%%%%%%%%%%%%%%%%%%%%%%%%%%%%%%%
\newcommand{\ApdaAAA}{
\subsection{Cardinality Bound on Auxiliary Random Variables}
\label{sub:ApdaAAA}

We first prove the following lemma.
\begin{lm}
\label{lm:CardLm}
\begin{align*}
& \underline{R}^{(\mu)}(p_{XY})
\defeq \min_{\scs p \in {\cal P}(p_{XY})}
\left\{{\prmtA} I_p(X;U)+{\prmtB}H_p(Y|U)\right\}
\\
&=R^{(\mu)}(p_{XY})
\defeq \min_{\scs p \in {\cal P}_{\rm sh}(p_{XY})}
\left\{ {\prmtA} I_p(X;U)+ {\prmtB} H_p(Y|U)\right\}.
\end{align*}
\end{lm}

{\it Proof:} We bound the cardinality $|{\cal U}|$ of $U$ 
to show that the bound $|{\cal U}|\leq |{\cal X}|$
is sufficient to describe $\underline{R}^{(\mu)}(p_{XY})$. 
Observe that 
\begin{align}
& p_{X}(x)
=\sum_{u\in {\cal U}}p_U(u)
p_{{X}|{U}}(x|u),
\label{eqn:asdfq}
\\
& {\prmtA} I_p(X;U)+ {\prmtB}H_p(Y|U)=\sum_{u\in {\cal U}}p_{U}(u)
\pi(p_{X|U}(\cdot|u)),
\label{eqn:aqqqaq}
\end{align}
where 
\begin{align*}
&\pi(p_{{X}|U}(\cdot|u))
\defeq \sum_{(x,y)\in{\cal X}\times{\cal Y}}
p_{X|U}(x|u)p_{Y|X}(y|x)
\\
& \quad \times \log \left\{\frac{p^{{\prmtA}}_{X|U}(x|u)}
{p^{{\prmtA}}_X(x)}
\left[
\ds \sum_{\tilde{x} \in {\cal X} }
p_{Y|X}(y|\tilde{x}) p_{X|U}(\tilde{x}|u)
\right]^{-{\prmtB}}\right\}.
\end{align*}
For each  $u\in {\cal U}$, $\pi(p_{{X}|U}(\cdot|u))$ 
is a continuous function of 
$p_{X|U}(\cdot|u)$. Then by the support lemma,
$$
|{\cal U}| \leq |{\cal X}|-1 +1= |{\cal X}| 
$$
is sufficient to express $|{\cal X}|-1$ values 
of (\ref{eqn:asdfq}) and 
one value of (\ref{eqn:aqqqaq}). 
\hfill \IEEEQED
}%%%%%%%%%%%%%%%%%%%%%%%%%%%%%%%%%%%%%%%%%%%%%%%%%%%%%%%%%%%%%%%%%%%%%%%%%%%%%%
%%%%%%%%%%%%%%%%%%%%%%%%%%%%%%%%%%%%%%%%%%%%%%%%%%%%%%%%%%%%%%%%%%%%%%%%%%%%%%%
%%%%%%%%%%%%%%%%%%%%%%%%%%%%%%%%%%%%%%%%%%%%%%%%%%%%%%%%%%%%%%%%%%%%%%%%%%%%%%%
%%%%%%%%%%%%%%%%%%%%%%%%%%%%%%%%%%%%%%%%%%%%%%%%%%%%%%%%%%%%%%%%%%%%%%%%%%%%%%%
%========================== \End of \ApdaAAA==================================%
%%%%%%%%%%%%%%%%%%%%%%%%%%%%%%%%%%%%%%%%%%%%%%%%%%%%%%%%%%%%%%%%%%%%%%%%%%%%%%%
%%%%%%%%%%%%%%%%%%%%%%%%%%%%%%%%%%%%%%%%%%%%%%%%%%%%%%%%%%%%%%%%%%%%%%%%%%%%%%%
Proofs of Property \ref{pr:pro0z} parts b) and c) are given 
in Appendix \ref{sub:ApdaAABz}.
%%%%%%%%%%%%%%%%%%%%%%%%%%%%%%%%%%%%%%%%%%%%%%%%%%%%%%%%%%%%%%%%%%%%%%%%%%%%%%%
%------------------------------- Start of \ApdaAABz---------------------------%
%%%%%%%%%%%%%%%%%%%%%%%%%%%%%%%%%%%%%%%%%%%%%%%%%%%%%%%%%%%%%%%%%%%%%%%%%%%%%%%
\newcommand{\ApdaAABz}{
\subsection{
%Proof of Property \protect{\ref{pr:pro0z}}
Supporting Hyperplain Expressions of ${\cal R}(p_{XY})$
}
\label{sub:ApdaAABz}

%}{%%% Not Forward Appendix %%%
In this appendix we prove Property \ref{pr:pro0z} parts 
b), c). We first prove the part b).

{\it Proof of Property \ref{pr:pro0z} part b):} \ 
For any $\mu \geq 0,$ we have the following chain 
of inequalities:
\begin{align}
&\min_{(R_1,R_2)\in {\cal R}(p_{XY})}
\{\prmtA R_1+\prmtB R_2\} 
\notag\\
&=
\min_{p \in {\cal P}_{\rm }(p_{XY})}
\{ \prmtA I_p(X;U) + \prmtB H_p(Y|U) \}
\notag\\
&\MEq{a}
\min_{p \in {\cal P}_{\rm sh}(p_{XY})}
\{  I_p(X;U) + \prmtB H_p(Y|U) \}
\notag\\
&=R^{(\mu)}(p_{XY}).
%\label{eqn:SDAzzp}
\notag
\end{align}
Step (a) follows from Lemma \ref{lm:CardLm} stating that the 
cardinality bound $|{\cal U}|\leq |{\cal X}|+1$ 
in ${\cal P}_{\rm }(p_{XY})$ can be reduced 
to that $|{\cal U}|\leq |{\cal X}|$ 
in ${\cal P}_{\rm sh}(p_{XY})$. 
\hfill\IEEEQED

We next prove the part c). We first prepare a lemma useful to prove this property.  
From the convex property of the region ${\cal R}(p_{XY})$, 
we have the following lemma. 
\begin{lm}\label{lm:asgsq}
Suppose that $(\hat{R}_1,\hat{R}_2)$ does not belong 
to ${\cal R}(p_{XY})$. Then there exist 
$\epsilon>0$ and $\mu_0 \geq 0$ such that 
for any $(R_1,R_2)\in {\cal R}(p_{XY})$
we have 
\begin{align*}
&{\prmtA}_0(R_1-\hat{R}_1)
+\overline{\mu_0}(R_2-\hat{R}_2)-\epsilon \geq 0.
\end{align*}
\end{lm} 

Proof of this lemma is omitted here. Lemma \ref{lm:asgsq} 
is equivalent to the fact that if the region ${\cal R}(p_{XY})$ 
is a convex set, then for any point $(\hat{R}_1,\hat{R}_2)$ 
outside the region ${\cal R}(p_{XY})$, 
there exists a line which separates the point $(\hat{R}_1,\hat{R}_2)$ 
from the region ${\cal R}(p_{XY})$. 
%Lemma \ref{lm:asgsq} 
%will be used to prove Property \ref{pr:pro0z} part b).

{\it Proof of Property \ref{pr:pro0z} part c):} \ 
We first prove $\underline{\cal R}_{\rm sh}(p_{XY})$ 
$\subseteq {\cal R}(p_{XY})$. We assume that 
$(\hat{R}_1,\hat{R}_2)\notin {\cal R}(p_{XY})$. 
Then by Lemma \ref{lm:asgsq}, there exist 
$\epsilon>0$ and $\mu_0 \geq 0$ such that for 
any $(R_1,R_2)\in {\cal R}(p_{XY})$, we have
\begin{align*}
&     {\prmtA}_0 \hat{R}_1 + \overline{\mu_0} \hat{R}_2 
\leq {\prmtA}_0       R_1 + \overline{\mu_0} R_2-\epsilon.
\end{align*}
Then we have
\begin{align}
&  {\prmtA}_0 \hat{R}_1+ \overline{\mu_0} \hat{R}_2
\leq 
   \min_{(R_1,R_2)\in {\cal R}(p_{XY})} 
   \left\{{\prmtA}_0 R_1+\overline{\mu_0} R_2 \right\} 
   -\epsilon
\nonumber\\ 
&\MEq{a}\min_{p\in {\cal P}(p_{XY})}
\left\{ {\prmtA}_0 I_p(U;X) + \overline{\mu_0} H_p(Y|U) \right\}
-\epsilon
\nonumber\\ 
&\leq\min_{p\in {\cal P}_{\rm sh}(p_{XY})}
\left\{ {\prmtA}_0 I_p(U;X) + \overline{\mu_0} H_p(Y|U) \right\}
-\epsilon
\nonumber\\ 
&=R^{(\mu_0)}(p_{XY})-\epsilon.
\label{eqn:sddsd}
\end{align}
Step (a) follows from the definition of ${\cal R}(p_{XY})$. 
The inequality (\ref{eqn:sddsd}) implies that 
$(\hat{R}_1,\hat{R}_2)$ $\notin {\cal R}_{\rm sh}(p_{XY})$.
Thus ${\cal R}_{\rm sh}(p_{XY})$ $\subseteq {\cal R}(p_{XY})$ 
is concluded. 
\hfill\IEEEQED
}%%%%%%%%%%%%%%%%%%%%%%%%%%%%%%%%%%%%%%%%%%%%%%%%%%%%%%%%%%%%%%%%%%%%%%%
Set
\begin{align*}
{\cal Q}(p_{Y|X}) &\defeq \{q=q_{UXY}: 
\pa {\cal U} \pa \leq \pa {\cal X} \pa,
{U} \markov {X} \markov {Y}, 
\\
&  p_{Y|X}=q_{Y|X}\}.
\end{align*}
For $(\mu,\alpha) \in [0,1]^2$, 
and for $q=q_{UXY}\in {\cal Q}(p_{Y|X})$, 
define 
\begin{align*}
& \omega_{q|p_X}^{(\mu,\alpha)}(x,y|u)
\\
& \defeq \bar{\alpha}\log \frac{q_{X}(x)}{p_{X}(x)}
+ \alpha\left[{\prmtA}\log \frac{q_{X|U}(x|u)}{p_{X}(x)}\right.
%\\
%& 
\left. +{\prmtB}\log\frac{1}{q_{Y|U}(y|u)}\right],
\\
& f^{(\mu,\alpha)}_{q|p_X}(x,y|u)
  \defeq \exp\left\{-
\omega^{(\mu,\alpha)}_{q|p_X}(x,y|u)
\right\},
\\
& \Omega^{(\mu,\alpha)}(q|p_X)
%\\
%& 
\defeq 
-\log 
{\rm E}_{q}
\left[\exp\left\{-\omega^{(\mu,\alpha)}_{q |p_X}(X,Y|U)\right\}\right],
\\
&\Omega^{(\mu,\alpha)}(p_{XY})
%\\
%&
\defeq 
%&
\min_{\scs 
   \atop{\scs 
    q \in {{\cal Q}}(p_{Y|X})
   }
}
\Omega^{(\mu,\alpha)}(q|p_X),
\\
& F^{(\mu,\alpha)}( {\prmtA}R_1+{\prmtB} R_2 |p_{XY})
\\
&\defeq
%=====================================================================%
\newcommand{\ZapA}{
%}{
\min_{\scs 
%   p\in {{\cal P}}(p_{XY}),
   \atop{\scs 
   q \in {{\cal Q}}(p_{Y|X})
   }
}
F^{(\mu,\lambda)}({\prmtA}R_1+{\prmtB} R_2|p_X,q)}
%====================================================================%
%\\
%&\defeq &
\frac{\Omega^{(\mu,\alpha)}(p_{XY})
-\alpha({\prmtA}R_1 +{\prmtB} R_2)}
{2+\alpha{\prmtB}},
\\
& F(R_1,R_2|p_{XY})
%\\
%&
\defeq \sup_{\scs (\mu,\alpha) \in [0,1]^2}
F^{(\mu,\alpha)}({\prmtA}R_1+{\prmtB} R_2|p_{XY}).
\end{align*}
We next define a function serving as a lower 
bound of $F(R_1,R_2|p_{XY})$. 
For $\lambda \geq 0$ and for  
$p_{UXY} \in {\cal P}_{\rm sh}(p_{XY})$, define
\begin{align*}
& \tilde{\omega}_{p}^{(\mu)}(x,y|u)
\defeq {\prmtA}
  \log \frac{p_{X|U}(x|u)}{p_{X}(x)}
  +{\prmtB} \log \frac{1}{p_{Y|U}(Y|U)},
\\
& \tilde{\Omega}^{(\mu,\lambda)}(p)
\defeq 
-\log 
{\rm E}_{p}
\left[\exp\left\{-\lambda
\tilde{\omega}^{(\mu)}_p (X,Y|U)\right\}\right].
\end{align*}
Furthermore, set
\begin{align*}
& \tilde{\Omega}^{(\mu,\lambda)}(p_{XY})
 \defeq \min_{\scs \atop{\scs 
p \in {{\cal P}_{\rm sh}(p_{XY})}}}
\tilde{\Omega}^{(\mu,\lambda)}(p),
\\
& {\loF}^{(\mu,\lambda)}({\prmtA}R_1 + {\prmtB} R_2 |p_{XY}) 
\\
&\defeq 
\frac{\tilde{\Omega}^{(\mu,\lambda)}(p_{XY})
-\lambda({\prmtA}R_1+{\prmtB} R_2)}{2+\lambda(5-{\prmtA})},
\\
& {\loF}(R_1,R_2|p_{XY})
\defeq \sup_{ \lambda \geq 0,\mu \in [0,1]} 
{\loF}^{(\mu,\lambda)}({\prmtA} R_1+{\prmtB} R_2|p_{XY}).
\end{align*}
\newcommand{\OmiTz}{%%%%%%%%%%%%%%%%%%%%%%%%%%%%%%%%%%%%%%%%%%%%%%%%
%%%%%%%%%%%%%%%%%%%%%%%%%%%%%%%%%%%%%%%%%%%%%%%%%%%%%%%%%%%%%%%%%%%%
%We define 
%\begin{align*}
\\
& \tilde{F}^{(\mu,\lambda)}({\prmtA}R_1 + {\prmtB} R_2 |p_{XY}) 
\\
&\defeq 
\frac{\tilde{\Omega}^{(\mu,\lambda)}(p_{XY})
-\lambda({\prmtA}R_1+{\prmtB} R_2)}{1+2\lambda},
\\
& \tilde{F}(R_1,R_2|p_{XY})
\defeq \sup_{ \lambda \geq 0,\mu \in [0,1]} 
\tilde{F}^{(\mu,\lambda)}({\prmtA} R_1+{\prmtB} R_2|p_{XY}).
\end{align*}
}%%%%%%%%%%%%%%%%%%%%%%%%%%%%%%%%%%%%%%%%%%%%%%%%%%%%%%%%%%%%%%%%%
%%%%%%%%%%%%%%%%%%%%%%%%%%%%%%%%%%%%%%%%%%%%%%%%%%%%%%%%%%%%%%%%%%
We can show that the above functions satisfy the 
following property. 
\begin{pr}\label{pr:pro1}  
$\quad$
\begin{itemize}
\item[a)] 
The cardinality bound 
$|{\cal U}|\leq |{\cal X}|$ in ${\cal Q}(p_{Y|X})$
is sufficient to describe the quantity
$\Omega^{(\mu,\beta,\alpha)}(p_{XY})$. 
Furthermore, the cardinality bound 
$|{\cal U}|\leq |{\cal X}|$ in ${\cal P}_{\rm sh}(p_{XY})$
is sufficient to describe the quantity
$\tilde{\Omega}^{(\mu,\lambda)}(p_{XY})$. 

\item[b)] For any $R_1,R_2\geq0$, we have 
\begin{align*}
& &F(R_1,R_2|p_{XY})\geq {\loF}(R_1,R_2|p_{XY}).
\end{align*}

\item[c)] For any $p=p_{UXY} \in {\cal P}_{\rm sh}(p_{XY})$
and any $(\mu,\lambda)\in [0,$ $1]^2$, we have 
\beq
0\leq \tilde{\Omega}^{(\mu,\lambda)}(p) 
 \leq \prmtA \log |{\cal X}|
    + \prmtB \log |{\cal Y}|.
\label{eqn:Asddx}
\eeq
\item[d)] Fix any $p=p_{UXY}\in {\cal P}_{\rm sh}(p_{XY})$ and $\mu\in [0,1]$.
For $\lambda \in [0,1]$, we define a probability distribution 
$p^{(\lambda)}=p_{UXY}^{(\lambda)}$ by
\begin{align*} 
& p^{(\lambda)}(u,x,y)
\defeq
\frac{
p(u,x,y)
\exp\left\{
-\lambda \tilde{\omega}^{(\mu)}_p(x,y|u)
\right\}
}
{{\rm E}_{p}
\left[\exp\left\{-\lambda
\tilde{\omega}^{(\mu)}_p(X,Y|U)\right\}\right]}.
\end{align*}
\irr{Then for $\lambda \in [0,1/2]$, 
$\tilde{\Omega}^{(\mu,\lambda)}(p)$ is twice differentiable.
Furthermore, for $\lambda \in [0,1/2]$, we have}
\begin{align*}
&  \frac{\rm d}{{\rm d}\lambda}
\tilde{\Omega}^{(\mu,\lambda)}(p)
={\rm E}_{p^{(\lambda)}}
\left[\tilde{\omega}^{(\mu)}_{p}(X,Y|U)\right],
\\
&  \frac{\rm d^2}{{\rm d}\lambda^2} 
\tilde{\Omega}^{(\mu,\lambda)}(p)
=-{\rm Var}_{p^{(\lambda)}}
\left[\tilde{\omega}^{(\mu)}_{p}(X,Y|U)\right].
\end{align*}
The second equality implies that 
$\tilde{\Omega}^{(\mu,\lambda)}(p$$|p_{XY})$ 
is a concave function of $\lambda \in [0,1/2]$. 
\item[e)] 
For every $(\mu,\lambda)\in [0,1]\times [0,1/2]$, 
define
\begin{align*}
& \rho^{(\mu,\lambda)}(p_{XY})
\\
&\defeq \irr{\max_{\scs (\nu, p) \in [0,\lambda] 
    \atop{\scs \times {\cal P}_{\rm sh}(p_{XY}):
        \atop{\scs \tilde{\Omega}^{(\mu,\lambda)}(p) 
             \atop{\scs 
             =\tilde{\Omega}^{(\mu,\lambda)}(p_{XY})
             }}}}
}
{\rm Var}_{\irr{p^{(\nu)}}}\left[\tilde{\omega}^{(\mu)}_{p}(X,Y|U)\right],
\end{align*}
and set
\begin{align*}
& \rho =\rho(p_{XY})
\defeq \max_{\scs (\mu,\lambda)
\in [0,1]\times [0,1/2]}
\rho^{(\mu,\lambda)}(p_{XY}).
\end{align*}
Then, we have $\rho(p_{XY})< \infty.$ 
Furthermore, for any $(\mu,\lambda) \in [0,1]\times [0,1/2]$, 
we have 
\beq
\tilde{\Omega}^{(\mu,\lambda)}(p_{XY}) 
\geq \lambda R^{(\mu)}(p_{XY})
-\frac{\lambda^2}{2}\rho(p_{XY}).
\label{eqn:ZssPi}
\eeq
\item[f)] For every $\tau\in (0,(1/2)\rho(p_{XY}))$,
the condition 
$(R_1+\tau,$ $R_2+\tau) \notin {\cal R}(p_{XY})$
implies 
$$
{\loF}(R_1,R_2|p_{XY}) > \frac{\rho(p_{XY})}{4}\cdot g^2
\left(\frac{\tau}{\rho(p_{XY})}\right)>0,
$$
where $g$ is the inverse function of 
$\vartheta(a) \defeq a+(5/4)a^2, a \geq 0$.
\newcommand{\OmiTxx}{%%%%%%%%%%%%%%%%%%%%%%%%%%%%%%%%%%%%%%%%%%%%%
%%%%%%%%%%%%%%%%%%%%%%%%%%%%%%%%%%%%%%%%%%%%%%%%%%%%%%%%%%%%%%%%%%
\item[f)] For every $\eta\in (0,(1/2)\rho(p_{XY}))$,
the condition 
$(R_1,$ $R_2+3\eta) \notin {\cal R}(p_{XY})$
implies 
$$
\tilde{F}(R_1+\eta,R_2 +\eta |p_{XY})> \frac{\rho(p_{XY})}{4}\cdot \tilde{g}^2
\left(\frac{\eta}{\rho(p_{XY})}\right)>0,
$$
where $\tilde{g}$ is the inverse function of 
$\tilde{\vartheta}(a) \defeq a+a^2, a \geq 0$.
}%%%%%%%%%%%%%%%%%%%%%%%%%%%%%%%%%%%%%%%%%%%%%%%%%%%%%%%%%%%%%%%%
%%%%%%%%%%%%%%%%%%%%%%%%%%%%%%%%%%%%%%%%%%%%%%%%%%%%%%%%%%%%%%%%%
\end{itemize}
\end{pr}

Property \ref{pr:pro0z} part a) is stated as Lemma \ref{lm:CardLmA} 
in Appendix \ref{sub:ApdaAAA}. Proof of this lemma is given 
in this appendix. %%%%%%%%%%%%%%%%%%%%%%%%%%%%%%%%%%%%%%%%%%%%%%%%%%%%%%%%%%%%%%
\newcommand{\ApdaAAAb}{

%%%%%%%%%%%%%%%%%%%%%%%%%%%%%%%%%%%%%%%%%%%%%%%%%%%%%%%%%%%
%}{%-------------- Not Forward Appendix ------------------%
%%%%%%%%%%%%%%%%%%%%%%%%%%%%%%%%%%%%%%%%%%%%%%%%%%%%%%%%%%%
%%%%%%%%%%%%%%%%%%%%%%%%%%%%%%%%%%%%%%%%%%%%%%%%%%%%%%%%%%%
Next we prove the following lemma. 
\begin{lm}\label{lm:CardLmA}
The cardinality bound 
$|{\cal U}|\leq |{\cal X}|$ in ${\cal Q}(p_{Y|X})$
is sufficient to describe the quantity
$\Omega^{(\mu,\alpha)}(p_{XY})$. 
The cardinality bound $|{\cal U}|\leq |{\cal X}|$ 
in ${\cal P}_{\rm sh}(p_{XY})$ is sufficient 
to describe the quantity $\tilde{\Omega}^{(\mu,\lambda)}(p_{XY})$. 
\end{lm}

{\it Proof:} We first bound  the cardinality $|{\cal U}|$ of ${U}$ in 
${\cal Q}(p_{Y|X})$ to show that the bound 
$|{\cal U}|\leq |{\cal X}|$ is sufficient to describe 
${\Omega}^{(\mu,\alpha)}$ $(p_{XY})$. Observe that 
\begin{align}
& q_{X}(x)
=\sum_{u \in {\cal U}}q_{U}(u)
q_{X|{U}}(x|u),
\label{eqn:asdfs}
\\
& \exp
\left\{-\Omega^{(\mu,\alpha)}(q|p_{X})
\right\}
\nonumber\\
& \qquad\quad
=\sum_{u\in {\cal U}}q_{U}(u)
\Pi^{(\mu,\alpha)}
(q_X, q_{XY|U}(\cdot,\cdot|u)),
\label{eqn:aqqqas}
\end{align}
where 
\begin{align*}
& \Pi^{(\mu,\alpha)}
(q_X, q_{XY|{U}}(\cdot,\cdot|u)) 
\\
& \defeq 
\sum_{\scs (x,y) 
\atop{  \in{\cal X}
        \times{\cal Y}
     }
}
q_{XY|U}(x,y|u)
\exp\left\{-\omega^{(\mu,\alpha)}_{q|p_X}
(x,y|u) \right\}.
\end{align*}
The value of $q_X$ included in $\Pi^{(\mu,\alpha)}
(q_X, q_{{XY}|{U}}(\cdot,\cdot|u))$
must be preserved under the reduction of ${\cal U}$.
For each $u\in {\cal U}$, 
$\Pi^{(\mu,\alpha)}(q_X,q_{XY|U}(\cdot,\cdot|u))$ 
is a contineous function of $q_{XY|U}($ $\cdot,\cdot|u)$.
Then by the support lemma, 
$$|{\cal U}|\leq |{\cal X}| -1 +1 = |{\cal X}|$$
is sufficient to express $|{\cal X}|-1$ values of (\ref{eqn:asdfs}) 
and one value of (\ref{eqn:aqqqas}). We next bound  the cardinality 
$|{\cal U}|$ of ${U}$ in ${\cal P}_{\rm sh}(p_{XY})$ to show 
that the bound $|{\cal U}|\leq |{\cal X}|$ is sufficient 
to describe ${\Omega}^{(\mu,\lambda)}$ $(p_{XY})$. Observe that 
\begin{align}
& p_{X}(x)
=\sum_{u \in {\cal U}}p_{U}(u)p_{X|U}(x|u),
\label{eqn:Zasdfs}
\\
& \exp
\left\{-\Omega^{(\mu,\lambda)}(p)
\right\}
\nonumber\\
& \qquad\quad 
=\sum_{u\in {\cal U}}p_{{U}}(u)
\tilde{\Pi}^{(\mu,\lambda)}
(p_X,p_{XY|U}(\cdot,\cdot|u)),
\label{eqn:Zaqqqas}
\end{align}
where 
\begin{align*}
& \tilde{\Pi}^{(\mu,\lambda)}
(p_X,p_{XY|{U}}(\cdot,\cdot|u)) 
\\
& \defeq 
\sum_{\scs (x,y) 
\atop{  \in{\cal X}
         \times{\cal Y}
     }
}
p_{XY|U}(x,y|u)
\exp\left\{-\lambda \tilde{\omega}^{(\mu)}_{p}
(x,y|u) \right\}.
\end{align*}
The value of $p_X$ included in 
$\tilde{\Pi}^{(\mu,\lambda)}(p_X,p_{{XY}|{U}}(\cdot,\cdot|u))$
must be preserved under the reduction of ${\cal U}$. For each 
$u\in {\cal U}$, 
$\tilde{\Pi}^{(\mu,\lambda)}(p_X,p_{XY|U}(\cdot,\cdot|u))$ 
is a contineous function of $p_{XY|U}(\cdot,\cdot|u)$. Then by the 
support lemma, 
$$
|{\cal U}|\leq |{\cal X}| -1 +1 = |{\cal X}|$$
is sufficient to express $|{\cal X}|-1$ 
values of (\ref{eqn:Zasdfs}) and one value of (\ref{eqn:Zaqqqas}).
\hfill \IEEEQED
}%%%%%%%%%%%%%%%%%%%%%%%%%%%%%%%%%%%%%%%%%%%%%%%%%%%%%%%%%%%%%%%%%%%%%%%%%%
%%%%%%%%%%%%%%%%%%%%%%%%%%%%%%%%%%%%%%%%%%%%%%%%%%%%%%%%%%%%%%%%%%%%%%%%%%%
%%%%%%%%%%%%%%%%%%%%%%%%%%%%%%%%%%%%%%%%%%%%%%%%%%%%%%%%%%%%%%%%%%%%%%%%%%%
%%%%%%%%%%%%%%%%%%%%%%%%%%%%%%%%%%%%%%%%%%%%%%%%%%%%%%%%%%%%%%%%%%%%%%%%%%%
%========================= End of \ApdaAAAb ==============================%
%%%%%%%%%%%%%%%%%%%%%%%%%%%%%%%%%%%%%%%%%%%%%%%%%%%%%%%%%%%%%%%%%%%%%%%%%%%
%%%%%%%%%%%%%%%%%%%%%%%%%%%%%%%%%%%%%%%%%%%%%%%%%%%%%%%%%%%%%%%%%%%%%%%%%%%
%%%%%%%%%%%%%%%%%%%%%%%%%%%%%%%%%%%%%%%%%%%%%%%%%%%%%%%%%%%%%%%%%%%%%%%%%%%
%%%%%%%%%%%%%%%%%%%%%%%%%%%%%%%%%%%%%%%%%%%%%%%%%%%%%%%%%%%%%%%%%%%%%%%%%%%
%%%%%%%%%%%%%%%%%%%%%%%%%%%%%%%%%%%%%%%%%%%%%%%%%%%%%%%%%%%%%%%%%%%%%%%%%%%
%%%%%%%%%%%%%%%%%%%%%%%%%%%%%%%%%%%%%%%%%%%%%%%%%%%%%%%%%%%%%%%%%%%%%%%%%%%
%%%%%%%%%%%%%%%%%%%%%%%%%%%%%%%%%%%%%%%%%%%%%%%%%%%%%%%%%%%%%%%%%%%%%%%%%%%
%%%%%%%%%%%%%%%%%%%%%%%%%%%%%%%%%%%%%%%%%%%%%%%%%%%%%%%%%%%%%%%%%%%%%%%%%%%
%%%%%%%%%%%%%%%%%%%%%%%%%%%%%%%%%%%%%%%%%%%%%%%%%%%%%%%%%%%%%%%%%%%%%%%%%%%
%%%%%%%%%%%%%%%%%%%%%%%%%%%%%%%%%%%%%%%%%%%%%%%%%%%%%%%%%%%%%%%%%%%%%%%%%%%
%%%%%%%%%%%%%%%%%%%%%%%%%%%%%%%%%%%%%%%%%%%%%%%%%%%%%%%%%%%%%%%%%%%%%%%%%%%
%%%%%%%%%%%%%%%%%%%%%%%%%%%%%%%%%%%%%%%%%%%%%%%%%%%%%%%%%%%%%%%%%%%%%%%%%%%
%%%%%%%%%%%%%%%%%%%%%%%%%%%%%%%%%%%%%%%%%%%%%%%%%%%%%%%%%%%%%%%%%%%%%%%%%%%
%%%%%%%%%%%%%%%%%%%%%%%%%%%%%%%%%%%%%%%%%%%%%%%%%%%%%%%%%%%%%%%%%%%%%%%%%%%
%%%%%%%%%%%%%%%%%%%%%%%%%%%%%%%%%%%%%%%%%%%%%%%%%%%%%%%%%%%%%%%%%%%%%%%%%%%
Proof of Property \ref{pr:pro1} part b) is given in %%%%%%%%%%%%%%%%%%%%%%%
Appendix \ref{sub:ApdaAACa}. %%%%%%%%%%%%%%%%%%%%%%%%%%%%%%%%%%%%%%%%%%%%%%
Proofs of Property \ref{pr:pro1} parts c), d), e), and f) %%%%%%%%%%%%%%%%%
are given in Appendix \ref{sub:ApdaAAC}. %%%%%%%%%%%%%%%%%%%%%%%%%%%%%%%%%%
%%%%%%%%%%%%%%%%%%%%%%%%%%%%%%%%%%%%%%%%%%%%%%%%%%%%%%%%%%%%%%%%%%%%%%%%%%%
%%%%%%%%%%%%%%%%%%%%%%%%%%%%%%%%%%%%%%%%%%%%%%%%%%%%%%%%%%%%%%%%%%%%%%%%%%%
%------------------------- Start of ApdaAACa -----------------------------%
%%%%%%%%%%%%%%%%%%%%%%%%%%%%%%%%%%%%%%%%%%%%%%%%%%%%%%%%%%%%%%%%%%%%%%%%%%%
\newcommand{\ApdaAACa}{
\subsection{
Proof of Property \ref{pr:pro1} Part b)
} 
\label{sub:ApdaAACa}
%}{

In this appendix we prove Property \ref{pr:pro1} part b). 
Fix $q=q_{UXY}\in{\cal Q}(p_{Y|X})$ and 
$p=p_{UXY}=(p_{U|X},p_{XY}) \in {\cal P}_{\rm sh}($ $p_{XY})$ arbitrary.  
For 
$\beta \geq 0$, 
$p \in {\cal P}_{\rm sh}(p_{XY})$, and $q_{Y|U}$ 
induced by $q$, define
\begin{align*}
& \hat{\omega}_{p,q_{Y|U}}^{(\mu)}(x,y|u)
\\
&\defeq \biggl[
           {\prmtA} \log \frac{p_{X|U}(x|u)}{p_{X}(x)}
+{{\prmtB}}\log \frac{1}{q_{Y|U}(y|u)}\biggr],
\\
& \hat{\Omega}^{(\mu,\pmt)}(p,q_{Y|U})
\defeq -\log {\rm E}_{p}
\left[\exp\left\{-\pmt
\hat{\omega}^{(\mu)}_{p,q_{Y|U}}(X,Y|U)\right\}\right].
\end{align*}

Then we have the following two lemmas.  
\begin{lm}\label{lm:lemmaSS}
For any 
$\mu$ $\in [0,1]$, $\alpha \in [0,1)$, 
and any $q=q_{UXY}\in {\cal Q}(p_{Y|X})$, 
there exists $p=p_{UXY}\in {\cal P}_{\rm sh}(p_{XY})$ such that 
\beq
{\Omega}^{(\mu,\alpha)}(q|p_{X})
\geq \bar{\alpha}\hat{\Omega}^{(\mu,\frac{\alpha}{\bar{\alpha}})}(p,q_{Y|U}).
\label{eqn:Zsss}
\eeq
\end{lm}

\begin{lm}\label{lm:lemmaSSz}
For any $\mu,\alpha$ satisfying $\mu$ $\in [0,1]$, $\alpha \in [0,1/2)$,  
any $p=p_{UXY}\in {\cal P}_{\rm sh}(p_{XY})$, and any stochastic matrix 
$q_{Y|U}$ induced by $q_{UXY}\in {\cal Q}(p_{Y|X})$, we have 
\begin{align}
& \hat{\Omega}^{(\mu, \frac{\alpha}{\bar{\alpha}} )}(p,q_{Y|U})
\geq \frac{1-{2\alpha}}{\bar{\alpha}}
\tilde{\Omega}^{(\mu,\frac{\alpha}{1-{2\alpha}})}(p).
\label{eqn:ZsssZZZ}
\end{align}
\end{lm}

From Lemmas \ref{lm:lemmaSS} and \ref{lm:lemmaSSz} we have 
the following corollary.
\begin{co}\label{co:CoOne} For any $\mu,\alpha$ 
satisfying $\mu$ $\in [0,1]$, $\alpha \in [0,1/2)$, 
and any $q=q_{UXY} \in {\cal Q}(p_{Y|X})$, 
there exists $p=p_{UXY}\in {\cal P}_{\rm sh}(p_{XY})$
such that 
\begin{align}
& \Omega^{(\mu,\alpha)}(q|p_{X})
 \geq (1-2\alpha)
\tilde{\Omega}^{(\mu,\frac{\alpha}{1-2\alpha})}(p).
\label{eqn:ZsssZZZqq}
\end{align}
From (\ref{eqn:ZsssZZZqq}), we have that for any 
$\mu$ $\in [0,1]$, $\alpha\in [0,1/2)$, we have 
\begin{align}
&\Omega^{(\mu,\alpha)}(p_{XY})
\geq (1-2\alpha)
\tilde{\Omega}^{(\mu,\frac{\alpha}{1-2\alpha})}
(p_{XY}).
\label{eqn:ZsssAZZqq}
\end{align}
\end{co}

{\it Proof of Lemma \ref{lm:lemmaSS}:}
We fix $(\mu,\alpha)\in [0,1]^2$ arbitrary. 
For each $q=q_{UXY} \in {\cal Q}(p_{Y|X})$, 
we choose 
$p=p_{UXY} \in {\cal P}_{\rm sh}(p_{XY})$ so that
$p_{U|X}=q_{U|X}$. Then we have the following: 
\begin{align}
&\exp\left\{-{\Omega}^{(\mu,\alpha)}(q|p_{X})\right\}
\nonumber\\
&
={\rm E}_q\left[
\frac{p^{\bar{\alpha}}_{X}(X)}{q^{\bar{\alpha}}_{X}(X)}
\left\{
\frac{p^{\prmtA \alpha}_{X}(X)
      q^{\prmtB \alpha}_{Y|U}(Y|U)}
{q^{{\prmtA}\alpha}_{X|U}(X|U)}
\right\}
\right]
\nonumber\\
&={\rm E}_q\left[\left\{
\frac{p_{UX}(U,X)}{q_{UX}(U,X)}\right\}^{\bar{\alpha}}
\left\{
\frac{p^{\prmtA \frac{\alpha}{\bar{\alpha}} }_{X}(X)
      q^{\prmtB \frac{\alpha}{\bar{\alpha}} }_{Y|U}(Y|U)}
{p^{{\prmtA} \frac{\alpha}{\bar{\alpha}}}_{X|U}(X|U)}
\right\}^{\bar{\alpha}}\right.
\nonumber\\
&  \quad \times\left\{
\frac{ p^{\prmtA }_{X|U}(X|U)}
     { q^{\prmtA }_{X|U}(X|U)}
              \right\}^{\alpha}
\Hugebr
\nonumber\\
&\MLeq{a}\left({\rm E}_q\left[
\frac{p_{UX}(U,X)}{q_{UX}(U,X)}
\frac{p^{{\prmtA} \frac{\alpha}{\bar{\alpha}} }_{X}(X)
      q^{{\prmtB} \frac{\alpha}{\bar{\alpha}} }_{Y|U}(Y|U)}
{p^{{\prmtA} \frac{\alpha}{\bar{\alpha}}}_{X|U}(X|U)}
\right]\right)^{\bar{\alpha}}
\nonumber\\
&  \quad \times
\left({\rm E}_q 
\left[
\frac{p^{ {\prmtA}}_{X|U}(X|U)}{q^{{\prmtA}}_{X|U}(X|U)}
\right]\right)^{{\alpha}}
\nonumber
\nonumber\\
&=\exp\left\{
-\bar{\alpha} \hat{\Omega}^{(\mu,\frac{\alpha}{\bar{\alpha}})}(p,q_{Y|U})\right\}
A^{{\alpha}},
\label{eqn:EddxSSS}
\end{align}
where we set
$$
A\defeq 
{\rm E}_q 
\left[
\frac{p^{ {\prmtA}}_{X|U}(X|U)}{q^{ {\prmtA}}_{X|U}(X|U)}
\right].
$$
Step (a) follows from H\"older's inequality. From (\ref{eqn:EddxSSS}), 
we can see that it suffices to show $A \leq 1$ to 
complete the proof. When $\mu=1$,  we have $A=1$. 
When $\mu \in [0,1)$, 
%%%%%%%%%%%%%%%%%%%%%%%%%%%%%%%%%%%%%%%%%%%%%%%%%%%%%%%%%%%%%%%%%%%%%
%$\pmt\frac{\alpha}{\bar{\alpha}}\leq 1$ or equivalent to
%$\alpha \in [0,\frac{1}{\beta+1}]$, 
%we have ${\prmtA}\beta \frac{\alpha}{\bar{\alpha}}\leq 1$.  
%%%%%%%%%%%%%%%%%%%%%%%%%%%%%%%%%%%%%%%%%%%%%%%%%%%%%%%%%%%%%%%
we apply H\"older's inequality to $A$ to obtain 
\begin{align*}
A&={\rm E}_q \left[
\frac{ p_{X|U}^{ {\prmtA}  }(X|U)}
     { q_{X|U}^{ {\prmtA}  }(X|U)}\right]
\leq \left({\rm E}_q\left[
\frac{p_{X|U}(X|U)}{q_{X|U}(X|U)}\right]
     \right)^{{\prmtA}}
=1.
\end{align*}
Hence we have (\ref{eqn:Zsss}) in Lemma \ref{lm:lemmaSS}.
\hfill \IEEEQED 

{\it Proof of Lemma \ref{lm:lemmaSSz}:} We fix $\mu$ $\in [0,1]$,  
$\alpha\in [0,1/2)$, arbitrary.
For any $p=p_{UXY}\in {\cal P}_{\rm sh}(p_{XY})$, 
and any $q=q_{UXY}\in {\cal Q}(p_{Y|X})$, we have the following 
chain of inequalities: 
\beqa
& &\exp \left\{ 
-\hat{\Omega}^{(\mu,\frac{\alpha}{\bar{\alpha}})}(p,q_{Y|U})
\right\}
\nonumber\\
&=&
{\rm E}_{p}
\HUgebl
\left\{
\frac{p^{\prmtA \frac{\alpha}{1-2\alpha} }_{X|U}(X|U)
      p^{\prmtB \frac{\alpha}{1-2\alpha} }_{Y|U}(Y|U)}
     {p^{\mu \frac{\alpha}{1-2\alpha}}_{X}(X)}
\right\}^{\frac{1-2\alpha}{\bar{\alpha}}}
\nonumber\\
& &\qquad \times 
\left\{\frac{q^{\prmtB}_{Y|U}(Y|U)}
            {p^{\prmtB}_{Y|U}(Y|U)}\right\}^{\frac{\alpha}{\bar{\alpha}}}
\HUgebr
\nonumber\\
&\MLeq{a}&
\exp \left\{-\frac{1-2\alpha}{\bar{\alpha}}
\tilde{\Omega}^{(\mu, \frac{\alpha}{1-2\alpha})}(p)\right\}
 \left({\rm E}_{p}\left[
                 \frac{q^{\prmtB}_{Y|U}(Y|U)}
                      {p^{\prmtB}_{Y|U}(Y|U)}
                \right]
\right)^{\frac{\alpha}{\bar{\alpha}}}
\nonumber\\
&=&
\exp \left\{-\frac{1-2\alpha}{\bar{\alpha}}
\tilde{\Omega}^{(\mu,\frac{\alpha}{1-2\alpha})}
  (p)\right\}B,
\label{eqn:EddxSSSb} 
\eeqa
where we set
$$
B\defeq 
{\rm E}_q 
\left[
\frac{q^{ {\prmtB}}_{Y|U}(Y|U)}{
      p^{ {\prmtB}}_{Y|U}(Y|U)}
\right].
$$
Step (a) follows from H\"older's inequality. 
From (\ref{eqn:EddxSSSb}), we can see that it suffices to show $B \leq 1$ to 
complete the proof. In a manner quite smilar to the proof 
of $A\leq 1$ in the proof of (\ref{eqn:Zsss}) in Lemma \ref{lm:lemmaSS}, 
we can show that $B \leq 1$. 
Thus we have (\ref{eqn:ZsssZZZ}) in 
Lemma \ref{lm:lemmaSSz}. 
\hfill \IEEEQED 

{\it Proof of Property \ref{pr:pro1} part b):} 
We evaluate lower bounds 
of $F(R_1,R_2|p_{XY})$ to obtain 
the following chain of inequalities:
\begin{align}
& F(R_1,R_2|p_{XY})
\notag\\
&\MGeq{a}
\sup_{ \scs  \mu \in [0,1],
     \atop{
      \scs  \alpha \in [0,1/2)
         }
}
\frac{(1-2\alpha)
\tilde{\Omega}^{(\mu,\frac{ \alpha }{1-2\alpha})}(p_{XY})
  -\alpha({\prmtA}R_1+ {\prmtB}R_2)}{2+ \alpha \prmtB}
\notag\\
&=\sup_{\scs \mu \in [0,1],
       \atop{\scs
       \alpha \in [0,1/2),
       \atop{\scs \lambda=\frac{\alpha}{1-2\alpha}
       }
   }
}
\frac{(1-2\alpha)
\tilde{\Omega}^{(\mu,\lambda)}(p_{XY})
  -\alpha({\prmtA}R_1+ {\prmtB}R_2)}{2+\alpha{\prmtB}}
\notag\\
&\MEq{b}
\sup_{ \scs \mu \in [0,1],
     \atop{\scs  \alpha=\frac{\lambda}{1+2\lambda},
           \lambda\geq 0
          }
     }
\frac{(1-2\alpha)
   \tilde{\Omega}^{(\mu,\lambda)}(p_{XY})
  -\alpha ( {\prmtA} R_1+ {\prmtB}R_2)}
{2+\alpha{\prmtB}}
\notag\\
&\MEq{c}
\sup_{\scs  \mu \in [0,1],\lambda\geq 0}
\frac{\tilde{\Omega}^{(\mu,\lambda)}(p_{XY})
  -\lambda({\prmtA}R_1+ {\prmtB}R_2)}{2+\lambda(5-\mu)}
\notag\\
&=\sup_{\scs \mu \in [0,1],\lambda \geq 0}
{\loF}^{(\mu,\pmt)}({\prmtA}R_1+{\prmtB}{R_2}|p_{XY}).
\label{eqn:ZZi}
\end{align}
Step (a) follows from the definition of $F(R_1,R_2|p_{XY})$ and 
$(\ref{eqn:ZsssAZZqq})$ in Corollary \ref{co:CoOne}. 
Steps (b) and (c) follow from that
$$
\alpha\in [0,1/2),
\lambda=\frac{\alpha}{1-2\alpha}\: 
\Leftrightarrow\:
\lambda \geq 0,\alpha=\frac{\lambda}{1+2\lambda}. 
$$
From (\ref{eqn:ZZi}), we have
\begin{align*}
& F(R_1,R_2|p_{XY})
\geq \sup_{\scs \mu\in [0,1],\lambda \geq 0} 
{\loF}^{(\mu,\lambda)}({\prmtA}R_1+ {\prmtB} R_2|p_{XY})
\\
&={\loF}(R_1,R_2|p_{XY}),
\end{align*}
completing the proof. 
\hfill\IEEEQED

}%%%%%%%%%%%%%%%%%%%%%%%%%%%%%%%%%%%%%%%%%%%%%%%%%%%%%%%%%%%%%%%%%%%%
%%%%%%%%%%%%%%%%%%%%%%%%%%%%%%%%%%%%%%%%%%%%%%%%%%%%%%%%%%%%%%%%%%%%%
%--------------------- End of \ApdaAACa ----------------------------%
%%%%%%%%%%%%%%%%%%%%%%%%%%%%%%%%%%%%%%%%%%%%%%%%%%%%%%%%%%%%%%%%%%%%%
%%%%%%%%%%%%%%%%%%%%%%%%%%%%%%%%%%%%%%%%%%%%%%%%%%%%%%%%%%%%%%%%%%%%%

%%%%%%%%%%%%%%%%%%%%%%%%%%%%%%%%%%%%%%%%%%%%%%%%%%%%%%%%%%%%%%%%%%%%%%%%%%%
%------------------------- Start of ApdaAAC ------------------------------%
%%%%%%%%%%%%%%%%%%%%%%%%%%%%%%%%%%%%%%%%%%%%%%%%%%%%%%%%%%%%%%%%%%%%%%%%%%%
\newcommand{\ApdaAAC}{
%}{
%%
\subsection{
Proof of Property \ref{pr:pro1} parts c), d), e, and f) 
}
\label{sub:ApdaAAC}

%}{

In this appendix we prove Property \ref{pr:pro1}
parts  c), d), e), and f). 
We first prove the part c).
and next prove the parts d) and e).
We finally prove the part f). 

{\it Proof of Property \ref{pr:pro1} part c):} \ 
We first prove the second inequality 
in (\ref{eqn:Asddx}) in the part c). 
We frist observe that 
\begin{align}
&\exp[-\tilde{\Omega}^{(\mu,\lambda)}(p)] 
=
{\rm E}_p\left[
\frac{ p_{X}^{\mu\lambda}(X) p_{Y|U}^{ \prmtB \lambda} (Y|U)}
{p_{X|U}^{\mu\lambda}(X|U)}
\right].
\label{eqn:Zspp}
\end{align}
Let $\bar{p}_{X}$ be the uniform distribution 
on ${\cal X}$ and 
let $\bar{p}_{Y}$ be the uniform distribution 
on ${\cal Y}$. 
On lower bound of 
$\exp[-\tilde{\Omega}^{(\mu,\lambda)}(p)]$ 
for $p\in {\cal P}_{\rm sh}(p_{XY})$ 
and $(\mu,\lambda)\in [0,1]^2$, 
we have the following chain of inequalities: 
\begin{align}
&\exp[-\tilde{\Omega}^{(\mu,\lambda)}(p)] 
=\frac{1}{|{\cal X}|^{\mu\lambda}|{\cal Y}|^{\prmtB\lambda} }
{\rm E}_{p}\Hugebl 
p_{X|U}^{-\mu\lambda}(X|U)
\notag\\
&\qquad \times 
\left\{\frac{{p}_{X}(X)}{\bar{p}_{X}(X)}\right\}^{\mu\lambda}
\left\{\frac{{p}_{Y|U}(Y|U)}{\bar{p}_{Y}(Y)}
\right\}^{\prmtB\lambda}
\Hugebr 
\notag\\
&\MGeq{a}\frac{1}{|{\cal X}|^{\mu}|{\cal Y}|^{\prmtB}}
{\rm E}_{p}\Hugebl 
\left\{\frac{\bar{p}_{X}(X)}{{p}_{X}(X)}\right\}^{-\mu\lambda}
\left\{\frac{\bar{p}_{Y} (Y)}{{p}_{Y|U}(Y|U)}
\right\}^{-\prmtB\lambda}
\Hugebr 
\notag\\
&\MGeq{b}\frac{1}{|{\cal X}|^{\prmtA} 
|{\cal Y}|^{\prmtB}}
\left(
{\rm E}_{p}\Hugebl \frac{\bar{p}_{X}(X)}{p_{X}(X)}\Hugebr
\right)^{-\mu\lambda}
\notag\\
&\qquad \times 
\left({\rm E}_{p}\Hugebl 
\frac{\bar{p}_{Y}(Y)}{p_{Y|U}(Y|U)}\Hugebr
\right)^{-\prmtB\lambda}
=\frac{1}{|{\cal X}|^{\prmtA} |{\cal Y}|^{\prmtB}}.
\label{eqn:Sdppx}
\end{align}
Step (a) follows from that $\lambda \in [0,1]$ and 
$p_{X|U}(x|u)\leq 1$ for 
any $(u,$ $x)\in {\cal U} \times {\cal X}$. 
Step (b) follows from the reverse H\"older's inequality.   
The bound (\ref{eqn:Sdppx})
implies the second inequality in (\ref{eqn:Asddx}). 
We next show that $\tilde{\Omega}^{(\mu,\lambda)}(p)
\geq 0$ for $\lambda\in [0,1]$. 
On upper bounds 
of $\exp[-\tilde{\Omega}^{(\mu,\lambda)}(p)]$ 
for $p\in {\cal P}_{\rm sh}(p_{XY})$ and $\lambda\in [0,1]$, 
we have the following chain of inequalities: 
\begin{align}
&\exp[-\tilde{\Omega}^{(\mu,\lambda)}(p)] 
\MLeq{a}
{\rm E}_{p}\Hugebl 
\left\{\frac{p_{X}(X)}{p_{X|U}(X|U)}\right\}^{\mu\lambda}
\Hugebr
\notag\\
&\MLeq{b}
\left\{{\rm E}_{p}\Hugebl 
\frac{p_{X}(X)}{p_{X|U}(X|U)}
\Hugebr
\right\}^{\mu\lambda}=1.
\end{align}
Step (a) follows from (\ref{eqn:Zspp}) and 
$p_{Y|U}(y|u) \leq 1$ for any 
$(u,y) \in {\cal U} \times {\cal Y}$. 
Step (b) follows from $\mu\lambda \in [0,1]$ and H\"older's 
inequality. 
\hfill\IEEEQED

{\it Proof of Property \ref{pr:pro1} parts  d) and e):} \ 
We first prove that for each 
$p\in {\cal P}_{\rm sh}(p_{XY})$ and $\mu \in[0,1]$, 
$\tilde{\Omega}^{(\mu,\lambda)}(p)$ is twice differentiable
for $\lambda\in [0,1/2]$. 
For simplicity of notations, set 
\begin{align*}
& 
\underline{a} \defeq (u,x,y), 
\underline{A}\defeq (U,X,Y),
\underline{\cal A} \defeq 
 {\cal U} \times 
 {\cal X} \times 
 {\cal Y},
\\
& \tilde{\omega}^{(\mu)}_{p}(x,y|u) 
\defeq \varsigma(\underline{a}),
\tilde{\Omega}^{(\mu,\lambda)}(p)
\defeq \xi(\lambda).
\end{align*}
Then we have 
\beq
\tilde{\Omega}^{(\mu,\lambda)}(p)=\xi(\lambda)=-\log
\left
[\sum_{\underline{a}\in \underline{\cal A} }
p_{\underline{A}}(\underline{a})
{\rm e}^{-\lambda \varsigma(\underline{a})}\right].
\label{eqn:SdfV}
\eeq
The quantity 
$p^{(\lambda)}(\underline{a})
=p_{\underline{A}}^{(\lambda)}(\underline{a}),
                             \underline{a}\in {\cal A}$
has the following form:
\beq
p%_{\underline{A}}
^{(\lambda)}(\underline{a})
={\rm e}^{\xi(\lambda)}
p%_{\underline{A}}
(\underline{a})
{\rm e}^{-\lambda \varsigma(\underline{a})}.
\label{eqn:aazP011}
\eeq
By simple computations we have 
\begin{align}
&\xi^{\prime}(\lambda)=
{\rm e}^{\xi(\lambda)}
\left[
\sum_{\underline{a}\in \underline{\cal A}} 
p%_{\underline{A}}
(\underline{a})
{\varsigma}(\underline{a}) 
{\rm e}^{-\lambda{\varsigma}(\underline{a})}\right]
%\nonumber\\
%&
=%& 
\sum_{\underline{a}\in \underline{\cal A}} 
p%_{\underline{A}}
^{(\lambda)}(\underline{a})
{\varsigma}(\underline{a}),
\nonumber
%\label{eqn:aaz11} 
%\eeqa
%\beqa
\\
&\xi^{\prime\prime}(\lambda)=-{\rm e}^{2\xi(\lambda)}
\nonumber\\
&\quad \times 
\left[
\sum_{\underline{a}, \underline{b}\in \underline{\cal A}}
 p%_{\underline{A}}
(\underline{a})p_{\underline{A}}(\underline{b})
\frac{\left\{{\varsigma}(\underline{a})-{\varsigma}(\underline{b})\right\}^2}{2}
{\rm e}^{-\lambda\left\{{\varsigma}(\underline{a})
+{\varsigma}(\underline{b})\right\}}\right]
\nonumber\\
&=-
\sum_{\underline{a}, \underline{b}\in \underline{\cal A}}
p%_{\underline{A}}
^{(\lambda)}(\underline{a})
p%_{\underline{A}}
^{(\lambda)}(\underline{b})
\frac{\left\{{\varsigma}(\underline{a})
-{\varsigma}(\underline{b})\right\}^2}{2}
\nonumber\\
&=-
\sum_{\underline{a}\in \underline{\cal A}}
p%_{\underline{A}}
^{(\lambda)}(\underline{a})
{\varsigma}^2(\underline{a})
+\left[
\sum_{\underline{a}\in \underline{\cal A}} 
p%_{\underline{A}}
^{(\lambda)}(\underline{a})
{\varsigma}(\underline{a})\right]^2 \leq 0.
\label{eqn:aaz011}
\end{align}
On upper bound of $-\xi^{\prime\prime}(\lambda) \geq 0$ 
for $\lambda\in [0,1/2]$, we have the following chain 
of inequalities:
\begin{align}
&-\xi^{\prime\prime}(\lambda)
\MLeq{a}\sum_{\underline{a}\in \underline{\cal A}}
p^{(\lambda)}(\underline{a})\varsigma^2(\underline{a})
\MEq{b}
\sum_{\underline{a}\in \underline{\cal A}}
p(\underline{a})\varsigma^2(\underline{a})
{\rm e}^{-\lambda{\varsigma}(\underline{a})+\xi(\lambda)}
\nonumber\\
&={\rm e}^{\xi(\lambda)}
\sum_{\underline{a}\in \underline{\cal A}}
p(\underline{a})\sqrt{{\rm e}^{-2\lambda{\varsigma}(\underline{a})}}
\sqrt{\varsigma^4(\underline{a})}
\nonumber\\
& \MLeq{c} 
\sqrt{{\rm e}^{2\xi(\lambda)-\xi(2\lambda)}}
%\left[
\sqrt{
\sum_{\underline{a}\in \underline{\cal A}}
p(\underline{a})\varsigma^4(\underline{a})}
\nonumber\\
& \MLeq{d} 
\sqrt{{\rm e}^{2\xi(\lambda)}}
\sqrt{
\sum_{\underline{a}\in \underline{\cal A}}
p(\underline{a})\varsigma^4(\underline{a})}.
\label{eqn:Sdppi}
\end{align}
Step (a) follows from (\ref{eqn:aaz011}).
Step (b) follows from (\ref{eqn:aazP011}).
Step (c) follows from Cauchy-Schwarz inequality and 
(\ref{eqn:SdfV}). 
Step (d) follows from that $\xi(2\lambda)\geq 0$ 
for $2\lambda\in [0,1]$.
Note that $\xi(\lambda)$ exists for 
$\lambda \in [0,1/2]$. 
Furtheomore we have the following:
$$
\sum_{\underline{a}\in \underline{\cal A}}
p(\underline{a})\varsigma^4(\underline{a})< \infty.
$$
Hence, by (\ref{eqn:Sdppi}), $\xi^{\prime\prime}(\lambda)$ exists
for $\lambda \in [0,1/2]$. 
\irr{
We next prove the part e). 
We derive the lower bound (\ref{eqn:ZssPi}) of 
$\tilde{\Omega}^{(\mu,\lambda)}(p_{XY})$. 
Fix any $(\mu,\lambda) \in [0,1]$ $\times [0,1/2]$ 
and any $p \in {\cal P}_{\rm sh}(p_{XY})$.} 
By the Taylor expansion of 
$\xi(\lambda)=\tilde{\Omega}^{(\mu,\lambda)}(p)$
with respect to $\lambda$ around $\lambda=0$,
we have \irr{that for any $p\in {\cal P}_{\rm sh}(p_{XY})$ 
and for some $ \nu \in [0,\lambda]$}
\begin{align}
& \tilde{\Omega}^{(\mu,\lambda)}(p)
%\notag\\
%& 
=\xi(0)+ \xi^{\prime}(0)\lambda
+\frac{1}{2}\xi^{\prime\prime}(\irr{\nu})\irr{\lambda^2}
\notag\\
& =\lambda {\rm E}_{p}
\left[\tilde{\omega}^{(\mu)}_p(X,Y|U)\right]
%\notag\\
%&\quad 
-\frac{\lambda^2}{2}{\rm Var}_{\irr{p^{(\nu)}}}
\left[\tilde{\omega}^{(\mu)}_{p}(X,Y|U)\right]
\notag\\
&\irr{\MGeq{a}}
\lambda {R}^{(\mu)}(p_{XY})
-\frac{\lambda^2}{2}{\rm Var}_{\irr{p^{(\nu)}}}
\left[\tilde{\omega}^{(\mu)}_{p}(X,Y,Z|U)\right].
\label{eqn:ZsddAA}
\end{align}
\irr{ Step (a) follows from $p\in {\cal P}_{\rm sh}(p_{XY})$, 
$$
{\rm E}_{p}
\left[\tilde{\omega}^{(\mu)}_p(X,Y|U)\right]
= \prmtA I_p(X;U) +\prmtB H_p(Y|U),
$$
and the definition of ${R}^{(\mu)}(p_{XY})$.
Let $(\nu_{\rm opt},p_{\rm opt})$ 
$\in [0, \lambda] \times $ ${\cal P}_{\rm sh}(p_{XY})$ 
be a pair which attains $\rho^{(\mu,\lambda)}(p_{XY})$.
By this definition %of $(\gamma_{\rm opt}, q_{\rm opt})$,
we have that
\begin{align}
&\tilde{\Omega}^{(\mu, \lambda)}(p_{\rm opt})
=\tilde{\Omega}^{(\mu, \lambda)}(p_{XY}) 
\label{eqn:asWWWd}
\end{align}
and that for any $\nu \in [0,\lambda],$
\begin{align}
&{\rm Var}_{ p_{\rm opt}^{(\nu)}}
\left[\omega_{ p_{\rm opt} }^{(\mu)}(X,Y|U)\right]
\notag\\
&\leq {\rm Var}_{p_{\rm opt}^{(\nu_{\rm opt})}}
   \left[\omega_{p_{\rm opt}}^{(\mu)}(X,Y|U)\right]
=\rho^{(\mu,\lambda)}(p_{XY}).
\label{eqn:asWxxd}
\end{align}
On lower bounds of $\Omega^{(\mu, \lambda)}(p_{XY})$,} 
we have the following chain of inequalities:
\begin{align*}
& \tilde{\Omega}^{(\mu,\lambda)}(p_{XY})\MEq{a} 
   \tilde{\Omega}^{(\mu,\lambda)}(\irr{p_{\rm opt}}) 
\nonumber\\
&\MGeq{b} \lambda R^{(\mu)}(p_{XY})
-\frac{\lambda^2}{2}
{\rm Var}_{\irr{p_{\rm opt}^{(\nu)}}}
\left[\tilde{\omega}^{( \mu)}_{\irr{p_{\rm opt}}}(X,Y|U)\right]
\\
&\MGeq{c} \lambda {R}^{(\mu)}(p_{XY})
-\frac{\lambda^2}{2}\rho^{(\mu,\lambda)}(p_{XY})
\\
&\MGeq{d} \lambda R^{(\mu)}(p_{XY})-\frac{\lambda^2}{2}\rho(p_{XY}).
\end{align*}
Step (a) follows from \irr{(\ref{eqn:asWWWd})}.
Step (b) follows from (\ref{eqn:ZsddAA}).
Step (c) follows from \irr{(\ref{eqn:asWxxd})}.
Step (d) follows from the definition of 
$\rho(p_{XY})$.\hfill \IEEEQED

To prove the part f), we use the following lemma.
\begin{lm}\label{lm:LemSaS}
When $\tau \in (0, (1/2)\rho]$, the maximum of 
$$
\frac{1}{2+5\lambda}
\left\{-\frac{\rho}{2}\lambda^2 + {\tau} \lambda
\right\}
$$
for $\lambda \in (0,1/2]$ is attained 
by the positive $\lambda_0$ satisfying
\beq
\vartheta(\lambda_0)\defeq \lambda_0+\frac{5}{4}\lambda_0^2
=\frac{\tau}{\rho}.
\label{eqn:AsssD}
\eeq
Let $g(a)$ be the inverse function of 
$\vartheta(a)$ for $a\geq0$. Then the condition 
of (\ref{eqn:AsssD}) is equivalent to 
$\lambda_0=g(\frac{\tau}{\rho})$. 
The maximum is given by 
$$
\frac{1}{2+5\lambda_0}
\left\{
-\frac{\rho}{2}\lambda_0^2+{\tau} \lambda_0
\right\}=\frac{\rho}{4}\lambda^2_0
=\frac{\rho}{4}g^2\left(\frac{\tau}{\rho}\right).
$$
\end{lm}   

By an elementary computation we can prove this lemma. 
We omit the detail.

{\it Proof of Property \ref{pr:pro1} part f):}
By the hyperplane expression ${\cal R}_{\rm sh}(p_{XY})$ of 
${\cal R}(p_{XY})$ stated Property \ref{pr:pro0z} part b)
we have that when 
$(R_1+\tau, R_2+\tau) \notin {\cal R}(p_{XY})$, 
we have 
\beqa 
& &R^{(\mu_0)}(p_{XY})-(\mu_0 R_1+\overline{\mu_0}{R_2})>\tau 
\label{eqn:Xddcc}
\eeqa 
for some $ \mu_0 \in [0,1]$. Then for each positive $\tau$,
we have the following chain of inequalities: 
\begin{align*}
& {\loF}(R_1,R_2|p_{XY})
\geq \sup_{\lambda \in (0,1/2]}
{\loF}^{(\mu_0,\lambda)}
( \mu_0 R_1 + \overline{\mu_0} R_2|p_{XY})
\\
&=
\sup_{\lambda \in (0,1/2]}
\frac{\tilde{\Omega}^{(\mu_0,\lambda)}(p_{XY})
  -\lambda({\mu_0} R_1+ \overline{\mu_0}{R_2})}
{2+\lambda(5-\mu_0)}
\\
&\MGeq{a}  
\sup_{\lambda \in (0,1/2]}
\frac{1}{2 + 5 \lambda}
\left\{ -\frac{\rho}{2} \lambda^2\right.
\\
&\qquad \qquad +\lambda {R}^{({\mu_0})}(p_{XY})
-\lambda( \mu_0 R_1+ \overline{\mu_0} R_2)\biggr\}
\\
&\MG{b} 
\sup_{\lambda \in (0,1/2]}
\frac{1}{2+ 5 \lambda}
\left\{
-\frac{\rho}{2}\lambda^2+ \tau\lambda
\right\}
\MEq{c}
\frac{\rho}{4}g^2\left(\frac{\tau}{\rho}\right).
\end{align*}
Step (a) follows from Property \ref{pr:pro1} part d).
Step (b) follows from (\ref{eqn:Xddcc}).
Step (c)  follows from Lemma \ref{lm:LemSaS}. 
\hfill\IEEEQED
}%%%%%%%%%%%%%%%%%%%%%%%%%%%%%%%%%%%%%%%%%%%%%%%%%%%%%%%%%%%%%%%%%%%%
Our main result is the following.
\begin{Th}\label{Th:main} 
For any $R_1,R_2\geq 0$, any $p_{XY}$, and 
for any $(\varphi^{(n)}_1,$ $\varphi^{(n)}_1,$ 
$\psi^{(n)})$ satisfying 
$
(1/n)\log ||\varphi_i^{(n)}|| \leq R_i, i=1,2, 
$
we have 
\beq
{\rm P}_{\rm c}^{(n)}(\varphi_1^{(n)},
\varphi_2^{(n)},\psi^{(n)})
\leq 5\exp \left\{-n F(R_1,R_2|p_{XY})\right\}.
\label{eqn:mainIeq}
\eeq
\end{Th}

It follows from Theorem \ref{Th:main} and Property \ref{pr:pro1} 
part d) that if $(R_1,R_2)$ is outside the capacity region, then 
the error probability of decoding goes to one exponentially 
and its exponent is not below $F(R_1,R_2|p_{XY})$. It immediately 
follows from Theorem \ref{Th:main} that 
we have the following corollary. 
\begin{co}\label{co:mainCo}
\begin{align*}
& G(R_1,R_2|p_{XY})\geq F(R_1,R_2|p_{XY}),
\\
& {\cal G}(p_{XY})
\subseteq 
\overline{\cal G}(p_{XY})
\\
& =\left\{(R_1,R_2,G):
G\geq 
F(R_1,R_2|p_{XY})
\right\}.
\end{align*}
\end{co}

Proof of Theorem \ref{Th:main} will be given in the next section. The 
exponent function at rates outside the rate region was derived by Oohama 
and Han \cite{OhHan94} for the separate source coding problem for 
correlated sources \cite{sw}. The techniques used by them is a method of 
types \cite{ckBook81}, which is not useful to prove Theorem \ref{Th:main}. 
Some novel techniques based on the information spectrum method 
introduced by Han \cite{han} are necessary to prove this theorem.

From Theorem \ref{Th:main} and Property \ref{pr:pro1} part e), 
we can obtain an explicit outer bound of 
${\cal R}_{\rm AKW}(\varepsilon |p_{XY})$ 
with an asymptotically vanishing deviation from 
${\cal R}_{\rm AKW}(p_{XY})$ $={\cal R}(p_{XY})$. 
The strong converse theorem established by Ahlswede {\it et 
al.} \cite{agk76} immediately follows from this corollary. 
%%%%%%%%%%%%%%%%%%%%%%%%%%%%%%%%%%%%%%%%%%%%%%%%%%%%%%%%%%%%%%%%%%%%%%%%%%
To discribe this outer bound, for $\kappa>0$, we set 
\begin{align*}
&  {\cal R}(p_{XY})-\kappa(1,1)
\\
& \defeq  \{(R_1-\kappa, R_2-\kappa): (R_1,R_2) 
\in {\cal R}(p_{XY})\}, 
\end{align*}
which serves as an outer bound of ${\cal R}(p_{XY})$. 
%%%%%%%%%%%%%%%%%%%%%%%%%%%%%%%%%%%%%%%%%%%%%%%%%%%%%%%%%%%%%%%%%%%%%%%%%%
%%%%%%%%%%%%%%%%%%%%%%%%%%%%%%%%%%%%%%%%%%%%%%%%%%%%%%%%%%%%%%%%%%%%%%%%%%
For each fixed $\varepsilon\in(0,1)$, we define  
$\kappa_n$$=\kappa_n(\varepsilon,\rho(p_{XY}))$ by
\beqa
\kappa_n&\defeq&
\rho(p_{XY}) \vartheta\left(
\sqrt{ \frac{4}{n\rho(p_{XY})} 
\log\left(\frac{5}{1-\varepsilon}\right)} 
\right)
\label{eqn:zdd}\\
&\MEq{a}&
2\sqrt{%\ts
\frac{\rho(p_{XY})}{n}
\log\left(\frac{5}{1-\varepsilon}\right)}
+ \frac{5}{n}
\log\left(\frac{5}{1-\varepsilon}\right).
\nonumber
\eeqa
Step (a) follows from $\vartheta(a)=a+(5/4)a^2$.  
Since $\kappa_n \to 0$ as $n\to \infty$, we have the 
smallest positive integer $n_0=n_0(\varepsilon,\rho(p_{XY}))$
such that $\kappa_n \leq \irr{(1/2)}\rho(p_{XY})$ for $n\geq n_0$.
From Theorem \ref{Th:main} and Property \ref{pr:pro1} 
part e), we have the following corollary.
\begin{co}
\label{co:StConv} 
For each fixed $\varepsilon$ $ \in (0,1)$, we choose the above 
positive integer $n_0=$$n_0(\varepsilon,\rho(p_{XY}))$. 
Then, for any $n\geq n_0$, we have
\begin{align*}
& &{\cal R}_{\rm AKW}(n,\varepsilon |p_{XY})
\subseteq 
{\cal R}(p_{XY})-\kappa_n(1,1). 
\end{align*}
The above result together with 
\begin{align*}
%&&
{\cal R}_{\rm AKW}(\varepsilon | p_{XY})
%\\
&={\rm cl}\left(\bigcup_{m\geq 1}
\bigcap_{n \geq m}{\cal R}_{\rm AKW}(n,\varepsilon | p_{XY})
\right)
\end{align*}
yields that for each fixed $\varepsilon \in (0,1)$, we have 
\begin{align*}
& {\cal R}_{\rm AKW}(\varepsilon | p_{XY})
  ={\cal R}_{\rm AKW}(p_{XY})
  ={\cal R}(p_{XY}).
\end{align*}
This recovers the strong converse theorem proved by 
Ahlswede {\it et al.} \cite{agk76}.
\end{co}

Proof of this corollary will be given in the next section.
%%%%%%%%%%%%%%%%%%%%%%%%%%%%%%%%%%%%%%%%%%%%%%%%%%%%%%%%%%%%
%%%%%%%%%%%%%%%%%%%%%%%%%%%%%%%%%%%%%%%%%%%%%%%%%%%%%%%%%%%%
%%%%%%%%%%%%%%%%%% Start of \ProofCor %%%%%%%%%%%%%%%%%%%%%%
\newcommand{\ProofCor}{  
%}{

{\it Proof of Corollary \ref{co:StConv}:} 
Since $g$ is an inverse function of $\vartheta$, the definition 
(\ref{eqn:zdd}) of $\kappa_n$ is equivalent to 
\beq
g\left(
\frac{\kappa_n}{\rho(p_{XY})}\right)
=\sqrt{%\ts 
\frac{4}{n\rho(p_{XY})} \log\left(\frac{5}{1-\varepsilon}\right)}.
\label{eqn:zddQ}
\eeq
By the definition of $n_0=n_0(\varepsilon,\rho(p_{XY}))$, 
we have that $\kappa_n \leq (1/2)\rho(p_{XY})$ 
for $n\geq n_0$. We assume that for $n\geq n_0$,
$(R_1,R_2) \in {\cal R}_{\rm AKW}(n,\varepsilon| p_{XY}).$
Then there exists a sequence 
$\{(\varphi_1^{(n)},$ 
   $\varphi_2^{(n)},$ 
        $\psi^{(n)})$ 
$\}_{n\geq n_0}$ such that for $n\geq n_0$, we have
\begin{align}
& \frac{1}{n}\log ||\varphi_i^{(n)}||\leq R_i, i=1,2,
\notag\\
& 
1-\varepsilon
\leq {\rm P}_{\rm c}^{(n)}(\varphi^{(n)}_1,
\varphi^{(n)}_2,\psi^{(n)}).
\label{eqn:ZaaBB}
\end{align}
Then by Theorem \ref{Th:main}, we have
\beqa
1-\varepsilon&\leq& 
{\rm P}_{\rm c}^{(n)}(\varphi_1^{(n)},\varphi_2^{(n)},\psi^{(n)})
\nonumber\\
&\leq& 5\exp \left\{-n F(R_1,R_2|p_{XY})\right\}
\label{eqn:Zsddd}
\eeqa
for any $n\geq n_0(\varepsilon,\rho(p_{XY}))$. 
From (\ref{eqn:Zsddd}), we have that 
for $n \geq n_0(\varepsilon,\rho($ $p_{XY}))$, 
\begin{align} 
&F(R_1,R_2|p_{XY}) 
\notag \\
&\leq \frac{1}{n}\log\left(\frac{5}{1-\varepsilon}\right)
\MEq{a}\frac{\rho(p_{XY})}{4}\cdot 
g^2\left(\frac{\kappa_{n}}{\rho(p_{XY})}\right).
\label{eqn:ZsddP}
\end{align} 
Step (a) follows from (\ref{eqn:zddQ}). Hence 
by Property \ref{pr:pro1} part e), we have that 
under $\kappa_n \leq (1/2)\rho(p_{XY})$, 
the inequality (\ref{eqn:ZsddP}) implies 
\beq
(R_1,R_2)\in {\cal R}(p_{XY})+\kappa_n(1,1).
\label{eqn:ZsDDc}
\eeq
Since (\ref{eqn:ZsDDc}) holds for any $n \geq n_0$ and 
$(R_1,R_2) \in {\cal R}_{\rm AKW}($ 
$n,\varepsilon| p_{XY})$, we have  
$$
{\cal R}_{\rm AKW}(n,\varepsilon |p_{XY}) 
\subseteq {\cal R}(p_{XY})+\kappa_n(1,1)
\mbox{ for }n\geq n_0, 
$$ 
completing the proof.
\hfill\IEEEQED
}%%%%%%%%%%%%%%%%%%%%%%%%%%%%%%%%%%%%%%%%%%%%%%%%%%%%%%%%%%%
%%%%%%%%%%%%%%%%%%%%%%%%%%%%%%%%%%%%%%%%%%%%%%%%%%%%%%%%%%%%
%%%%%%%%%%%%%%%%%%%% End of \ProofCor %%%%%%%%%%%%%%%%%%%%%%

\section{Proof of the Main Result}
\label{sec:Secaa}
% If the distribution of random variables 
% is obvious from the context,
% we omit it in the subscript 
% of information measure on this randome variable.

Let $(X^n,Y^n)$ be a pair of random variables from 
the information source. We set $S=\varphi_1^{(n)}(X^n)$. 
Joint distribution $p_{SX^nY^n}$ of $(S, X^n,Y^n)$
is given by
\begin{align*}
& p_{SX^nY^n}(s,x^n,y^n)=p_{S|X^n}(s|x^n)
\prod_{t=1}^n p_{X_tY_t}(x_t,y_t). 
\end{align*}
It is obvious that 
$S \markov X^n \markov Y^n$. Then we have the following. 
\begin{lm}\label{lm:Ohzzz}
For any $\eta>0$ and for any $(\varphi_1^{(n)}$, 
$\varphi_2^{(n)},\psi^{(n)})$ satisfying 
$
(1/n)\log {||\varphi_i^{(n)}||}\leq R_i, i=1,2,
$
we have 
\begin{align}
& {\rm P}_{\rm c}^{(n)}(\varphi_1^{(n)},\varphi_2^{(n)},\psi^{(n)})
%\nonumber\\
\leq 
p_{SX^nY^n}\biggl\{
\nonumber\\
%%%%%%%%%%%%%%%%%%%%%%%%%%%%%%%%%%%%%%%%%%%%%%%%%%%%%%%%%%%%%%%
%\sum_{t=1}^n
%\log
%\frac {p_{Y_t|SX^{t-1}Y^{t-1}}(Y_t|S,X^{t-1},Y^{t-1})}
%      {p_{Y_t|SY^{t-1}}(Y_t|S,Y^{t-1})}
%\right.
%\nonumber\\
%& &\quad\:\:\geq -\eta,
%%%%%%%%%%%%%%%%%%%%%%%%%%%%%%%%%%%%%%%%%%%%%%%%%%%%%%%%%%%%%%%
& \:\:\,\ds 
0 \geq \frac{1}{n}\log
\frac{\hat{q}_{{SX^nY}^n}(S,X^n,Y^n)}{p_{SX^nY^n}(S,X^n,Y^n)}-\eta,
\label{eqn:asppa}\\
& \:\:\,\ds 
0 \geq \frac{1}{n}\log \frac{ Q_{{X}^n}(X^n)}{p_{X^n}(X^n)}-\eta,
\label{eqn:asppb}\\
& \ds 
R_1\geq \frac{1}{n}\log
\frac{\tilde{Q}_{X^n|S}(X^n|S)}{p_{X^n}(X^n)}-\eta,
\label{eqn:asppc}\\
& \ds 
R_2 \geq \left.
\frac{1}{n}\log\frac{1}{p_{Y^n|S}(Y^n|S)}-\eta
\right\}
+4{\rm e}^{-n\eta}. 
\label{eqn:azsad}
\end{align}
The probability distributions appearing 
in the three inequalities 
(\ref{eqn:asppa}), 
(\ref{eqn:asppb}), and (\ref{eqn:asppc}) 
in the right members of (\ref{eqn:azsad}) have 
a property that we can select them arbitrary. 
In (\ref{eqn:asppa}), we can choose any probability 
distribution $\hat{q}_{S{X}^nY^n}$ on 
${\cal S}$$\times{\cal X}^n$$\times{\cal Y}^n$. 
In (\ref{eqn:asppb}), we can choose any 
distribution ${Q}_{{X}^n}$ on ${\cal X}^n$. 
In (\ref{eqn:asppc}), we can choose any 
stochastic matrix $\tilde{Q}_{X^n|U^n}$: ${\cal X}^n$ 
$\to {\cal U}^n$. 
\end{lm}

Proof of this lemma is given in Appendix \ref{sub:Apda}.
\newcommand{\Apda}{%---------------------------------------------%
\subsection{
Proof of Lemma \ref{lm:Ohzzz}
}\label{sub:Apda}

To prove Lemma \ref{lm:Ohzzz}, we prepare a lemma. Set
$$
{\cal A}_n
\defeq 
\left\{(s,x^{n},y^{n}): 
\frac{1}{n}\log 
\frac {p_{SX^nY^n}(s,x^n,y^n)}{\hat{q}_{SX^nY^n}(s,x^n,y^n)}
\geq -\eta
\right\}.
$$
Furthermore, set
\begin{align*}
& \tilde{\cal B}_n
\defeq 
\left\{x^{n}: 
\frac{1}{n}\log 
\frac {p_{X^n}(x^n)}{Q_{{X}^n}(x^n)}
\geq -\eta
\right\},
\\
&{\cal B}_n \defeq \tilde{\cal B}_n \times {\cal M}_1\times {\cal Y}^n,
  {\cal B}_n^{\rm c}
  \defeq \tilde{\cal B}_n^{\rm c}\times {\cal M}_1\times {\cal Y}^n,
\\
&\tilde{\cal C}_n\defeq
\{(s,{x^n}): 
\ba[t]{l}
s=\varphi_1^{(n)}({x^n}),\\ 
p_{X^n|S}({ x^n}|s)\leq M_1 {\rm e}^{n\eta}p_{X^n}({ x^n})\}, 
\ea
\\
&{\cal C}_n\defeq \tilde{\cal C}_n\times {\cal Y}^n,
  {\cal C}_n^{\rm c}
  \defeq \tilde{\cal C}_n^{\rm c}\times {\cal Y}^n,
\\
&{\cal D}_n\defeq
\{(s,{x^n},{y^n}): 
\ba[t]{l}
s=\varphi_1^{(n)}({x}^n),\\ 
p_{Y^n|S}({ y^n}|s)\geq (1/M_2) {\rm e}^{-n\eta}\},
\ea
\\
&{\cal E}_n\defeq
\{(s,{x^n},{y^n}): 
\ba[t]{l}
s=\varphi_1^{(n)}({x^n}),\\ 
\psi^{(n)}(\varphi_1^{(n)}({ x^n}),
\varphi_2^{(n)}({y^n}))={y^n} \}. 
\ea
\end{align*}
Then we have the following lemma. 
\begin{lm}\label{lm:zzxa}{%For 
\begin{align*}
& 
p_{SX^nY^n}
\left({\cal A}_n^{\rm c}\right)\leq {\rm e}^{-n\eta}, 
p_{SX^nY^n}
\left({\cal B}_n^{\rm c}\right)\leq {\rm e}^{-n\eta}, 
\\
& p_{SX^nY^n}
\left({\cal C}_n^{\rm c} \right)\leq {\rm e}^{-n\eta},
%\\
%& &
p_{SX^nY^n}
\left( {\cal D}_n^{\rm c}\cap {\cal E}_n \right)
\leq {\rm e}^{-n\eta}.
\end{align*}
}
\end{lm}

{\it Proof:} We first prove the first inequality. 
\begin{align*}
& p_{SX^nY^n}
({\cal A}_n^{\rm c})
=\sum_{(s,x^n,y^n) \in {\cal A}_n^{\rm c}}
      p_{SX^nY^n}(s,x^n,y^n)
\\
&\MLeq{a}\sum_{(s,x^n,y^n)\in {\cal A}_n^{\rm c}}
{\rm e}^{-n\eta}\hat{q}_{SX^nY^n}(s,x^n,y^n)
\\
&={\rm e}^{-n\eta}\hat{q}_{SX^nY^n}
\left( {\cal A}_n^{\rm c}\right)
\leq {\rm e}^{-n\eta}.
\end{align*}
Step (a) follows from the definition of ${\cal A}_n$.   
On the second inequality we have   
\begin{align*}
& p_{SX^nY^n}
({\cal B}_n^{\rm c})
= p_{X^n}
(\tilde{\cal B}_n^{\rm c})
=\sum_{x^n\in \tilde{\cal B}_n^{\rm c} } p_{X_n}(x^n)
\\
&\MLeq{a}\sum_{x^n\in \tilde{\cal B}_n^{\rm c}}
{\rm e}^{-n\eta}Q_{{X}^n }(x^n)
={\rm e}^{-n\eta}
Q_{X^n}\left( \tilde{\cal B}_n^{\rm c}\right)
\leq {\rm e}^{-n\eta}.
\end{align*}
Step (a) follows from the definition of ${\cal B}_n$.   
We next prove the third inequality. 
\begin{align*}
& p_{SX^nY^n}
   ({\cal C}_n^{\rm c})
   =p_{SX^n}(\tilde{\cal C}_n^{\rm c})
\\
&=\sum_{s\in{\cal M}_1}
\sum_{\scs x^n: \varphi_1^{(n)}({x^n})=s
\atop{\scs
      p_{X^n}(x^n)\leq (1/M_1){\rm e}^{-n\eta}
      \atop{\scs
           \qquad\qquad \times \tilde{Q}_{X^n|S}(x^n|s)
           }
     }
}
p_{X^n}(x^n)
\\
&\leq 
\frac{1}{M_1}{\rm e}^{-n\eta}
\sum_{s\in{\cal M}_1}
\sum_{\scs x^n: \varphi_1^{(n)}({x^n})=s
\atop{\scs
      p_{X^n}(x^n)\leq (1/M_1){\rm e}^{-n\eta}
      \atop{
            \scs \qquad\qquad\times \tilde{Q}_{X^n|S}(x^n|s)
           }
     }
}
\tilde{Q}_{X^n|S}(x^n|s)
\\
&\leq \frac{1}{M_1}{\rm e}^{-n\eta}|{\cal M}_1|
={\rm e}^{-n\eta}.
\end{align*}
Finally we prove the fourth inequality. We first observe that 
\begin{align*}
& &p_S(s)=\sum_{\scs x^n: \varphi_1^{(n)}({x^n})=s}p_{X^n}(x^n),\:
%\\
%& &
p_{X^n|S}(x^n|s)=\frac{p_{X^n}(x^n)}{p_{S}(s)}.
\end{align*}
We have the following chain of inequalities:
\begin{align*}
& p_{SX^nY^n}
\left({\cal D}_n^{\rm c}\cap {\cal E}_n\right)
\\
&=\sum_{s\in{\cal M}_1}p_S(s)
\sum_{\scs x^n: \varphi_1^{(n)}({x^n})=s}
p_{X^n|S}(x^n|s)
\\
&\quad \times
\sum_{\scs y^n: 
      \psi^{(n)}(s,\varphi_2^{(n)}(y^n))=y^n  
      \atop{\scs
           p_{Y^n|S}(y^n|s)\leq (1/M_2){\rm e}^{-n\eta}
     }
}
p_{Y^n|X^n}(y^n|x^n)
\\
&=\sum_{s\in{\cal M}_1}p_S(s)
\sum_{\scs y^n: 
      \psi^{(n)}(s,\varphi_2^{(n)}(y^n))=y^n  
      \atop{\scs
           p_{Y^n|S}(y^n|s)\leq (1/M_2){\rm e}^{-n\eta}
     }
}
p_{Y^n|S}(y^n|s)
\\
&\leq 
\sum_{s\in{\cal M}_1}p_S(s)
\frac{1}{M_2}{\rm e}^{-n\eta}
\left|\left\{ 
y^n:\psi^{(n)}(s,\varphi_2^{(n)}(y^n))=y^n
\right\}\right|
\\
&\MLeq{a}\sum_{s\in{\cal M}_1}p_S(s) 
\frac{1}{M_2}{\rm e}^{-n\eta}M_2
={\rm e}^{-n\eta}.
\end{align*}
Step (a) follows from that the number of 
$y^n$ correctly decoded does not exceed $M_2$.   
\hfill\IEEEQED

{\it Proof of Lemma \ref{lm:Ohzzz}:} 
By definition we have
\begin{align*}
& p_{SX^nY^n}
\left(
{\cal A}_n\cap {\cal B}_n\cap {\cal C}_n
\cap {\cal D}_n
\right)
\\
&=
p_{SX^nY^n}
\ba[t]{l}
\left\{
\ds 
\frac{1}{n}\log \frac{p_{SX^nY^n}(S,X^n,Y^n)} 
{\hat{q}_{{SX^nY}^n}(S,X^n,Y^n)}
\geq -\eta,
\right.
\\
\quad \ds 
0 \geq \frac{1}{n}\log \frac{ q_{X^n}(X^n)}{p_{X^n}(X^n)}-\eta,
\\
\:\:\ds 
\frac{1}{n}\log M_1 \geq \frac{1}{n}\log
\frac{p_{X^n|S}(X^n|S)}{p_{X^n}(X^n)}-\eta,
\\
\:\:\ds 
\frac{1}{n}\log M_2 \geq \left.
\frac{1}{n}\log\frac{1}{p_{Y^n|S}(Y^n|S)}-\eta
\right\}.
\ea
\end{align*}
%---------------------------------------------------------------%
%\newcommand{\emPty}
%
%
{
Then for any $(\varphi_1^{(n)}$, $\varphi_2^{(n)},\psi^{(n)})$ 
satisfying 
$
(1/n)\log {||\varphi_i^{(n)}||}\leq R_i, i=1,2,
$
we have 
\begin{align*}
& p_{SX^nY^n}
\left(
{\cal A}_n\cap {\cal B}_n\cap {\cal C}_n\cap {\cal D}_n
\right)
\\
&\leq
p_{SX^nY^n}
\ba[t]{l}
\left\{
\ds 
\frac{1}{n}\log
\frac{p_{SX^nY^n}(S,X^n,Y^n)}{\hat{q}_{{SX^nY}^n}(S,X^n,Y^n)}
\geq -\eta, \right.
\\
\quad \ds 
0 \geq \frac{1}{n}\log \frac{ q_{X^n}(X^n)}{p_{X^n}(X^n)}-\eta,
\\
\:\:\ds 
R_1\geq \frac{1}{n}\log
\frac{p_{X^n|S}(X^n|S)}{p_{X^n}(X^n)}-\eta,
\\
\:\:\ds 
R_2 \geq \left.
\frac{1}{n}\log\frac{1}{p_{Y^n|S}(Y^n|S)}-\eta
\right\}.
\ea
\end{align*}
}
%
%============================================================%
%
Hence, it suffices to show 
\begin{align*}
{\rm P}_{\rm c}^{(n)}(\varphi_1^{(n)},\varphi_2^{(n)},\psi^{(n)})
&\leq
p_{SX^nY^n}\left({\cal A}_n
 \cap {\cal B}_n
 \cap {\cal C}_n \cap {\cal D}_n\right)
\\
& +4{\rm e}^{-n\eta}
\end{align*}
to prove Lemma \ref{lm:Ohzzz}. By definition 
we have
\begin{align*} 
&{\rm P}_{\rm c}^{(n)}(\varphi_1^{(n)},\varphi_2^{(n)},\psi^{(n)})
=p_{SX^nY^n}\left({\cal E}_n\right).
\end{align*}
Then we have the following.
\begin{align*}
&{\rm P}_{\rm c}^{(n)}(\varphi_1^{(n)},\varphi_2^{(n)},\psi^{(n)})
=p_{SX^nY^n}\left({\cal E}_n\right)
\\
&=
p_{SX^nY^n}\left(
     {\cal A}_n
\cap {\cal B}_n
\cap {\cal C}_n
\cap {\cal D}_n
\cap {\cal E}_n
\right)
\\
&\quad +p_{SX^nY^n}
\left(
\left[{\cal A}_n
 \cap {\cal B}_n
 \cap {\cal C}_n 
 \cap {\cal D}_n\right]^{\rm c}
 \cap {\cal E}_n 
\right)
\\
&\leq
p_{SX^nY^n}
\left({\cal A}_n
 \cap {\cal B}_n
 \cap {\cal C}_n
 \cap {\cal D}_n
\right)
\\
&\quad +p_{SX^nY^n}
\left(
{\cal A}_n^{\rm c}
\right)
+p_{SX^nY^n}
\left(
{\cal B}_n^{\rm c}
\right)
\\
&\quad +p_{SX^nY^n}\left({\cal C}_n^{\rm c}\right)
+p_{SX^nY^n}\left({\cal D}_n^{\rm c}\cap {\cal E}_n 
\right)
\\
&\MLeq{a}
p_{SX^nY^n}\left(
     {\cal A}_n
\cap {\cal B}_n
\cap {\cal C}_n
\cap {\cal D}_n
\right)
+4{\rm e}^{-n\eta}. 
\end{align*}
Step (a) follows from Lemma \ref{lm:zzxa}.
\hfill\IEEEQED
%
%============================== End of \Apda ==================================%
%
}
From Lemma \ref{lm:Ohzzz}, we obtain the following lemma.
\begin{lm}\label{lm:OhzzzB}
For any $\eta>0$ and for any $(\varphi_1^{(n)}$, $\varphi_2^{(n)},\psi^{(n)})$ 
satisfying 
$
(1/n)\log {||\varphi_i^{(n)}||}\leq R_i, i=1,2,
$
we have 
\begin{align}
& {\rm P}_{\rm c}^{(n)}(\varphi_1^{(n)},\varphi_2^{(n)},\psi^{(n)})
\notag\\
&\leq p_{SX^nY^n}
\left\{
0 \geq \frac{1}{n}
\sum_{t=1}^n \log \frac{ Q_{{X}_t}(X_t)}{p_{X_t}(X_t)}-\eta,
\right.
\label{eqn:sOne}\\
&R_1\geq \frac{1}{n}
  \sum_{t=1}^n
  \log \frac{\tilde{Q}_{X_t|SX^{t-1}}
  (X_t|S,X^{t-1})}{p_{X_t}(X_t)}-\eta,
  \label{eqn:sTwo}\\
&R_2\geq \left. 
  \frac{1}{n}
  \sum_{t=1}^n\log
  \frac{1}{p_{Y_t|SX^{t-1}Y^{t-1}}(Y_t|S,X^{t-1},Y^{t-1})}
  -2\eta
  \right\}
\notag\\
&\quad +4{\rm e}^{-n\eta}, 
\notag
\end{align}
where for each $t=1,2,\cdots,n$, the probability 
distribution $Q_{X_t}$ on ${\cal X}$ appearing 
in (\ref{eqn:sOne}) and the stochastic matrix 
$\tilde{Q}_{X_t|SX^{t-1}}:$ ${\cal M}_1 \times {\cal X}^{t-1}$ $\to{\cal X}$ 
appearing in (\ref{eqn:sTwo}) have a property that we can choose 
their values arbitrary. 
\end{lm}

{\it Proof:} In (\ref{eqn:asppa}) in Lemma \ref{lm:Ohzzz}, 
we choose $\hat{q}_{S{X}^nY^n}$ having the form 
\begin{align*}
& \hat{q}_{S{X}^nY^n}(S,X^n,Y^n)
\\
&=p_S(S) \prod_{t=1}^n
\left\{
p_{X_t|SX^{t-1}Y^{t}}(X_t|S,X^{t-1},Y^{t})\right.
\\
& \qquad \qquad \times 
\left. 
p_{Y_t|SY^{t-1}}(Y_t|S,Y^{t-1})
\right\}.
\end{align*}
In (\ref{eqn:asppb}) in Lemma \ref{lm:Ohzzz}, 
we choose $Q_{{X}^n}$
having the form 
$$
Q_{{X}^n}(X^n)=\prod_{t=1}^n Q_{{X}_t }(X_t).
$$
We further note that
\begin{align*}
& \frac{\tilde{Q}_{X^n|S}(X^n|S)}{p_{X^n}(X^n)}
=\prod_{t=1}^n \frac{
\tilde{Q}_{X_t|SX^{t-1}}(X_t|S,X^{t-1})}{p_{{X}_t }(X_t)},
\\
& p_{Y^n|S}(Y^n|S)=\prod_{t=1}^n p_{Y_t|SY^{t-1}}(Y_t|S,Y^{t-1}).
\end{align*}
Then the bound (\ref{eqn:azsad}) in Lemma \ref{lm:Ohzzz} becomes
\begin{align*}
& {\rm P}_{\rm c}^{(n)}(\varphi_1^{(n)},\varphi_2^{(n)},\psi^{(n)})
\leq 
p_{SX^nY^n}\biggl\{
\nonumber\\
& \:\:\,\ds 
0 \geq \frac{1}{n}\sum_{t=1}^n
\log \frac {p_{Y_t|SY^{t-1}}(Y_t|S,Y^{t-1})}
     {p_{Y_t|SX^{t-1}Y^{t-1}}(Y_t|S,X^{t-1},Y^{t-1})}
-\eta,
\nonumber\\
& \:\:\,\ds 
0 \geq \frac{1}{n}
\sum_{t=1}^n \log \frac{ Q_{X_t}(X_t)}{p_{X_t}(X_t)}-\eta,
\nonumber\\
& \ds 
R_1\geq \frac{1}{n}\sum_{t=1}^n\log
\frac{
\tilde{Q}_{X_t|SX^{t-1}}(X_t|S,X^{t-1})}{p_{{X}_t}(X_t)}
-\eta,
\nonumber\\
& \ds 
R_2 \geq \left.
\frac{1}{n}\sum_{t=1}^n
\frac{1}{p_{Y_t|SY^{t-1}}(Y_t|S,Y^{t-1})}-\eta \right\}
+4{\rm e}^{-n\eta} 
\\
& 
\leq p_{SX^nY^n}
\left\{
0 \geq \frac{1}{n}
\sum_{t=1}^n \log \frac{ Q_{{X}_t}(X_t)}{p_{X_t}(X_t)}-\eta,
\right.
\\
& 
R_1\geq \frac{1}{n}
\sum_{t=1}^n
\log \frac{\tilde{Q}_{X_t|SX^{t-1}}
(X_t|S,X^{t-1})}{p_{X_t}(X_t)}-\eta,
\\
& 
R_2\geq \left. 
\frac{1}{n}
\sum_{t=1}^n\log
\frac{1}{p_{Y_t|SX^{t-1}Y^{t-1}}(Y_t|S,X^{t-1},Y^{t-1})}
-2\eta
\right\}
\\
&  \quad +4{\rm e}^{-n\eta}, 
\end{align*}
completing the proof. \hfill\IEEEQED
\begin{lm}\label{lm:Mchain} 
Suppose that for each $t=1,2,\cdots,n$, 
the joint distribution $p_{SX^tY^t}$ 
of the random vector $SX^tY^t$ is a marginal distribution 
of $p_{SX^nY^n}$. Then we have the following 
Markov chain:
\beq
SX^{t-1} \leftrightarrow X_t \leftrightarrow Y_t
\label{eqn:sssq} 
\eeq
or equivalently that $I(Y_t;SX^{t-1}|X_t)=0$. 
Furthermore, we have the following Markov chain:
\beq
Y^{t-1} \leftrightarrow SX^{t-1} \leftrightarrow (X_t,Y_t)
\label{eqn:sssq2} 
\eeq
or equivalently that $I(X_tY_t;Y^{t-1}|SX^{t-1})=0$.
The above two Markov chains are equivalent to
the following one long Markov chain:
\beq
Y^{t-1} \leftrightarrow SX^{t-1} 
\leftrightarrow X_t \leftrightarrow Y_t.
\label{eqn:sssq3} 
\eeq
\end{lm}

Proof of this lemma is given in Appendix \ref{sub:Apdb}. 
%%%%%%%%%%%%%%%%%%%%%%%%%%%%%%%%%%%%%%%%%%%%%%%%%%%%%%%%%%%%%%%%%%%%
\newcommand{\Apdb}{%-----------------------------------------------%
%%%%%%%%%%%%%%%%%%%%%%%%%%%%%%%%%%%%%%%%%%%%%%%%%%%%%%%%%%%%%%%%%%%%
\subsection{
Proof of Lemma \ref{lm:Mchain} 
}\label{sub:Apdb}

In this appendix we prove Lemma \ref{lm:Mchain}. 

{\it Proof of Lemma \ref{lm:Mchain}:} 
We first prove the following Markov chain (\ref{eqn:sssq}) 
in Lemma \ref{lm:Mchain}: 
$$
SX^{t-1} \leftrightarrow X_t \leftrightarrow Y_t.
$$
We have the following chain of inequalities:
\begin{align*}
& I(Y_t;SX^{t-1}|X_t)=H(Y_t|X_t)-H(Y_t|SX^{t-1}X_t) 
\\
& \leq H(Y_t|X_t)-H(Y_t|SX^n) \MEq{a}H(Y_t|X_t)-H(Y_t|X^n)
\\
& \MEq{b} H(Y_t|X_t)-H(Y_t|X_t)=0.
\end{align*}
Step (a) follows from that $S=\varphi_1^{(n)}(X^n)$ is 
a function of $X^n$. Step (b) follows 
from the memoryless property of 
the information source $\{(X_t,Y_t)\}_{t=1}^{\infty}$.  
Next we prove the following Markov chain (\ref{eqn:sssq2}) 
in Lemma \ref{lm:Mchain}:
$$
Y^{t-1} \leftrightarrow SX^{t-1} \leftrightarrow (X_t,Y_t).
$$
We have the following chain of inequalities:
\begin{align*}
&    I(X_tY_t;Y^{t-1}|SX^{t-1})
\\ 
&=   H(Y^{t-1}|SX^{t-1})
-H(Y^{t-1}|SX^{t-1}X_tY_t) 
\\
&\leq H(Y^{t-1}|X^{t-1})-H(Y^{t-1}|X^nSY_t)
\\
&\MEq{a} H(Y^{t-1}|X^{t-1})-H(Y^{t-1}|X^nY_t)
\\
&\MEq{b} H(Y^{t-1}|X^{t-1})-H(Y^{t-1}|X^{t-1}Y_t)=0.
\end{align*}
Step (a) follows from that $S=\varphi_1^{(n)}(X^n)$ 
is a function of $X^n$. Step (b) follows 
from the memoryless property of 
the information source 
$\{(X_t,Y_t)\}_{t=1}^{\infty}$.  
\hfill\IEEEQED
%%%%%%%%%%%%%%%%%%%%%%%%%%%%%%%%%%%%%%%%%%%%%%%%%%%%%%%%%%%%%%%%%%%%%%%%%
%====================== End of \Apdb ===================================%
%%%%%%%%%%%%%%%%%%%%%%%%%%%%%%%%%%%%%%%%%%%%%%%%%%%%%%%%%%%%%%%%%%%%%%%%%
}
For $t=1,2,\cdots,n$, set 
${\cal U}_t\defeq {\cal M}_1$ $\times {\cal X}^{t-1}$.
Define a random variable 
$U_t \in {\cal U}_t$ by $U_t \defeq (S,X^{t-1})$. 
From Lemmas \ref{lm:OhzzzB} and \ref{lm:Mchain}, we have 
the following. 
\begin{lm}\label{lm:Ohzzzc}
For any $\eta>0$ and for any 
$(\varphi_1^{(n)}$, 
$\varphi_2^{(n)},\psi^{(n)})$ 
satisfying
$
(1/n)\log ||\varphi_i^{(n)}|| \leq R_i,i=1,2,
$
we have 
\begin{align}
& {\rm P}_{\rm c}^{(n)}(\varphi_1^{(n)},\varphi_2^{(n)},\psi^{(n)})
\notag\\
&\leq p_{SX^nY^n}
\left\{
0\geq \frac{1}{n}
\sum_{t=1}^n
\log \frac{Q_{{X}_t}(X_t)}{p_{{X}_t}(X_t)}-\eta,
\right.
\label{eqn:sdsOne}\\
& R_1\geq \frac{1}{n}
\sum_{t=1}^n
\log \frac{\tilde{Q}_{X_t|U_t}
(X_t|U_t)}{p_{X_t}(X_t)}-\eta,
%\right.
\label{eqn:sdsTwo} \\
& \left. R_2 \geq 
\frac{1}{n}
\sum_{t=1}^n\log
\frac{1}{p_{Y_t|U_t}(Y_t|U_t)}-2\eta
\right\}
+4{\rm e}^{-n\eta},
\notag
\end{align}
where for each $t=1,2,\cdots,n$, the probability 
distribution $Q_{X_t}$ on ${\cal X}$ appearing 
in (\ref{eqn:sdsOne}) and the stochastic matrix 
$\tilde{Q}_{X_t|U_t}:$ ${\cal U}_t$ $ \to {\cal X}$ 
appearing in (\ref{eqn:sdsTwo}) have a property 
that we can choose their values arbitrary. 
\end{lm}

For each $t=1,2,\cdots,n$, set 
$\underline{Q}_t \defeq (Q_{X_t},\tilde{Q}_{X_t|U_t}).$
Let $\underline{\cal Q}_t$ be a set of all $\underline{Q}_t$.
We define a quantity which serves as an exponential
upper bound of 
${\rm P}_{\rm c}^{(n)}(\varphi_1^{(n)},$ 
$\varphi_2^{(n)},\psi^{(n)})$. 
Let ${\cal P}^{(n)}(p_{XY})$ be a 
set of all probability distributions 
${p}_{SX^nY^n}$ on 
${\cal M}_1$
$\times {\cal X}^n$
$\times {\cal Y}^n$
%Let $p^{(n)}=p_{SX^nY^n}$ 
having a form:
\begin{align*}
& p_{SX^nY^n}(s,x^n,y^n)=p_{S|X^n}(s|x^n)
 \prod_{t=1}^n p_{XY}(x_t,y_t)\\
&\mbox{ for }(s,x^n,y^n)
\in {\cal M}_1 \times {\cal X}^n \times {\cal Y}^n.
\end{align*}
For simplicity of notation we use the notation $p^{(n)}$ 
for $p_{SX^nY^n}$ $\in {\cal P}^{(n)}$ $(p_{XY})$. 
For each $t=1,2,\cdots, n$,
$
p_{U_tX_tY_t}=p_{SX^tY_t}
$
is a marginal distribution of $p^{(n)}$. 
For $t=1,2,\cdots, n$, we simply write $p_t=$ 
$p_{U_tX_tY_t}$. 
For 
$\mu\in [0,1]$, 
$\alpha\in [0,1)$, 
$p^{(n)}$ $\in {\cal P}^{(n)}(p_{XY})$,
and $\underline{Q}^n$ $\in {\cal Q}^n$, we define
\begin{align*}
&
{\OMega}\ARgRv
\\
&\defeq
-\log
{\rm E}_{p^{(n)}}
\left[
\prod_{t=1}^n
\frac{p_{X_t}^{\bar{\alpha}}(X_t)}{Q_{X_t}^{\bar{\alpha} }(X_t)}
\frac 
{p^{ {\prmtA}\alpha}_{X_t}(X_t)
p^{{\prmtA}\alpha}_{Y_t|U_t}(Y_t|U_t)}
{\tilde{Q}^{{\prmtA}\alpha}_{X_t|U_t}(X_t|U_t)}
\right],
\end{align*}
where for each $t=1,2,\cdots,n$, 
the probability distribution 
${Q}_{X_t}$ and 
the conditional probability 
distribution $\tilde{Q}_{X_t|U_t}$ appearing 
in the definition of 
$\Omega^{(\mu,\theta)}(p^{(n)},\underline{Q}^{n})$ 
can be chosen arbitrary.

The following is well known as the Cram\`er's bound 
in the large deviation principle.
\begin{lm}
\label{lm:Ohzzzb}
For any real valued random variable 
$Z$ and any $\alpha\geq 0$, we have
$$
\Pr\{Z \geq a \}\leq 
\exp
\left[
-\left(
\alpha a -\log {\rm E}[\exp(\alpha Z)]
\right) 
\right].
$$
\end{lm}

By Lemmas \ref{lm:Ohzzzc} and \ref{lm:Ohzzzb}, 
we have the following proposition.
\begin{pro}
\label{pro:Ohzzp}
For any $(\mu,\alpha)\in [0,1]^2$ 
any $\underline{Q}^n \in \underline{\cal Q}^n$,
and any 
$(\varphi_1^{(n)}, $ $\varphi_2^{(n)},\psi^{(n)})$ 
satisfying 
$
(1/n)\log {||\varphi_i^{(n)}||}\leq R_i, i=1,2, 
$
there exists $p^{(n)} \in {\cal P}^{(n)}(W_1,W_2)$ 
such that
\begin{align*}
& {\rm P}_{\rm c}^{(n)}(\varphi_1^{(n)},\varphi_2^{(n)},\psi^{(n)})
\leq 5\exp
\biggl\{
-n\left[2+\alpha{\prmtB}\right]^{-1}
\\
& \times 
\left. \left[
\frac{1}{n}
{\OMega}{\ARgRv}
-\alpha({\prmtA}R_1+{\prmtB} R_2)\right]
\right\}.
\end{align*}
\end{pro}

{\it Proof:} By Lemma \ref{lm:Ohzzzc}, 
for $(\mu,\alpha) \in [0,1]^2$, we have 
the following chain of inequalities: 
\begin{align}
& 
{\rm P}_{\rm c}^{(n)}
(\varphi_1^{(n)},
 \varphi_2^{(n)},
 \psi^{(n)})
\nonumber\\
&\leq p_{SX^nY^n}
\left\{
0 \geq \left[
\frac{1}{n}
\sum_{t=1}^n
\log \frac{Q^{\bar{\alpha}}_{X_t}(X_t)}{p^{\bar{\alpha}}_{{X}_t}(X_t)}
-\bar{\alpha} \eta \right],
\right.
\nonumber\\
&\quad \alpha {\prmtA} R_1 \geq \frac{1}{n}
\sum_{t=1}^n
\log \frac{\tilde{Q}^{\alpha\prmtA}_{X_t|U_t}
(X_t|U_t)}{p^{\alpha\prmtA}_{X_t}(X_t)}-\alpha{\prmtA}\eta,
\nonumber\\
&\quad \left. \alpha {\prmtB} R_2 \geq 
\frac{1}{n}
\sum_{t=1}^n\log
\frac{1}{p^{\alpha{\prmtB}}_{Y_t|U_t}(Y_t|U_t)}-2\alpha {\prmtB} \eta
\right\}
+4{\rm e}^{-n\eta} 
\nonumber\\ 
&\leq p_{SX^nY^n}\hugel
\alpha({\prmtA}R_1+{\prmtB} R_2)+(1+\alpha{\prmtB})\eta 
\nonumber\\
&\quad 
\left. \geq -\frac{1}{n}
\sum_{t=1}^n
\log \left[
\frac{p^{\bar{\alpha}}_{X_t}(X_t)}
     {Q^{\bar{\alpha}}_{X_t}(X_t)}
\frac{p^{{\prmtA}\alpha}_{X_t}(X_t)p^{{\prmtB}\alpha}_{Y_t|U_t}(Y_t|U_t)}
     {\tilde{Q}^{{\prmtA}\alpha}_{X_t|U_t}(X_t|U_t)}\right]\right\}
\nonumber\\
&\quad +4{\rm e}^{-n\eta} 
\nonumber\\
&=p_{SX^nY^n}\left\{
\frac{1}{n}
\sum_{t=1}^n
\log \left[
\frac{p^{\bar{\alpha}}_{X_t}(X_t)}
     {Q^{\bar{\alpha}}_{X_t}(X_t)}
     \right.\right.
\nonumber\\
&\quad \qquad\qquad\qquad\qquad \times
\left.
\frac{p^{{\prmtA}{\alpha}}_{X_t}(X_t)p^{\alpha}_{Y_t|U_t}(Y_t|U_t)}
     {\tilde{Q}^{{\prmtA}{\alpha}}_{X_t|U_t}(X_t|U_t)}\right]
\nonumber\\
& \quad 
\quad\quad
\geq -\left[\alpha({\prmtA}R_1+{\prmtB} R_2)
    +(1+\alpha\prmtB)\eta\right] \biggr\}
+4{\rm e}^{-n\eta} 
\nonumber\\
&\MLeq{a} 
\exp \biggl[n\biggl\{\alpha({\prmtA}R_1+{\prmtB} R_2)
+(1+\alpha{\prmtB})\eta 
\nonumber\\
&\quad \qquad \left.\left.-\frac{1}{n}
{\OMega}{\ARgRv}\right\}\right] +4{\rm e}^{-n\eta}. 
\label{eqn:aaabv}
\end{align}
Step (a) follows from Lemma \ref{lm:Ohzzzb}. 
When ${\OMega}{\ARgRv}\leq n\alpha({\prmtA}R_1+{\prmtB} R_2)$, 
the bound we wish to prove is obvious. In the following argument 
we assume that    
${\OMega}{\ARgRv}> $ 
$n\alpha({\prmtA}R_1+{\prmtB} R_2)$.
We choose $\eta$ so that 
\beqa
-\eta&=& \alpha({\prmtA}R_1+{\prmtB} R_2)+(1 + \alpha\prmtB)\eta
\nonumber\\ 
&  &-\frac{1}{n}{\OMega}{\ARgRv}.
\label{eqn:aaappp}
\eeqa
Solving (\ref{eqn:aaappp}) with respect to $\eta$, we have 
\begin{align*}
\eta=
\frac{(1/n){\OMega}{\ARgRv}
-\alpha({\prmtA}R_1+{\prmtB} R_2)}
{2+\alpha\prmtB}.
\end{align*}
For this choice of $\eta$ and (\ref{eqn:aaabv}), we have
\begin{align*}
& {\rm P}_{\rm c}^{(n)}(\varphi_1^{(n)},\varphi_2^{(n)},\psi^{(n)})
\leq 5{\rm e}^{-n\eta}
=5\exp
\biggl\{
-n\left[2+\alpha{\prmtB}\right]^{-1}
\\
& \qquad \times 
\left. \left[
\frac{1}{n}
{\OMega}{\ARgRv}
-\alpha({\prmtA}R_1+{\prmtB} R_2)\right]
\right\},
\end{align*}
completing the proof. 
\hfill \IEEEQED

Set 
\begin{align*}
& \underline{\Omega}^{(\mu,\alpha)}(p_{XY})
\\
&\defeq  
\inf_{n\geq 1}
\min_{\scs p^{(n)}\in {\cal P}^{(n)}}
\max_{\scs \underline{Q}^n \in \underline{\cal Q}^n}
\frac{1}{n}
{\OMega}{\ARgRv}.
\end{align*}
By Proposition \ref{pro:Ohzzp} we have the following corollary.  
\begin{co} 
\label{co:corOne}
For any $(\mu,\alpha)\in [0,1]^2$ and any 
$(\varphi_1^{(n)}, $ $\varphi_2^{(n)},\psi^{(n)})$ 
satisfying 
$
(1/n)\log {||\varphi_i^{(n)}||}\leq R_i, i=1,2, 
$
we have 
\begin{align*}
& {\rm P}_{\rm c}^{(n)}(\varphi_1^{(n)},\varphi_2^{(n)},\psi^{(n)})
\\
&\leq 5\exp
\left\{-n\left[
\frac{\underline{\Omega}^{(\mu,\alpha)}(p_{XY})
-\alpha({\prmtA}R_1+{\prmtB} R_2)}
{2+\alpha{\prmtB}}\right]\right\}.
\end{align*}
\end{co}
%%%%%%%%%%%%%%%%%%%%%%%%%%%%%%%%%%%%%%%%%%%%%%%%%%%%%%%%%%%%%%%%%%%%
%%%%%%%%%%%%%%%%%%%%%%% \newcommand %%%%%%%%%%%%%%%%%%%%%%%%%%%%%%%%
\newcommand{\Ft}{ {\cal F}_t }
\newcommand{\Ftt}{ {\cal F}^t}
\newcommand{\Fttmo}{ {\cal F}^{t-1}}
\newcommand{\Fi}{ {\cal F}_i }
\newcommand{\Fiimo}{ {\cal F}^{i-1}}
%%%%%%%%%%%%%%%%%%%%%%%%%%%%%%%%%%%%%%%%%%%%%%%%%%%%%%%%%%%%%%%%%%%
%%%%%%%%%%%%%%%%%%%%%%%%%%%%%%%%%%%%%%%%%%%%%%%%%%%%%%%%%%%%%%%%%%%

We shall call $\underline{\Omega}^{(\mu,{}\alpha)}(p_{XY})$ 
the communication potential. The above corollary implies that 
the analysis of $\underline{\Omega}^{(\mu,\alpha)}($ $p_{XY})$ 
leads to an establishment of a strong converse 
theorem for the one helper source coding problem.     
In the following argument we drive an explicit 
lower bound of $\underline{\Omega}^{(\mu,{}\alpha)}(p_{XY})$. 
For each $t=1,2,\cdots,n$, set 
$u_t=(s,x^{t-1})$ $\in {\cal U}_t$ and 
$$
\Ft\defeq (p_{X_t},p_{X_tY_t|U_t},\underline{Q}_t),\quad
\Ftt \defeq \{{\cal F}_i\}_{i=1}^{t}.
$$
For $t=1,2,\cdots,n$, define a function of 
$(u_t,x_t,y_t)$
$\in {\cal U}_t$
$\times {\cal X}$
$\times {\cal Y}$
by 
\begin{align*}
&
f_{\Ft}^{(\mu,{}\alpha)}
(x_t,y_t|u_t)%}
\defeq 
\frac{p_{X_t}^{\bar{\alpha}}(x_t)}
     {Q_{X_t}^{\bar{\alpha}}(x_t)}
\frac{p^{{\prmtA} \alpha}_{X_t}(x_t)
      p^{\alpha}_{Y_t|U_t}(y_t|u_t)}
     {\tilde{Q}^{{\prmtA} \alpha}_{X_t|U_t}(x_t|u_t)}.
\end{align*}
By definition we have
\begin{align*}
&
\exp\left\{-{\OMega}{\ARgRv}
\right\}
\\
&=\sum_{s,x^n, y^n}p_{S X^n Y^n}(s,x^n, y^n)
\prod_{t=1}^n
f_{\Ft}^{(\mu,{}\alpha)}(x_t,y_t|u_t).
\end{align*}
For each $t=1,2,\cdots,n$, we define the probability 
distribution
$$
{p}_{SX^tY^t;\Ftt}^{(\mu,{}\alpha)}\defeq
\left\{
{p}_{SX^tY^t;\Ftt}^{(\mu,{}\alpha)}(s,x^t,y^t)
\right\}_{(s,x^t, y^{t})
\in {\cal M}_1\times {\cal X}^t\times {\cal Y}^t}
$$
by
\begin{align*}
& p_{SX^tY^t;\Ftt}^{(\mu,{}\alpha)}(s,x^t,y^t) 
\defeq 
C_t^{-1}
p_{SX^tY^t}(s,x^t,y^t)
\\ 
& \times
\prod_{i=1}^t
f_{\Fi}^{(\mu,{}\alpha)}(x_i,y_i|u_i)
\end{align*} 
where
\begin{align*}
C_t&\defeq 
\sum_{s,x^t,y^t}
p_{SX^tY^t}(s,x^t,y^t) 
%\\
%& & \qquad \times 
\prod_{i=1}^t
f_{\Fi}^{(\mu,\alpha)}(x_i,y_i)
\end{align*}
are constants for normalization. For $t=1,2,\cdots,n$, define 
\beq 
\Phi_t^{(\mu,{}\alpha)} \defeq C_tC_{t-1}^{-1}, 
\label{eqn:defa}
\eeq
where we define $C_0=1$. Then we have the following lemma.
\begin{lm}\label{lm:aaa}
For each $t=1,2,\cdots,n$, and for any 
$(s,$ $x^t, y^t)\in {\cal M}_1$ 
$\times {\cal X}^t$ 
$\times {\cal Y}^t$, 
we have
\begin{align}
& {p}_{SX^t Y^t;\Ftt}^{(\mu,{}\alpha)}(s,x^t,y^t)
\nonumber\\
&=(\Phi_t^{(\mu,{}\alpha)})^{-1}
p_{SX^{t-1}Y^{t-1};\Fttmo}^{(\mu,{}\alpha)}(s,x^{t-1},y^{t-1})
\nonumber\\
& \quad \times 
p_{X_tY_t|SX^{t-1}Y^{t-1}}(x_t,y_t|s,x^{t-1},y^{t-1})
\nonumber\\
& \quad\times 
f_{\Ft}^{(\mu,\alpha)}(x_t,y_t|u_t).
\label{eqn:satt}
\end{align}
Furthermore, we have  
\begin{align}
&\Phi_t^{(\mu,{}\alpha)}
%&
= \sum_{s,x^t,y^t} 
p_{SX^{t-1}Y^{t-1};\Fttmo}^{(\mu,\alpha)}(s,x^{t-1},y^{t-1})
\nonumber\\
& \quad\times
p_{X_tY_t|SX^{t-1}Y^{t-1}}(x_t,y_t|s,x^{t-1},y^{t-1})
\nonumber\\
& \quad\times 
f_{\Ft}^{(\mu,{}\alpha)}
(x_t,y_t|u_t).
\label{eqn:sattbz}
\end{align}
\end{lm}

Proof of this lemma is given in Appendix \ref{sub:sdfa}. %%%%%%%%%
%%%%%%%%%%%%%%%%%%%%%%%%%%%%%%%%%%%%%%%%%%%%%%%%%%%%%%%%%%%%%%%%%%
\newcommand{\Apdc}{
%
%---------------------------\Apdc--------------------------------%
%
\subsection{Proof of Lemma \ref{lm:aaa}}\label{sub:sdfa}

In this appendix we prove Lemma \ref{lm:aaa}.  

%}{

{\it Proof of  Lemma \ref{lm:aaa} :} By the definition 
of ${p}_{SX^tY^t;\Ftt}^{(\mu,\alpha)}(s,$ $x^t,y^{t})$, 
for $t=1,2,\cdots,n$, we have 
\begin{align}
& p_{SX^tY^{t};\Ftt}^{(\mu,\alpha)}(s,x^t,y^{t})
=C_t^{-1}
p_{SX^tY^{t}}(s,x^t,y^{t})
\nonumber\\
&\quad \times 
\prod_{i=1}^t
f_{\Fi}^{(\mu,{}\alpha)}(x_i,y_i|u_i).
\label{eqn:azaq}
\end{align} 
Then we have the following chain of equalities:
\begin{align} 
& p_{SX^tY^{t};\Ftt}^{(\mu,\alpha)}(s,x^t,y^t)
\MEq{a}%&
C_t^{-1} 
p_{SX^tY^{t}}(s,x^t,y^t)
\nonumber\\
& \quad\times
\prod_{i=1}^t
f_{\Fi}^{(\mu,{}\alpha)}
(x_i,y_i|u_i)
\nonumber\\
&=C_t^{-1}
p_{SX^{t-1}Y^{t-1}}(s,x^{t-1},y^{t-1})
\nonumber\\
& \quad\times
\prod_{i=1}^{t-1}
f_{\Fi}^{(\mu,{}\alpha)}(x_i,y_i|u_i)
\nonumber\\
&  
\quad \times 
p_{X_tY_t|SX^{t-1}Y^{t-1}}(x_t,y_t|s,x^{t-1},y^{t-1})
\nonumber\\
&  
\quad \times 
f_{ \Ft }^{(\mu,{}\alpha)}
(x_t,y_t|u_t)
\nonumber\\
&\MEq{b}
C_t^{-1}C_{t-1}p_{SX^{t-1}Y^{t-1}}^{(\mu,{}\alpha)}(s,x^{t-1},y^{t-1})
\nonumber\\
&  
\quad \times 
p_{X_tY_t|SX^{t-1}Y^{t-1}}(x_t,y_t|s,x^{t-1},y^{t-1})
\nonumber\\
&  
\quad \times 
f_{\Ft}^{(\mu,{}\alpha)}
(x_t,y_t|u_t)
\nonumber\\
&=(\Phi_t^{(\mu,{}\alpha)})^{-1}
p_{SX^{t-1}Y^{t-1}; \Fttmo }^{(\mu,{}\alpha)}(s,x^{t-1},y^{t-1})
\nonumber\\
& \quad\times
p_{X_tY_t|SX^{t-1}Y^{t-1}}(x_t,y_t|s,x^{t-1},y^{t-1})
\nonumber\\
& 
\quad \times f_{\Ft}^{(\mu,{}\alpha)}
(x_t,y_t|u_t).
\label{eqn:daaaq}
\end{align}
Steps (a) and (b) follow 
from (\ref{eqn:azaq}). From (\ref{eqn:daaaq}), we have 
\begin{align} 
& \Phi_t^{(\mu,{}\alpha)}
p_{SX^tY^{t}; \Ftt}^{(\mu,{}\alpha)}(s,x^t,y^{t})
\label{eqn:daxx}\\
&=
p_{SX^{t-1}Y^{t-1};\Fttmo}^{(\mu,{}\alpha)}(s,x^{t-1},y^{t-1})
\nonumber\\
& \quad\times
p_{X_tY_t|SX^{t-1}Y^{t-1}}(x_t,y_t|s,x^{t-1},y^{t-1})
\nonumber\\
& \quad\times
f_{\Ft}^{(\mu,{}\alpha)}(x_t,y_t|u_t).
\label{eqn:daaxx}
\end{align}
Taking summations of (\ref{eqn:daxx}) and 
(\ref{eqn:daaxx}) with respect to $s,x^t,y^t$, 
we obtain %(\ref{eqn:sattb}).
\begin{align*}
& \Phi_t^{(\mu,\alpha)}
%\nonumber\\
%&
= \sum_{s,x^t,y^{t}}
p_{SX^{t-1}Y^{t-1};\Fttmo}^{(\mu,{}\alpha)}(s,x^{t-1},y^{t-1})
\nonumber\\
& \quad\times 
p_{X_tY_t|SX^{t-1}Y^{t-1}}(x_t,y_t|s,x^{t-1},y^{t-1})
%\nonumber\\
%& 
%\quad\times 
f_{\Ft}^{(\mu,{}\alpha)}
(x_t,y_t|u_t),
\end{align*}
completing the proof.
\hfill \IEEEQED
}
%
%---------------------------End of \Apdc----------------------------%
%
Define
\begin{align*}
& p_{U_t;\Fttmo}^{(\mu,{}\alpha)}(u_t)
=p_{SX^{t-1};\Fttmo}^{(\mu,{}\alpha)}(s,x^{t-1})
\\
&\defeq 
\sum_{y^{t-1}} 
p_{SX^{t-1}Y^{t-1};\Fttmo}^{(\mu,{}\alpha)}(s,x^{t-1},y^{t-1}).
\end{align*}
Then we have the following lemma, which is a key result 
to derive a single-letterized lower bound of 
$\underline{\Omega}^{(\mu,\alpha)}(p_{XY})$. 

%We have the following.
%\end{lm}
%Then we have the following lemma.
%From Lemmas \ref{lm:Mchain} and \ref{lm:aaa}, 

%\begin{lm}\label{lm:aaaZZZ}
\begin{lm}\label{lm:keylm}
For any $p^{(n)}\in {\cal P}^{(n)}(p_{XY})$ and any 
$\underline{Q}^n \in \underline{\cal Q}^n$, we have 
\begin{align}
&{\OMega}{\ARgRv}
=(-1)\sum_{t=1}^n \log \Phi_t^{(\mu,{}\alpha)},
\label{eqn:defazzz}  
\\
&\Phi_t^{(\mu,{}\alpha)}
= \sum_{u_t,x_t,y_t} 
p_{U_t;\Fttmo}^{(\mu,\alpha)}(u_t)
p_{X_t|U_t}(x_t|u_t)
p_{Y_t|X_t}(y_t|x_t)
\nonumber\\
& \qquad\times
f_{\Ft}^{(\mu,{}\alpha)}
(x_t,y_t|u_t).
\label{eqn:sattbzz}
\end{align}
\end{lm}

{\it Proof}: We first prove (\ref{eqn:defazzz}).
From (\ref{eqn:defa}) we have
\beq
\log \Phi_t^{(\mu,\alpha)}
=-\log C_t + \log C_{t-1}. 
\label{eqn:aaap}
\eeq
Furthermore, by definition we have   
\beq
{\OMega}{\ARgRv}=-\log C_n, C_0=1. 
\label{eqn:aaapq}
\eeq
From (\ref{eqn:aaap}) and (\ref{eqn:aaapq}), 
(\ref{eqn:defazzz}) is obvious. 
We next prove (\ref{eqn:sattbzz}). We first observe that 
for $(s,x^t,y^t)$ 
$\in {\cal S}\times{\cal X}^t \times {\cal Y}^t
$
and for $t=1,2,\cdots,n$,
\begin{align*}
& p_{X_tY_t|SX^{t-1}Y^{t-1}}(x_t,y_t|s,x^{t-1},y^{t-1})
\\
&=p_{X_t|SX^{t-1}Y^{t-1}}(x_t|s,x^{t-1},y^{t-1})
\\
&\quad \times 
p_{Y_t|SX^tY^{t-1}}(y_t|s,x^{t},y^{t-1})
\\
&\MEq{a}p_{X_t|SX^{t-1}}(x_t|s,x^{t-1})
   p_{Y_t|X_t}(y_t|x_t).
\end{align*}
Step (a) follows from Lemma \ref{lm:Mchain}.
Then by Lemma \ref{lm:aaa}, we have 
\begin{align*}
& \Phi_t^{(\mu,{}\alpha)}
 = \sum_{s,x^t,y^{t}}
p_{SX^{t-1}Y^{t-1};\Fttmo}^{(\mu,{}\alpha)}(s,x^{t-1},y^{t-1})
\nonumber\\
& \qquad\times 
p_{X_tY_t|SX^{t-1}Y^{t-1}}(x_t,y_t|s,x^{t-1},y^{t-1})
\nonumber\\
& \qquad\times 
f_{\Ft}^{(\mu,{}\alpha)}
(x_t,y_t|u_t),
\nonumber\\
&= \sum_{s,x^t,y^{t}}
p_{SX^{t-1}Y^{t-1};\Fttmo}^{(\mu,{}\alpha)}(s,x^{t-1},y^{t-1})
\nonumber\\
& \quad\times 
p_{X_t|SX^{t-1}}(x_t|s,x^{t-1})p_{Y_t|X_t}(y_t|x_t)
%\nonumber\\
%& \quad\times 
f_{\Ft}^{(\mu,{}\alpha)}
(x_t,y_t|u_t)
\nonumber\\
&= \sum_{s,x^t,y_t}
p_{SX^{t-1}}^{(\mu,{}\alpha)}(s,x^{t-1})
\nonumber\\
& \quad\times 
p_{X_t|SX^{t-1}}(x_t|s,x^{t-1})p_{Y_t|X_t}(y_t|x_t)
%\nonumber\\
%& \quad\times 
f_{\Ft}^{(\mu,{}\alpha)}
(x_t,y_t|u_t),
\end{align*}
completing the proof.\hfill\IEEEQED

The following proposition is a mathematical core to prove 
our main result. 
\begin{pro}
\label{pro:mainpro} For any $\mu\in [0,1]$ and 
any ${}\alpha \geq 0$,
we have 
$$
\underline{\Omega}^{(\mu,{}\alpha)}(p_{XY})
\geq \Omega^{(\mu,{}\alpha)}(p_{XY}).
$$
\end{pro}

{\it Proof:} Set
\begin{align*}
{\cal Q}_n(p_{Y|X})
\defeq &
\{q=q_{UXY}: 
\pa {\cal U} \pa \leq 
\pa {\cal M}_1 \pa \pa {\cal X}^{n-1}\pa\pa {\cal Y}^{n-1}\pa,
\\
  & q_{Y|X}=p_{Y|X}, 
{U} \markov {X} \markov {Y} \},
\\
\hat{\Omega}_n^{(\mu,{}\alpha)}(p_{XY})
\defeq&
\min_{\scs 
     %U\in {{\cal P}}_n(p_{XY}),
     \atop{\scs 
     q\in {\cal Q}_n(p_{Y|X})
     }
}
\Omega^{(\mu,{}\alpha)}(q|p_{XY}).
\end{align*}
%Let ${U}_t$ be random variables 
%taking values in 
%${\cal M}_1 \times {\cal X}^{t-1}$. 
For each $t=1,2,\cdots,n$, we define 
$q_t=q_{U_tX_tY_tZ_t}$ by 
%We choose $q_{U_t}$ so that    
%\beq_
\begin{align}
&\left.
%ppp
\ba{l}
q_{U_t}(u_t)=p_{U_t;\Fttmo}^{(\mu,{}\alpha)}(u_t),
\\
q_{X_tY_t|U_t}(x_t,y_t|u_t)=p_{X_t|U_t}(x_t|u_t)
p_{Y|X}(y_t|x_t).
\ea
\right\}
\label{eqn:azo}
\end{align}
The equation (\ref{eqn:azo}) imply that
$
q_t=q_{U_tX_tY_t} \in {\cal Q}_n(p_{Y|X}).
$
Furthermore, for each $t=1,2,\cdots,n$, we choose 
$
\underline{Q}_t=(Q_{X_t},\tilde{Q}_{X_t|U_t})
$
appearing in
\begin{align*}
&
f_{\Ft}^{(\mu,{}\alpha)}
(x_t,y_t|u_t)%}
=\frac{p_{X_t}^{\bar{\alpha}}(x_t)}
     {Q_{X_t}^{\bar{\alpha}}(x_t)}
\frac{p^{{\prmtA} \alpha}_{X_t}(x_t)
      p^{\alpha}_{Y_t|U_t}(y_t|u_t)}
     {\tilde{Q}^{{\prmtA} \alpha}_{X_t|U_t}(x_t|u_t)}
\end{align*}
such that 
$
\underline{Q}_t=(Q_{X_t},\tilde{Q}_{X_t|U_t})
=(q_{X_t},q_{X_t|U_t}). 
$
For this choice of $\underline{Q}_t$, 
%we have 
%\begin{align}
%&f_{\Ft}^{(\mu,\theta)}(x_t,y_t|u_t)
%=f_{q_t|p_{X}}^{(\mu,\theta)}(x_t,y_t|u_t)
%\notag\\
%&\mbox{ for }(u_t,x_t,y_t) 
%\in {\cal U}_t \times {\cal X}\times {\cal Y}.
%\label{eqn:AsFFzz}
%\end{align} 
%Hence 
we have the following chain of inequalities: 
\begin{align}
&\Phi_t^{(\mu,\alpha)}
\MEq{a}{\rm E}_{q_t}\left[
f_{\Ft}^{(\mu,\theta)}(X_t,Y_t|U_t)
\right]
\nonumber\\
&\MEq{b}{\rm E}_{q_t}\left[
\frac{p_{X_t}^{\bar{\alpha}}(X_t)}
     {q_{X_t}^{\bar{\alpha}}(X_t)}
\frac{p^{{\prmtA} \alpha}_{X_t}(X_t)
      p^{\alpha}_{Y_t|U_t}(Y_t|U_t)}
     {q^{{\prmtA} \alpha}_{X_t|U_t}(X_t|U_t)}
\right]
\nonumber\\
&={\rm E}_{q_t}\left[
f_{q_t|p_{X_t}}^{(\mu,{}\alpha)}(X_t,Y_t|U_t)
\right]
=\exp\left\{-\Omega^{(\mu,{}\alpha)}(q_t|p_{X_t})\right\}
\nonumber\\
&\MEq{c}\exp\left\{-\Omega^{(\mu,{}\alpha)}(q_t|p_{X})\right\}
 \MLeq{d}\exp\left\{-\hat{\Omega}_n^{(\mu,{}\alpha)}(p_{XY})\right\}
\nonumber\\
&\MEq{e}\exp\left\{-{\Omega}^{(\mu,{}\alpha)}(p_{XY})\right\}.
\label{eqn:aSzaa}
\end{align}
Step (a) follows from Lemma \ref{lm:keylm} and (\ref{eqn:azo}).
Step (b) follows from 
the choice $(Q_{X_t}, \tilde{Q}_{X_t|U_t})=$
$(q_{X_t}, q_{X_t|U_t})$ 
of $(Q_{X_t}, \tilde{Q}_{X_t|U_t})$ 
for $t=1,2,\cdots,n$. %and (\ref{eqn:AsFFzz}).
Step (c) follows from $p_{X_t}=p_X$ 
for $t=1,2,\cdots,n$.
Step (d) follows from 
$q_t\in {\cal Q}_n(p_{Y|X})$ and 
the definition of 
$\hat{\Omega}_n^{(\mu,\alpha)}(p_{XY})$. 
Step (e) follows from Property \ref{pr:pro1} part a).
Hence we have the following:
\begin{align}
& \max_{\scs \underline{Q}^n \in \underline{\cal Q}^n} 
%q_{X^n}=
%\atop{ \scs  \prod_{t=1}^n q_{{X}_t}}}
\frac{1}{n}{\OMega}{\ARgRv}
\geq \frac{1}{n}
{\OMega}{\ARgRv}
\notag\\
& \MEq{a}-\frac{1}{n}\sum_{t=1}^n
\log \Phi_t^{(\mu,{}\alpha)}
\MGeq{b}\Omega^{(\mu,{}\alpha)}(p_{XY}).
\label{eqn:aSz}
\end{align}
Step (a) follows from Lemma \ref{lm:keylm}. 
Step (b) follows from (\ref{eqn:aSzaa}).
Since (\ref{eqn:aSz}) holds fo any $n\geq 1$ and any 
$p_{SX^nY^n}$ satisfying $S \markov X^n \markov Y^n$, we have 
that for any $(\mu,\alpha) \in [0,1]^2$,   
$$
\underline{\Omega}^{(\mu,{}\alpha)}(p_{XY}) 
\geq \Omega^{(\mu,{}\alpha)}(p_{XY}).
$$ 
Thus, Proposition \ref{pro:mainpro} is proved.
\hfill \IEEEQED

{\it Proof of Theorem \ref{Th:main}: }
For any $(\mu,\alpha)\in [0,1]^2$, 
for any $R_1,R_2\geq 0$ and 
for any $(\varphi_1^{(n)},$ $\varphi_2^{(n)},$ 
$\psi^{(n)})$ satisfying 
$(1/n)\log || \varphi_i^{(n)}|| \leq R_i,i=1,2,$
we have the following:
\begin{align*}
& %G(R,R_2|p_{XY})
\frac{1}{n}\log\left\{
\frac{5}{{\rm P}_{\rm c}^{(n)}
(\varphi_1^{(n)}, \varphi_2^{(n)},\psi^{(n)})}
\right\}
\\
&\MGeq{a} 
\frac{
\underline{\Omega}^{(\mu,{}\alpha)}(p_{XY})
-\alpha ({\prmtA}R_1 + {\prmtB}{R_2})
}
{2+\alpha{\prmtB}}
\\
&\MGeq{b} 
\frac{{\Omega}^{(\mu,{}\alpha)}(p_{XY})
-\alpha({\prmtA}R_1 + {\prmtB}{R_2})}
{2+\alpha{\prmtB}}
\\
&=F^{(\mu,{}\alpha)}({\prmtA}R_1+{\prmtB}{R_2}|p_{XY}). 
\end{align*}
Step (a) follows from Corollary \ref{co:corOne}. Step (b) follows from 
Proposition \ref{pro:mainpro}. 
Since the above bound holds for any $\mu \in [0,1]$ and 
any ${}\alpha \geq 0$, 
we have 
$$ 
\frac{1}{n}\log\left\{
\frac{5}{{\rm P}_{\rm c}^{(n)}(\varphi_1^{(n)},\varphi_2^{(n)},\psi^{(n)})}
\right\}
%G(R,R_2|p_{XY})
\geq F(R_1,R_2|p_{XY}). 
$$
Thus (\ref{eqn:mainIeq}) in Theorem \ref{Th:main} is proved.
%%%%%%%%%%%%%%%%%%%%%%%%%%%%%%%%%%%%%%%%%%%%%%%%%%%%%%%%%%%%%%%%%
%%%%%%%%%%%%%%%%%%%%%%%%%%%%%%%%%%%%%%%%%%%%%%%%%%%%%%%%%%%%%%%%%
%${\cal G}(p_{XY})$ $\subseteq \overline{\cal G}(p_{XY})$ %%%%%%%
%is obvious from this bound.%%%%%%%%%%%%%%%%%%%%%%%%%%%%%%%%%%%%%
\hfill\IEEEQED

\ProofCor

\section{One Helper Problem Studied by Wyner} 

We consider a communication system depicted in Fig. 2.
Data sequences $X^{n}$, $Y^{n}$, and $Z^{n}$, respectively 
are separately encoded to 
$\varphi_1^{(n)}(X^{n})$, $\varphi_2^{(n)}(Y^{n})$, 
and  $\varphi_3^{(n)}(Z^{n})$. 
The encoded data $\varphi_1^{(n)}(X^{n})$ 
and $\varphi_2^{(n)}(Y^{n})$ and are sent to 
the information processing center 1.
The encoded data $\varphi_1^{(n)}(X^{n})$ 
and $\varphi_3^{(n)}(Z^{n})$ and are sent to 
the information processing center 2.
At the center 1 the decoder function 
$\psi^{(n)}$ observes 
$(\varphi_1^{(n)}(X^{n}),$ $\varphi_2^{(n)}(Y^{n}))$ to output 
the estimation $\hat{Y}^{n}$ of ${Y}^{n}$. 
At the center 2 the decoder function 
$\phi^{(n)}$ observes 
$(\varphi_1^{(n)}(X^{n}),$ $\varphi_3^{(n)}(Z^{n}))$ to output 
the estimation $\hat{Z}^{n}$ of ${Z}^{n}$. 
The error probability of decoding is 
\begin{align*}
& {\rm P}_{\rm e}^{(n)}(\varphi_1^{(n)},\varphi_2^{(n)},
\varphi_3^{(n)},\psi^{(n)},\phi^{(n)})
\nonumber\\
&=\Pr\left\{\hat{Y}^{n}\neq Y^{n}\mbox{ or }
\hat{Z}^{n}\neq Z^{n}\right\}, 
\end{align*}
where $\hat{Y}^{n}
=\psi^{(n)}(
$ $\varphi_1^{(n)}(X^{n}),
\varphi_2^{(n)}(Y^{n}))$ and 
$\hat{Z}^{n}
=\psi^{(n)}(
$ $\varphi_1^{(n)}(X^{n}),
   \varphi_3^{(n)}(Z^{n}))$.

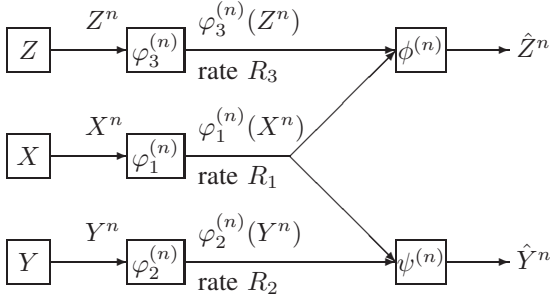
\begin{figure}[t]
\setlength{\unitlength}{0.94mm}
\begin{picture}(84,47)(4,4)

\put(10,40){\framebox(6,6){$Z$}}
\put(10,25){\framebox(6,6){$X$}}
\put(10,10){\framebox(6,6){$Y$}}

\put(21,46){${Z^n}$}
\put(16,43){\vector(1,0){11}}
\put(21,31){${X^n}$}
\put(16,28){\vector(1,0){11}}
\put(21,16){${Y^n}$}
\put(16,13){\vector(1,0){11}}

\put(27,40){\framebox(8,6){$\varphi_3^{(n)}$}}
\put(37,46){$ \varphi_3^{(n)}({Z^n})$}
\put(37,39){rate $R_3$}

\put(27,25){\framebox(8,6){$\varphi_1^{(n)}$}}
\put(37,31){$\varphi_1^{(n)}({X^n})$}
\put(37,24){rate $R_1$}

\put(27,10){\framebox(8,6){$\varphi_2^{(n)}$}}
\put(37,16){$\varphi_2^{(n)}({Y^n})$}
\put(37,9){rate $R_2$}

\put(50,28){\vector(1, 1){15}}
\put(50,28){\vector(1,-1){15}}

\put(35,43){\vector(1,0){30}}
\put(35,28){\line(1,0){15}}
\put(35,13){\vector(1,0){30}}

\put(65,40){\framebox(7,6){$\phi^{(n)}$}}
\put(65,10){\framebox(7,6){$\psi^{(n)}$}}

\put(72,43){\vector(1,0){9}}
\put(82,42){$\hat{Z}^n$}

\put(72,13){\vector(1,0){9}}
\put(82,12){$\hat{Y}^n$}

\end{picture}
\caption{
One helper source coding system investigated by Wyner.
}
\label{fig:wyner}
\noindent
\end{figure}

A rate triple $(R_1,R_2,R_3)$ is $\vep$-{\it achievable} if 
for any ${\delta}>0$, there exist a positve interger 
$n_0=n_0(\vep,\delta)$ and a sequence of three encoders 
and two decoders functions 
$\{(\varphi_1^{(n)},\varphi_2^{(n)},$ $\varphi_3^{(n)},\psi^{(n)},$
$\phi^{(n)})\}_{n \geq n_0}$ such that for 
$n\geq n_0(\vep,\delta)$, 
\begin{align*}
& 
\frac{1}{n} \log || {\varphi_i^{(n)}}||\leq R_i+\delta 
\mbox{ for }i=1,2,3, 
\\
& {\rm P}_{\rm e}^{(n)}
(\varphi_1^{(n)},
 \varphi_2^{(n)},
 \varphi_3^{(n)},\psi^{(n)},\phi^{(n)})\leq \vep.
\end{align*}
The rate region ${\cal R}_{\rm W}(\vep |p_{XYZ})$ 
is defined by 
\begin{align*}
& {\cal R}_{\rm W}(\vep | p_{XYZ})\defeq \{(R_1,R_2,R_3):
\\
&(R_1,R_2,R_3)\mbox{ is }{\vep}
\mbox{-achievable for }p_{XYZ}\}.
\end{align*}
Furthermore, define
$$
{\cal R}_{\rm W}(p_{XYZ})
\defeq \bigcap_{\varepsilon\in (0,1)}{\cal R}_{\rm W}(\varepsilon| p_{XYZ}).
$$
We can show that the two rate regions 
${\cal R}_{\rm W }( \varepsilon |$ $p_{XYZ})$, 
$\varepsilon \in (0,1)$ and 
${\cal R}_{\rm W }(p_{XYZ})$ 
satisfy the following property. 
%===================================================================%
%===================================================================%
\begin{pr}\label{pr:pro0a}
$\quad$
\begin{itemize}
\item[a)] 
The regions 
${\cal R}_{\rm W }(\varepsilon|p_{XYZ})$, 
$\varepsilon \in(0,1)$, 
and ${\cal R}_{\rm W }($ $p_{XYZ})$
are closed convex sets of 
$\mathbb{R}_{+}^3$.
\item[b)] 
We set 
\begin{align*}
& {\cal R}_{\rm W}(n,\varepsilon |p_{XYZ})
=\{(R_1,R_2,R_3): %\mbox{ There exists }
\\
& \mbox{ There exists }
(\varphi_1^{(n)},\varphi_2^{(n)},\varphi_3^{(n)},\psi^{(n)})
\mbox{ such that }
\\
& \frac{1}{n}\log ||\varphi_i^{(n)}||\leq R_i, i=1,2,3 
\\
& 
{\rm P}_{\rm e}^{(n)}
(\varphi_1^{(n)},\varphi_2^{(n)},\varphi_3^{(n)},\psi^{(n)})
\leq \varepsilon\},
\end{align*} 
which is called the $(n,\vep)$-rate region. 
Using ${\cal R}_{\rm W}(n,$ $\varepsilon| p_{XYZ})$, 
${\cal R}_{\rm W}(\varepsilon|p_{XYZ})$ can be 
expressed as 
\begin{align*}
%&&
{\cal R}_{\rm W}(\varepsilon | p_{XYZ})
%\\
&={\rm cl}\left(\bigcup_{m\geq 1}
\bigcap_{n \geq m}{\cal R}_{\rm W}(n,\varepsilon | p_{XYZ})
\right).
\end{align*}
%The rate pair $(R,\Delta)$ is an $(n,\vep)$-achievable %%%%%%%%%%%%%%%%%%%%%%%%%%
%rate pair if there exists a triple $(\varphi_1^{(n)},$ %%%%%%%%%%%%%%%%%%%%%%%%%%
%$\varphi_2^{(n)}, \psi^{(n)})$ such that %%%%%%%%%%%%%%%%%%%%%%%%%%%%%%%%%%%%%%%%   
%The set that consists of all $\varepsilon$-achievable rate pair %%%%%%%%%%%%%%%%%
%is denoted by ${\cal C}_{\rm GMAC}(n, \varepsilon,P_1,P_2|W)$, which is called %%
%the $(n, \varepsilon)$-capacity region of the GMAC. %%%%%%%%%%%%%%%%%%%%%%%%%%%%%
%%%%%%%%%%%%%%%%%%%%%%%%%%%%%%%%%%%%%%%%%%%%%%%%%%%%%%%%%%%%%%%%%%%%%%%%%%%%%%%%%%
\end{itemize}
\end{pr}

It is well known that ${\cal R}_{\rm W}(p_{XYZ})$ 
was determined by Wyner. To describe his result 
we introduce an auxiliary random variable 
$U$ taking values in a finite set ${\cal U}$.  
%%%%%%%%%%%%%%%%%%%%%%%%%%%%%%%%%%%%%%%%%%%%%%%%%%%%%%%%%%%%%%%%%%%%%%%%%%%
We assume that the joint distribution of $(U,X,Y,Z)$ is 
$$ 
p_{U{X}{Y}}(u,x,y,z)=p_{U}(u)
p_{{X}|U}(x|u)p_{YZ|X}(y,z|x).
$$
The above condition is equivalent to $U \markov X \leftrightarrow YZ$. 
Define the set of probability distribution on 
${\cal U}$ 
$\times {\cal X}$ 
$\times {\cal Y}$ 
$\times {\cal Z}$ 
by
\begin{align*}
{\cal P}(p_{XYZ})
\defeq &
\{p=p_{UXYZ}: \pa {\cal U} \pa \leq \pa {\cal X} \pa+2,
\\
 & U \markov X\markov YZ \}.
\end{align*}
Set 
\begin{align*}
{\cal R}(p)
&\defeq 
\ba[t]{l}
\{(R_1,R_2,R_3): R_1,R_2,R_3 \geq 0,
\vSpa\\
\ba{rcl}
R_1 &\geq & I_p({X};{U}), R_2  \geq H_p({Y}|{U}), 
\\
R_3 &\geq & H_p({Z}|{U})\},
\ea
\ea
\\
{\cal R}(p_{XYZ})&\defeq \bigcup_{p \in {\cal P}(p_{XYZ})}
{\cal R}(p).
\end{align*}
We can show that the region ${\cal R}(p_{XYZ})$ satisfies 
the following property.
\begin{pr}\label{pr:pro0zxxx}  
$\quad$
\begin{itemize}
\item[a)] 
The region ${\cal R}(p_{XYZ})$ is a closed convex 
subset of $\mathbb{R}_{+}^3$.
%%%%%%%%%%%%%%%%%%%%%%%%%%%%%%%%%%%%%%%%%%%%%%%%%%%%%%%%%%%%%%%%%%
%, where%%%%%%%%%%%%%%%%%%%%%%%%%%%%%%%%%%%%%%%%%%%%%%%%%%%%%%%%%%
%\begin{align*}%%%%%%%%%%%%%%%%%%%%%%%%%%%%%%%%%%%%%%%%%%%%%%%%%%%%%%%%%%%
%\mathbb{R}_{+}^2&\defeq &\{(R_1,R_2): R_1 \geq 0,R_2 \geq 0\}.%%%
%\end{align*}%%%%%%%%%%%%%%%%%%%%%%%%%%%%%%%%%%%%%%%%%%%%%%%%%%%%%%%%%%%
%%%%%%%%%%%%%%%%%%%%%%%%%%%%%%%%%%%%%%%%%%%%%%%%%%%%%%%%%%%%%%%%%%
\item[b)] For any $p_{XYZ}$, and any $\gamma \in [0,1]$, 
we have 
\begin{align}
&\min_{(R_1,R_2,R_3)\in {\cal R}(p_{XY})}
(R_1+\bar{\gamma} R_2+{\gamma}R_3)
\notag\\
& =\bar{\gamma} H_p(Y) + {\gamma} H_p(Z).
\label{eqn:SdEEExx}
\end{align}
The minimun 
is attained by $(R_1,R_2,R_3)=(0,H_p(Y),$ $H_p(Z))$. 
This result implies that 
\begin{align*}
& {\cal R}(p_{XYZ}) 
%\\
%& 
\subseteq 
\ba[t]{l}
\Biggl[\ds \bigcap_{\gamma\in [0,1]}\{
(R_1,R_2,R_3):R_1+\bar{\gamma} R_2+ \gamma R_3
\\ 
\quad \geq \bar{\gamma} H_p(Y)+\gamma H_p(Z)\}
\Biggr] \cap \mathbb{R}_{+}^3.
\ea
\end{align*} 
Furthermore, the point $(0,H_p(Y),H_p(Z))$ always belongs to 
${\cal R}(p_{XYZ})$. 
%In general, the lower boundary of 
%${\cal R}(p_{XYZ})$ contains a plane segment with normal vector 
%$(1,\bar{\gamma}, \gamma)$ going 
%through the point $(0,H_p(Y),H_p(Z))$ but 
%plane segment may reduce to the point $(0, H_p(Y),H_p(Z))$.
\end{itemize}
\end{pr}

The rate region ${\cal R}_{\rm W}(p_{XYZ})$ was determined by 
Wyner \cite{w0}. His result is the following.
\begin{Th}[Wyner \cite{w0}]
\label{th:akw}
\begin{align*}
& {\cal R}_{\rm W}(p_{XYZ})={\cal R}(p_{XYZ}).
\end{align*}
\end{Th}

On the strong converse theorem Csisz\'ar and K\"orner \cite{ckBook81}
obtained the following. 
\begin{Th}[Csisz\'ar and K\"orner \cite{ckBook81}]
\label{th:ckBook81}
For each fixed $\varepsilon $ $\in (0,1)$, 
we have 
\begin{align*}
& &{\cal R}_{\rm W}(\varepsilon | p_{XYZ})={\cal R}(p_{XYZ}).
\end{align*}
\end{Th}

To examine a rate of convergence for the error probability 
of decoding to tend to one as $n \to \infty$ for $(R_1,R_2,R_3)$ 
$\notin {\cal R}_{\rm W}(p_{XYZ})$, we define the following quantity. Set
\begin{align*}
&  {\rm P}_{\rm c}^{(n)}
(\varphi_1^{(n)},\varphi_2^{(n)},\varphi_3^{(n)},\psi^{(n)},\phi^{(n)})
\\
&\defeq  
1-{\rm P}_{\rm e}^{(n)}
(\varphi_1^{(n)},\varphi_2^{(n)},\varphi_3^{(n)},\psi^{(n)},\phi^{(n)}),
\\
&  
G^{(n)}(R_1,R_2,R_3|p_{XYZ})
\\
&\defeq 
\hspace*{-1mm}
\min_{\scs 
(\varphi_1^{(n)},\varphi_2^{(n)},\varphi_3^{(n)},
     \atop{\scs 
     \psi^{(n)}, \phi^{(n)}):
          \atop{\scs 
          (1/n)\log \|\varphi_i^{(n)}\|
            \atop{\scs \leq R_i,i=1,2,3
            } 
    }
}
}
\hspace*{-2mm}
\left(-\frac{1}{n}\right)
\log {\rm P}_{\rm c}^{(n)}
(\varphi_1^{(n)}
\hspace*{-1mm},
\varphi_2^{(n)}
\hspace*{-1mm},
\varphi_3^{(n)}
\hspace*{-1mm},
\psi^{(n)}
\hspace*{-1mm}, 
\phi^{(n)}),
\\
& 
G(R_1,R_2,R_3|p_{XYZ}) 
\defeq \lim_{n \to \infty}G^{(n)}(R_1,R_2,R_3|p_{XYZ}),
\\
& {\cal G}(p_{XYZ}) 
\\
&  \defeq \{(R_1,R_2,R_3,G): 
G\geq G(R_1,R_2,R_3|p_{XYZ})\}. 
\end{align*}
By time sharing we have that 
\begin{align}
& G^{(n+m)}\left(\left.
{\ts \frac{n R_1+m R_1^{\prime}}{n+m}},
{\ts \frac{n R_2+m R_2^{\prime}}{n+m}},
{\ts \frac{n R_2+m R_2^{\prime}}{n+m}}
\right|p_{XYZ}\right) 
\nonumber\\
&\leq \frac{nG^{(n)}(R_1,R_2,R_3|p_{XYZ}) 
+mG^{(m)}(R_1^{\prime},R_2^{\prime},R_3^{\prime}
|p_{XYZ})}{n+m}.
\nonumber\\
\label{eqn:aaZx} 
\end{align}
Choosing $R=R^\prime$ in (\ref{eqn:aaZx}), we 
obtain the following subadditivity property
on $\{G^{(n)}(R_1,R_2,R_3|p_{XYZ})$ 
$\}_{n\geq 1}$: 
\begin{align*}
& G^{(n+m)}(R_1,R_2,R_3|p_{XYZ}) 
\\
&\leq \frac{nG^{(n)}(R_1,R_2,R_3|p_{XYZ}) 
+mG^{(m)}(R_1,R_2,R_3|p_{XYZ})}{n+m},
\end{align*}
from which we have that $G(R_{1},R_2,R_3|p_{XYZ})$ exists and 
satisfies the following:  
\begin{align*}
&  G(R_{1},R_2,R_3|p_{XYZ}) 
\\
&=\inf_{n\geq 1}G^{(n)}(R_1,R_2,R_3|p_{XYZ}).
\end{align*}
The exponent function $G(R_1,R_2,R_3|p_{XYZ})$ is a convex function 
of $(R_1,R_2,R_3)$. In fact, by time sharing we have that 
\begin{align*}
& G^{(n+m)}\left(\left.
{\ts \frac{n R_1+m R_1^{\prime}}{n+m}},
{\ts \frac{n R_2+m R_2^{\prime}}{n+m}},
{\ts \frac{n R_2+m R_2^{\prime}}{n+m}}
\right|p_{XYZ}\right) 
\\
&\leq \frac{nG^{(n)}(R_1,R_2,R_3|p_{XYZ}) 
+mG^{(m)}(R_1^{\prime},R_2^{\prime},R_3^{\prime}
|p_{XYZ})}{n+m}, 
\end{align*}
from which we have that for any $\alpha \in [0,1]$
\begin{align*}
& G(\alpha R_1+\bar{\alpha}R_1^{\prime},
     \alpha R_2+\bar{\alpha}R_2^{\prime},
     \alpha R_3+\bar{\alpha}R_3^{\prime}|p_{XYZ})
\\
&\leq 
\alpha G(R_1,R_2,R_3|p_{XYZ})
+\bar{\alpha} G( R_1^{\prime},R_2^{\prime},R_3^{\prime}|p_{XYZ}).
\end{align*}
The region ${\cal G}(p_{XYZ})$ is also a closed convex set. 
Our main aim is to find an explicit characterization of 
${\cal G}(p_{XYZ})$. In 
this paper we derive an explicit outer bound of ${\cal G}$ 
$(p_{XYZ})$ whose section by the plane $G=0$ coincides 
with ${\cal R}_{\rm W}(p_{XYZ})$.
%\section{Main Result}
%Let ${\cal U}$ be a finite set and let $U\in {\cal U}$
%be an auxiliary random variable. We assume that 
%$U \markov X \leftrightarrow Y$. 
%
%-------------------------------------------------------------%
%
%In this section we state our main result. 
We first  explain that the region ${\cal R}(p_{XYZ})$ has another
expression using the supporting hyperplane. We define two 
sets of probability distributions 
on ${\cal U}$ 
$\times{\cal X}$ 
$\times{\cal Y}$ 
$\times{\cal Z}$ 
by
\begin{align*}
&{\cal P}_{\rm sh}(p_{XYZ})
\defeq 
\{p=p_{UXYZ}: \pa {\cal U} \pa \leq \pa {\cal X} \pa,
\\ 
& \quad U \markov  X\markov YZ \},
\\
&{\cal Q}(p_{YZ|X})
\defeq 
\{q=q_{UXYZ}: 
\pa {\cal U} \pa \leq \pa {\cal X} \pa,
\\
&\quad p_{YZ|X}=q_{YZ|X},
{U} \markov X \markov YZ \}.
\end{align*}
For $(\mu,\gamma) \in[0,1]^2$, set
\begin{align*}
& R^{(\mu,\gamma)}(p_{XYZ})
\defeq 
\ba[t]{l} \ds 
\max_{p \in {\cal P}_{\rm sh}(p_{XYZ})}
\left\{ {\prmtA} I_p(X;U) \right.
\vspace*{1mm}\\
\left. +{\prmtB}(\bar{\gamma} H_p(Y|U)+ \gamma H_p(Z|U))\right\}.
\ea
\end{align*}
Furthermore, define
\begin{align*}
& \underline{\cal R}_{\rm sh}(p_{XYZ})
\\
&=\bigcap_{(\mu,\gamma)\in [0,1]^2}
\ba[t]{l}
\{(R_1,R_2,R_3): {\prmtA} R_1+ {\prmtB}(\bar{\gamma} R_2+\gamma R_3) 
\\ \geq R^{(\mu,\gamma)}(p_{XYZ})\},
\ea
\\
& {\cal R}_{\rm sh}(p_{XYZ})
\\
&=\bigcap_{(\mu,\gamma)\in [0,1]^2}
\ba[t]{l} 
\{(R_1,R_2,R_3):{\prmtA}R_1+ {\prmtB}(\bar{\gamma} R_2+\gamma R_3)
\\
\geq R^{(\mu,\gamma)}(p_{XYZ})\}.
\ea
\end{align*}
Then we have the following property.
\begin{pr}\label{pr:pro4z} $\quad$
\begin{itemize}
\item[a)]
The bound $|{\cal U}|\leq |{\cal X}|$ is sufficient 
to describe ${R}^{(\mu)}($ $p_{XYZ})$.

\item[b)] For every $(\mu,\gamma)\in [0,1]^2$, 
we have 
\begin{align*}
&\min_{(R_1,R_2,R_3)\in {\cal R}(p_{XYZ})}
\{\mu R_1+\prmtB(\bar{\gamma} R_2+\gamma R_3)\} 
\notag\\
& =R^{(\mu,\gamma)}(p_{XYZ}).
\end{align*}

%\item[c)] For every $\mu \geq 1$ and 
%$\gamma \in [0,1]$, we have 
%$$
%R^{(\mu,\gamma)}(p_{XYZ})=\bar{\gamma} H_p(Y)+\gamma H_p(Z).
%$$
\item[c)] 
For any $p_{XYZ}$ we have
\beq
 %\underline{\cal R}_{\rm sh}(p_{XYZ})
%=
{\cal R}_{\rm sh}(p_{XYZ})={\cal R}(p_{XYZ}).
\label{eqn:PropEqBz}
\eeq
\end{itemize}
\end{pr}

%\begin{pr}\label{pr:pro4}  
%$\quad$
%$$
%{\cal R}_{\rm sh}(p_{XYZ})
%=\tilde{\cal R}_{\rm sh}(p_{XYZ})={\cal R}(p_{XYZ}).
%$$
%\end{pr}
%Since the proof of Property \ref{pr:pro4} is quite similar to that 
%of Property \ref{pr:pro0z}, we omit it. 
%%%%%%%%%%%%%%%%%%%%%%%%%%%%%%%%%%%%%%%%%%%%%%%%%%%%%%%%%%%%%%%%%%%%%%%%%%
%given in Appendix \ref{sub:ApdaAAB}. 
%On the cardinality bound of the auxiliary random variable $U$, 
%Property \ref{pr:pro0z} part c) seems to be new. This property is proved 
%via the hyperplane expression of the region ${\cal R}(p_{UXY})$. 
%When the strong converse theorem holds, we are interested 
%%%%%%%%%%%%%%%%%%%%%%%%%%%%%%%%%%%%%%%%%%%%%%%%%%%%%%%%%%%%%%%%%%%%%%%%%
For $(\mu,\gamma,\alpha) \in [0,1]^3$, 
and for $q=q_{UXYZ}\in {\cal Q}(p_{YZ|X})$, define 
\begin{align*}
& \omega_{q|p_X}^{(\mu,\gamma,\alpha)}(x,y,z|u)
\defeq \bar{\alpha}\log \frac{q_{X}(x)}{p_{X}(x)}
+ \alpha\left[{\prmtA}\log \frac{q_{X|U}(x|u)}{p_{X}(x)}\right.
\\
&\quad \left. +{\prmtB}\left( \bar{\gamma}\log\frac{1}{q_{Y|U}(y|u)}
                           +{\gamma}\log\frac{1}{q_{Z|U}(z|u)}
         \right)\right],
\\
& f^{(\mu,\gamma,{}\alpha)}_{q|p_X}(x,y,z|u)
  \defeq \exp\left\{-
\omega^{(\mu,\gamma,\alpha)}_{q|p_X}(x,y,z|u)
\right\},
\\
& \Omega^{(\mu,\gamma,{}\alpha)}(q|p_X)
%\\
%& 
\defeq -\log 
{\rm E}_{q}
\left[
f^{(\mu,\gamma,{}\alpha)}_{q|p_X}(X,Y,Z|U)
%\exp\left\{-\alpha
%\omega^{(\mu,\gamma,\beta)}_{q |p_X}(X,Y,Z|U)\right\}
\right],
\\
&\Omega^{(\mu,\gamma,{}\alpha)}(p_{XYZ})
%\\
%&
\defeq 
%&
\min_{\scs 
   \atop{\scs 
    q \in {{\cal Q}}(p_{YZ|X})
   }
}
\Omega^{(\mu,\gamma,{}\alpha)}(q|p_X),
\\
& F^{(\mu,\gamma,{}\alpha)}(
{\prmtA}R_1+ \bar{\gamma} R_2+\gamma R_3)
\\
&\defeq
\frac{\Omega^{(\mu,\gamma, {}\alpha)}(p_{XYZ})
-\alpha[{\prmtA}R_1+\prmtB(\bar{\gamma} R_2 +\gamma R_3)]}
{2+\alpha{\prmtB}},
\\
& F(R_1,R_2,R_3|p_{XYZ})
\\
& \defeq \sup_{\scs (\mu,\gamma,\alpha)\in [0,1]^3,
%\atop{\scs {}\alpha \geq 0%,
            %\alpha \geq 0
%            }
      }
F^{(\mu,\gamma,{}\alpha)}({\prmtA}R_1
+{\prmtB}( \bar{\gamma}R_2+ \gamma R_3) |p_{XYZ}).
\end{align*}
We next define a function serving as a lower 
bound of $F(R_1,R_2,R_3|p_{XYZ})$. For each 
$p=p_{UXYZ}\in {\cal P}_{\rm sh}(p_{XYZ})$, define
\begin{align*}
& \tilde{\omega}_{p}^{(\mu,\gamma)}(x,y,z|u)
\defeq {\prmtA}
  \log \frac{p_{X|U}(x|u)}{p_{X}(x)}
\\
&\quad +{\prmtB}\left(\bar{\gamma}\log \frac{1}{p_{Y|U}(y|u)} 
                + {\gamma}\log \frac{1}{p_{Z|U}(z|u)}\right),
\\
& \tilde{\Omega}^{(\mu,\gamma,\lambda)}(p)
\defeq 
-\log 
{\rm E}_{p}
\left[\exp\left\{-\lambda
\omega^{(\mu,\gamma)}_p (X,Y,Z|U)\right\}\right].
\end{align*}
Furthermore, set
\begin{align*}
& \tilde{\Omega}^{(\mu,\gamma,\lambda)}(p_{XYZ})
 \defeq \min_{\scs \atop{\scs 
p \in {{\cal P}_{\rm sh}(p_{XYZ})}}}
\tilde{\Omega}^{(\mu,\gamma,\lambda)}(p),
\\
& {\loF}^{(\mu,\gamma,\lambda)}(
{\prmtA}R_1+ \bar{\gamma} R_2 +\gamma R_3|p_{XYZ}) 
\\
&\defeq 
\frac{\tilde{\Omega}^{(\mu,\gamma,\lambda)}(p_{XYZ})
-\lambda[{\prmtA} R_1 + \prmtB(\bar{\gamma} R_2 +\gamma R_3)]}
{2+\lambda(5-{\prmtA})},
\\
& {\loF}(R_1,R_2,R_3|p_{XYZ})
\\
&\defeq \sup_{\scs (\mu,\gamma) \in [0,1]^2,
\atop{\scs \lambda\geq 0}
} 
{\loF}^{(\mu,\gamma,\lambda)}({\prmtA}R_1
+{\prmtB}\bar{\gamma} R_2+\gamma R_3|p_{XYZ}).
\end{align*}

We can show that the above functions and sets satisfy 
the following property. 
\begin{pr}\label{pr:pro1bz}  
$\quad$
\begin{itemize}

\item[a)] 
The cardinality bound $|{\cal U}|\leq |{\cal X}|$ 
in ${\cal Q}(p_{Y|X})$ is sufficient to describe the quantity
$\Omega^{(\mu,{}\alpha)}(p_{XY})$. Furthermore, the cardinality 
bound $|{\cal U}|\leq |{\cal X}|$ in ${\cal Q}(p_{YZ|X})$ is sufficient 
to describe the quantity $\tilde{\Omega}^{(\mu,\gamma,\lambda)}(p_{XYZ})$. 

\item[b)] For any $R_1,R_2,R_3\geq 0$, we have 
\begin{align*}
& F(R_1,R_2,R_3|p_{XYZ})\geq {\loF}(R_1,R_2,R_3|p_{XYZ}).
\end{align*}

\item[c)] For any $p=p_{UXY} \in {\cal P}_{\rm sh}(p_{XY})$
and any $(\mu,\gamma, \lambda$$) \in [0,$ $1]^3$, we have 
\beq
0\leq \tilde{\Omega}^{(\mu,\gamma, \lambda)}(p) 
 \leq \prmtA \log |{\cal X}|
    + \prmtB\log(|{\cal Y}|^{\bar{\gamma}} 
                  |{\cal Z}|^{{\gamma}}).
\label{eqn:Asddxzz}
\eeq

\item[d)] 
Fix any $p=p_{UXYZ}\in {\cal P}_{\rm sh}(p_{XYZ})$ and 
$(\mu,\gamma)\in [0,1]^2$.
We define a probability distribution 
$p^{(\lambda)}=p_{UXYZ}^{(\lambda)}$
by
\begin{align*} 
& p^{(\lambda)}(u,x,y,z)
\\
&\defeq
\frac{
p(u,x,y,z)
\exp\left\{-\lambda
\omega^{(\mu,\gamma)}_{p}(x,y,z|u)\right\}
}{
{\rm E}_{p}
\left[\exp\left\{-\lambda 
\omega^{(\mu,\gamma)}_{p}(X,Y,Z|U)\right\}\right]}.
\end{align*}
\irr{Then for $\lambda \in [0,1/2]$, 
$\tilde{\Omega}^{(\mu,\gamma, \lambda)}(p)$ is twice differentiable.
Furthermore, for $\lambda \in [0,1/2]$, we have}
\begin{align*}
&  \frac{\rm d}{{\rm d}\lambda} 
\tilde{\Omega}^{(\mu,\gamma,\lambda)}(p)
\\
& \qquad ={\rm E}_{p^{(\lambda)}}
\left[\omega^{(\mu,\gamma)}_{p}(X,Y,Z|U)\right],
\\
&  \frac{\rm d^2}{{\rm d}\lambda^2} 
\tilde{\Omega}^{(\mu,\gamma,\lambda)}(p)
\\
& \qquad =-{\rm Var}_{p^{(\lambda)}}
\left[\omega^{(\mu,\gamma)}_p(X,Y,Z|U)\right].
\end{align*}
The second equality implies that 
$\tilde{\Omega}^{(\mu,\gamma,\lambda)}(p)$ is a concave function 
of $\lambda\in [0,1/2]$. 

\item[e)]
For $(\mu,\gamma,\lambda)\in [0,1]^2\times [0,1/2]$, define 
\begin{align*}
& \rho^{(\mu,\gamma,\lambda)}(p_{XYZ})
\\
&\defeq \irr{\max_{\scs (\nu, p) \in [0,\lambda] 
    \atop{\scs \times {\cal P}_{\rm sh}(p_{XYZ}):
        \atop{\scs \tilde{\Omega}^{(\mu,\gamma,\lambda)}(p) 
             \atop{\scs 
             =\tilde{\Omega}^{(\mu,\gamma,\lambda)}(p_{XYZ})
             }}}}
}
{\rm Var}_{\irr{p^{(\nu)}}}
\left[\tilde{\omega}^{(\mu,\gamma)}_{p}(X,Y,Z|U)\right],
\end{align*}
and set
\begin{align*}
& \rho =\rho(p_{XYZ})
\defeq \max_{(\mu,\gamma,\lambda)\in [0,1]^2\times [0,1/2]}
\rho^{(\mu,\gamma,\lambda)}(p_{XYZ}).
\end{align*}
Then we have $\rho(p_{XYZ})<\infty $. 
\irr{Furthermore, for any $(\mu,\gamma,\lambda)$ 
$\in [0,1]^2\times [0,1/2]$, we have
$$
\tilde{\Omega}^{(\mu,\gamma,\lambda)}(p_{XYZ}) 
\geq \lambda R^{(\mu,\gamma)}(p_{XYZ})
-\frac{\lambda^2}{2}\rho(p_{XYZ}).
$$
}

\item[f)] For every $\tau \in (0,(1/2)\rho(p_{XYZ}))$,
the condition $(R_1+\tau,$ $R_2+\tau, R_3+\tau) 
\notin {\cal R}(p_{XYZ})$
implies 
\begin{align*}
& F(R_1,R_2,R_3|p_{XYZ})
\notag\\
& > \frac{\rho(p_{XYZ})}{4} \cdot g^2
\left( \frac{\tau}{\rho(p_{XYZ})}\right)>0,
\end{align*}
\end{itemize}
\end{pr}

Since proofs of the results stated in Property \ref{pr:pro1bz} are 
quite parallel with those of the results stated in 
Property \ref{pr:pro1}, we omit them.
Our main result is the following.
\begin{Th}\label{Th:main2}
For any $R_1,R_2,$ $R_3\geq 0$, any $p_{XYZ}$, and 
for any $(\varphi^{(n)}_1,$ $\varphi^{(n)}_2,$ $\varphi^{(n)}_3,$ 
$\psi^{(n)},\phi^{(n)})$ satisfying 
$
(1/n)\log ||\varphi_i^{(n)}||$ $\leq R_i,i=1,2,3, 
$
we have 
\begin{align*}
& {\rm P}_{\rm c}^{(n)}(\varphi_1^{(n)},
\varphi_2^{(n)},\varphi_3^{(n)},\psi^{(n)},\phi^{(n)})
\\
&\leq 7\exp \left\{-n F(R_1,R_2,R_3|p_{XYZ})\right\}.
\end{align*}
\end{Th}

It follows from Theorem \ref{Th:main2} and 
Property \ref{pr:pro1bz} part d) that if $(R_1,R_2,R_3)$ 
is outside the capacity region, then the error 
probability of decoding goes to one exponentially 
and its exponent is not below $F(R_1,R_2,R_3|p_{XYZ})$. 
It immediately follows from Theorem \ref{Th:main} 
that we have the following corollary. 
\begin{co}\label{co:mainCo2}
\begin{align*}
& G(R_1,R_2,R_3|p_{XYZ})\geq F(R_1,R_2,R_3|p_{XYZ}),
\\
& {\cal G}(p_{XYZ})
\subseteq 
\overline{\cal G}(p_{XYZ})
\\
& =\left\{(R_1,R_2,R_3,G):
G\geq 
F(R_1,R_2,R_3|p_{XYZ})
\right\}.
\end{align*}
\end{co}

Proof of Theorem \ref{Th:main2} will be given in the next section. 
The exponent function at rates outside the rate region was 
derived by Oohama and Han \cite{OhHan94} for the separate source coding 
problem for correlated sources \cite{sw}. The techniques used by them is 
a method of types \cite{ckBook81}, which is not useful to prove Theorem 
\ref{Th:main}. Some novel techniques based on the information spectrum 
method introduced by Han \cite{han} are necessary to prove this theorem.

%%%%%%%%%%%%%%%%%%%%%%%%%%%%%%%%%%%%%%%%%%%%%%%%%%%%%%%%%%%%%%%%%%%%%%%%%%
%established by 
%Ahlswede {\it et al.} \cite{agk76} 
%immediately follows from this corollary. 
%To discribe this outer bound, 
%for $\kappa>0$, we set 
%\begin{align*}
%& & {\cal R}(p_{XYZ})-\kappa(0,1)
%\\
%& \defeq & \{(R_1-\kappa, R_2-\kappa): (R_1,R_2) 
%\in {\cal R}(p_{XYZ})\}, 
%\end{align*}
%which serves as an outer bound of ${\cal R}(p_{XYZ})$. 
%From Theorem \ref{Th:main2} and Property \ref{pr:pro1bz} part e) 
%we have the following corollary.
%The strong converse theorem established by Ahlswede {\it et 
%al.} \cite{agk76} immediately follows from this corollary.
%we have the following corollary, which provides an explicit 
%outer bound of ${\cal R}_{\rm W}(\varepsilon |p_{XYZ})$ 
%with an asymptotically vanishing deviation from 
%${\cal R}_{\rm W}(p_{XYZ})$ $={\cal R}(p_{XYZ})$. 
%The strong converse theorem 
%%%%%%%%%%%%%%%%%%%%%%%%%%%%%%%%%%%%%%%%%%%%%%%%%%%%%%%%%%%%%%%%%%%%%%%%%%%%

From Theorem \ref{Th:main2} and Property \ref{pr:pro1bz} part e), 
we can obtain an explicit outer bound of 
${\cal R}_{\rm W}(\varepsilon |p_{XYZ})$ 
with an asymptotically vanishing deviation from 
${\cal R}_{\rm W}(p_{XYZ})$ $={\cal R}(p_{XYZ})$. 
The strong converse theorem established by 
Csisz\'ar and K\"orner \cite{ckBook81} 
immediately follows from this corollary. 
%%%%%%%%%%%%%%%%%%%%%%%%%%%%%%%%%%%%%%%%%%%%%%%%%%%%%%%%%%%%%%%%%%%%%%%%%%
To discribe this outer bound, for $\kappa>0$, we set 
\begin{align*}
&  {\cal R}(p_{XYZ})-\kappa(1,1,1)
\\
& \defeq  \{(R_1-\kappa, R_2-\kappa, R_3-\kappa): (R_1,R_2,R_3) 
\in {\cal R}(p_{XYZ})\}, 
\end{align*}
which serves as an outer bound of ${\cal R}(p_{XYZ})$. 
%%%%%%%%%%%%%%%%%%%%%%%%%%%%%%%%%%%%%%%%%%%%%%%%%%%%%%%%%%%%%%%%%%%%%%%%%%
%%%%%%%%%%%%%%%%%%%%%%%%%%%%%%%%%%%%%%%%%%%%%%%%%%%%%%%%%%%%%%%%%%%%%%%%%%
For each fixed $\varepsilon\in(0,1)$, we define  
$\tilde{\kappa}_n$$=\tilde{\kappa}_n(\varepsilon,\rho(p_{XYZ}))$ by
\beqa
\tilde{\kappa}_n&\defeq&
\rho(p_{XY}) \vartheta\left(
\sqrt{ \frac{4}{n\rho(p_{XY})} 
\log\left(\frac{7}{1-\varepsilon}\right)} 
\right)
\label{eqn:zddZZ}\\
&\MEq{a}&
2\sqrt{%\ts
\frac{\rho(p_{XY})}{n}
\log\left(\frac{7}{1-\varepsilon}\right)}
+ \frac{5}{n}
\log\left(\frac{7}{1-\varepsilon}\right).
\nonumber
\eeqa
Step (a) follows from $\vartheta(a)=a+(5/4)a^2$.  
Since $\tilde{\kappa}_n \to 0$ as $n\to \infty$, we have the 
smallest positive integer $n_1=n_1(\varepsilon,\rho(p_{XYZ}))$
such that $\tilde{\kappa}_n \leq \irr{(1/2)}\rho(p_{XYZ})$ 
for $n\geq n_1$.
From Theorem \ref{Th:main2} and Property \ref{pr:pro1bz} 
part e), we have the following corollary.
\begin{co}
\label{co:StConv2}
For each fixed $\varepsilon \in (0,1)$, 
we choose the above positive integer 
$n_1=$$n_1(\varepsilon,\rho(p_{XYZ}))$. 
Then, for any $n\geq n_1$, we have

\begin{align*}
{\cal R}_{\rm W}(\varepsilon|p_{XYZ})
&\subseteq 
{\cal R}(p_{XYZ})-\tilde{\kappa}_n(0,1,1). 
\end{align*}
The above result together with 
\begin{align*}
%&&
{\cal R}_{\rm W}(\varepsilon | p_{XYZ})
%\\
&={\rm cl}\left(\bigcup_{m\geq 1}
\bigcap_{n \geq m}{\cal R}_{\rm W}(n,\varepsilon | p_{XYZ})
\right)
\end{align*}
yields that for each fixed $\varepsilon \in (0,1)$, we have 
\begin{align*}
& {\cal R}_{\rm W}(\varepsilon | p_{XYZ})
  ={\cal R}_{\rm W}(p_{XYZ})
  ={\cal R}(p_{XYZ}).
\end{align*}
This recovers the strong converse theorem proved by 
Csisz\'ar and K\"orner \cite{ckBook81}.
%Ahlswede {\it et al.} \cite{agk76}.
\end{co}

%%%%%%%%%%%%%%%%%%%%%%%%%%%%%%%%%%%%%%%%%%%%%%%%%%%%%%%%%%
\newcommand{\ZaSSS}{%%%%%%%%%%%%%%%%%%%%%%%%%%%%%%%%%%%%%%
%%%%%%%%%%%%%%%%%%%%%%%%%%%%%%%%%%%%%%%%%%%%%%%%%%%%%%%%%%
\beqa
\tilde{\kappa}_n
&\defeq&
2\sqrt{%\ts
\frac{\rho(p_{XYZ})}{n}
\log\left(\frac{7}{1-\varepsilon}\right)}
+%\ts
\frac{6}{n}
\log\left(\frac{7}{1-\varepsilon}\right).
\nonumber
\eeqa
Since $\tilde{\kappa}_n\to 0$ as $ n \to \infty$, 
there exists the smallest integer 
$n_1=n_1(\varepsilon,\rho(p_{XYZ}))$ such that 
$\kappa_n\leq (1/2)\rho(p_{XYZ})$ for $n\geq n_1$. 

Then we have that 
for any $\varepsilon$ $ \in (0,1)$ 
and any $n\geq n_1(\varepsilon,\rho(p_{XYZ}))$,
It immediately follows from the above result that
for any $\varepsilon \in (0,1)$, we have 
\begin{align*}
& &{\cal R}_{\rm W}(\varepsilon | p_{XYZ})
  ={\cal R}_{\rm W}(p_{XYZ})
  ={\cal R}(p_{XYZ}).
\end{align*}
\end{co}
}%%%%%%%%%%%%%%%%%%%%%%%%%%%%%%%%%%%%%%%%%%%%%%%%%%%%%%%%
%%%%%%%%%%%%%%%%%%%%%%%%%%%%%%%%%%%%%%%%%%%%%%%%%%%%%%%%%

Proof of this corollary is quite parallel with that of 
Corollary \ref{co:StConv}. We omit the detail.

\section*{\empty}
\appendix

\newcommand{\Eop}{%-----------------------------------------------%
Set
\begin{align*}
\hat{\cal P}_n(p_{XY})
&\defeq 
\{U: \pa {\cal U} \pa \leq 
\pa {\cal M}_1 \pa \pa {\cal X}^{n-1}\pa,
U \markov  X\markov Y \},
\\
\tilde{\cal P}_n(p_{UXY})
&\defeq 
\{(\tilde{U},\tilde{X},\tilde{Y}): 
\pa {\cal U} \pa \leq 
\pa {\cal M}_1 \pa \pa {\cal X}^{n-1}\pa,
\\
&  p_{YX|U}=p_{\tilde{Y}\tilde{X}|\tilde{U}},
\tilde{U} \markov \tilde{X} \markov \tilde{Y} \},
\\
\hat{\Omega}_n^{(\mu,\lambda,\alpha)}(p_{XY})
&\defeq&
\min_{\scs 
U\in {{\cal P}}^{*}(p_{XY}),
    \atop{\scs 
     (\tilde{U},\tilde{X},\tilde{Y})\in \tilde{{\cal P}}(p_{UXY})
     }
}
\Omega^{(\mu,\lambda,\alpha)}(X,\tilde{X}\tilde{Y}|\tilde{Z})
\end{align*}
}%=============================================================%

\ApdaAaaa 
\ApdaAAA
\ApdaAAAb
\ApdaAABz

\ApdaAACa
\ApdaAAC

\Apda
\Apdb
\Apdc

\noindent
{\bf Acknowledgements}

We are very grateful to 
Dr. Shun Watanabe and 
Dr. Shigeaki Kuzuoka for their 
helpful comments.

\end{document}